\pdfoutput=1
\documentclass[twocolumn]{aastex63}

\usepackage{amsmath}
\usepackage{graphicx}
\usepackage{url}
\usepackage{tabularx}
\usepackage{footmisc}
\usepackage{verbatim}
\usepackage{appendix}
\usepackage{enumerate}

\graphicspath{{figures/}} 

\newcommand{\Kepler}{{Kepler}}

\submitjournal{The Astronomical Journal}
\accepted{2020 October 20}

\shorttitle{Habitable Occurrence from \Kepler}
\shortauthors{Bryson et al.}

\begin{document}

\title{The Occurrence of Rocky Habitable Zone Planets Around Solar-Like Stars from \Kepler\ Data}

\correspondingauthor{Steve Bryson}
\email{steve.bryson@nasa.gov}

\author[0000-0003-0081-1797]{Steve Bryson}
\affiliation{NASA Ames Research Center, Moffett Field, CA 94035, USA}

\author[0000-0001-9269-8060]{Michelle Kunimoto}
\affiliation{Department of Physics and Astronomy, University of British Columbia, 6224 Agricultural Road, Vancouver, BC V6T 1Z1, Canada}

\author[0000-0002-5893-2471]{Ravi K. Kopparapu}
\affiliation{NASA Goddard Space Flight Center, Greenbelt, MD, USA}

\author[0000-0003-1634-9672]{Jeffrey L. Coughlin}
\affiliation{NASA Ames Research Center, Moffett Field, CA 94035, USA}
\affiliation{SETI Institute, 189 Bernardo Ave, Suite 200, Mountain View, CA 94043, USA}

\author{William J. Borucki}
\affiliation{NASA Ames Research Center, Moffett Field, CA 94035, USA}

\author{David Koch}
\altaffiliation{deceased}
\affiliation{NASA Ames Research Center, Moffett Field, CA 94035, USA}

\author[0000-0002-6137-903X]{Victor Silva Aguirre}
\affiliation{Stellar Astrophysics Centre (SAC), Department of Physics and Astronomy, Aarhus University, Ny Munkegade 120, DK-8000 Aarhus C, Denmark}

\author{Christopher Allen}
\affiliation{Orbital Sciences Corporation, NASA Ames Research Center, Moffett Field, CA 94035, USA}

\author[0000-0002-3306-3484]{Geert Barentsen}
\affiliation{Bay Area Environmental Research Institute, 625 2nd St., Ste 209, Petaluma, CA 94952, USA}

\author{Natalie. M. Batalha}
\affiliation{University of California Santa Cruz, Santa Cruz, CA, USA}

\author[0000-0002-2580-3614]{Travis Berger}
\affiliation{Institute for Astronomy, University of Hawai`i, 2680 Woodlawn Drive, Honolulu, HI 96822, USA}

\author{Alan Boss}
\affiliation{Earth \& Planets Laboratory, Carnegie Institution for Science, USA}

\author[0000-0003-1605-5666]{Lars A. Buchhave}
\affiliation{DTU Space, National Space Institute, Technical University of Denmark, Elektrovej 328, DK-2800 Kgs. Lyngby, Denmark}

\author[0000-0002-7754-9486]{Christopher~J.~Burke}
\affiliation{Department of Physics and Kavli Institute for Astrophysics and Space Research, Massachusetts Institute of Technology, Cambridge, MA 02139, USA}

\author{Douglas A. Caldwell}
\affiliation{NASA Ames Research Center, Moffett Field, CA 94035, USA}
\affiliation{SETI Institute, 189 Bernardo Ave, Suite 200, Mountain View, CA 94043, USA}

\author{Jennifer R. Campbell}
\affiliation{NASA Ames Research Center, Moffett Field, CA 94035, USA}

\author[0000-0001-7980-4658]{Joseph Catanzarite}
\affiliation{Cross-Entropy Consulting, USA}

\author[0000-0002-9760-7735]{Hema Chandrasekaran}
\affiliation{Computational Engineering Division, Lawrence Livermore National Laboratory L-261 Livermore, CA-94550, USA}

\author[0000-0002-5714-8618]{William J. Chaplin}
\affiliation{School of Physics and Astronomy, University of Birmingham, Birmingham B15 2TT, UK}
\affiliation{Stellar Astrophysics Centre (SAC), Department of Physics and Astronomy, Aarhus University, Ny Munkegade 120, DK-8000 Aarhus C, Denmark}

\author[0000-0002-8035-4778]{Jessie L. Christiansen}
\affiliation{Caltech/IPAC-NASA Exoplanet Science Institute, Pasadena, CA 91125, USA}

\author[0000-0001-5137-0966]{Jørgen Christensen-Dalsgaard}
\affiliation{Stellar Astrophysics Centre (SAC), Department of Physics and Astronomy, Aarhus University, Ny Munkegade 120, DK-8000 Aarhus C, Denmark}

\author[0000-0002-5741-3047]{David R. Ciardi}
\affiliation{NASA Exoplanet Science Institute-Caltech/IPAC, Pasadena, CA 91125 USA}

\author{Bruce D. Clarke}
\affiliation{NASA Ames Research Center, Moffett Field, CA 94035, USA}
\affiliation{SETI Institute, 189 Bernardo Ave, Suite 200, Mountain View, CA 94043, USA}

\author[0000-0001-9662-3496]{William D. Cochran}
\affiliation{McDonald Observatory and Center for Planetary Systems Habitability, The University of Texas at Austin, Austin TX 78712, USA}

\author[0000-0003-4206-5649]{Jessie L. Dotson}
\affiliation{NASA Ames Research Center, Moffett Field, CA 94035, USA}

\author{Laurance R. Doyle}
\affiliation{SETI Institute, 189 Bernardo Ave, Suite 200, Mountain View, CA 94043, USA}

\author{Eduardo Seperuelo Duarte}
\affiliation{NASA Ames Research Center, Moffett Field, CA 94035, USA}
\affiliation{Instituto Federal de Educação Ciência e Tecnologia do Rio de Janeiro, Nilópolis, RJ, Brasil}

\author{Edward W. Dunham}
\affiliation{Lowell Observatory, 1400 W Mars Hill Rd, Flagstaff, AZ 86001, USA}

\author{Andrea K. Dupree}
\affiliation{Center for Astrophysics $\vert$ Harvard \& Smithsonian 60 Garden St., Cambridge, MA 02138, USA}

\author[0000-0002-7714-6310]{Michael Endl}
\affiliation{McDonald Observatory and Center for Planetary Systems Habitability, The University of Texas at Austin, Austin TX 78712, USA}

\author{James L. Fanson}
\affiliation{Jet Propulsion Laboratory, California Institute of Technology, USA}

\author[0000-0001-6545-639X]{Eric B. Ford}
\affiliation{Department of Astronomy \& Astrophysics, 525 Davey Laboratory, The Pennsylvania State University, University Park, PA 16802, USA}
\affiliation{Center for Exoplanets \& Habitable Worlds, 525 Davey Laboratory, The Pennsylvania State University, University Park, PA 16802, USA}
\affiliation{Center for Astrostatistics, 525 Davey Laboratory, The Pennsylvania State University, University Park, PA 16802, USA}
\affiliation{Institute for Computational \& Data Sciences, The Pennsylvania State University, University Park, PA 16802, USA}

\author{Maura Fujieh}
\affiliation{NASA Ames Research Center, Moffett Field, CA 94035, USA}

\author{Thomas N. Gautier III}
\affiliation{Jet Propulsion Laboratory, California Institute of Technology, USA}

\author{John C. Geary}
\altaffiliation{retired}
\affiliation{Smithsonian Astrophysical Observatory}

\author[0000-0002-1554-5578]{Ronald L Gilliland}
\affiliation{Space Telescope Science Institute, 3700 San Martin Drive, Baltimore, MD, 21218, USA}

\author{Forrest R. Girouard}
\altaffiliation{deceased}
\affiliation{NASA Ames Research Center, Moffett Field, CA 94035, USA}
\affiliation{Orbital Sciences Corporation, 2401 East El Segundo Boulevard, Suite 200, El Segundo, CA 90245, USA}

\author{Alan Gould}
\affiliation{Lawrence Hall of Science, University of California Berkeley, USA}

\author[0000-0003-2397-0045]{Michael R. Haas}
\altaffiliation{retired}
\affiliation{NASA Ames Research Center, Moffett Field, CA 94035, USA}

\author{Christopher E. Henze}
\affiliation{NASA Ames Research Center, Moffett Field, CA 94035, USA}

\author[0000-0002-1139-4880]{Matthew J. Holman}
\affiliation{Center for Astrophysics $\vert$ Harvard \& Smithsonian 60 Garden St., Cambridge, MA 02138, USA}

\author[0000-0001-8638-0320]{Andrew W. Howard}
\affiliation{Department of Astronomy, California Institute of Technology, Pasadena, CA 91125, USA}

\author[0000-0002-2532-2853]{Steve B. Howell}
\affiliation{NASA Ames Research Center, Moffett Field, CA 94035, USA}

\author[0000-0001-8832-4488]{Daniel Huber}
\affiliation{Institute for Astronomy, University of Hawai`i, 2680 Woodlawn Drive, Honolulu, HI 96822, USA}

\author{Roger C. Hunter}
\affiliation{NASA Ames Research Center, Moffett Field, CA 94035, USA}

\author[0000-0002-4715-9460]{Jon M. Jenkins}
\affiliation{NASA Ames Research Center, Moffett Field, CA 94035, USA}

\author[0000-0002-9037-0018]{Hans Kjeldsen}
\affiliation{Stellar Astrophysics Centre (SAC), Department of Physics and Astronomy, Aarhus University, Ny Munkegade 120, DK-8000 Aarhus C, Denmark}

\author{Jeffery Kolodziejczak}
\affiliation{NASA/Marshall Space Flight Center}

\author{Kipp Larson}
\affiliation{Ball Aerospace and Technologies Corp., Boulder, CO 80301, USA}

\author{David W. Latham}
\affiliation{Center for Astrophysics $\vert$ Harvard \& Smithsonian 60 Garden St., Cambridge, MA 02138, USA}

\author{Jie Li}
\affiliation{NASA Ames Research Center, Moffett Field, CA 94035, USA}
\affiliation{SETI Institute, 189 Bernardo Ave, Suite 200, Mountain View, CA 94043, USA}

\author{Savita Mathur}
\affiliation{Instituto de Astrof\'{\i}sica de Canarias, E-38200, La Laguna, Tenerife, Spain}
\affiliation{Universidad de La Laguna, Dpto. de Astrof\'{\i}sica, E-38205, La Laguna, Tenerife, Spain}

\author{Søren Meibom}
\affiliation{Center for Astrophysics $\vert$ Harvard \& Smithsonian 60 Garden St., Cambridge, MA 02138, USA}

\author{Chris Middour}
\affiliation{Millennium Engineering \& Integration Services, Moffett Field, CA 94035, USA}

\author[0000-0001-9303-3204]{Robert L. Morris}
\affiliation{NASA Ames Research Center, Moffett Field, CA 94035, USA}
\affiliation{SETI Institute, 189 Bernardo Ave, Suite 200, Mountain View, CA 94043, USA}

\author[0000-0002-8537-5711]{Timothy D. Morton}
\affiliation{Department of Physics and Astronomy, University of Southern California, Los Angeles, CA 90089-0484, USA}

\author{Fergal Mullally}
\affiliation{NASA Ames Research Center, Moffett Field, CA 94035, USA}
\affiliation{SETI Institute, 189 Bernardo Ave, Suite 200, Mountain View, CA 94043, USA}

\author[0000-0001-7106-4683]{Susan E. Mullally}
\affiliation{Space Telescope Science Institute, 3700 San Martin Drive, Baltimore, MD, 21218, USA}

\author[0000-0001-7691-2194]{David Pletcher}
\affiliation{NASA Ames Research Center, Moffett Field, CA 94035, USA}

\author[0000-0002-1913-0281]{Andrej Prsa}
\affiliation{Villanova University, Dept of Astrophysics and Planetary Science, 800 Lancaster Ave, Villanova PA 19085, USA}

\author[0000-0002-8964-8377]{Samuel N. Quinn}
\affiliation{Center for Astrophysics $\vert$ Harvard \& Smithsonian 60 Garden St., Cambridge, MA 02138, USA}

\author[0000-0003-1309-2904]{Elisa V. Quintana}
\affiliation{NASA Goddard Space Flight Center, Greenbelt, MD, USA}

\author[0000-0003-1080-9770]{Darin Ragozzine}
\affiliation{Brigham Young University, Department of Physics and Astronomy, N283 ESC, Provo, UT 84602, USA}

\author{Solange V. Ramirez}
\affiliation{Carnegie Observatories}

\author{Dwight T. Sanderfer}
\affiliation{NASA Ames Research Center, Moffett Field, CA 94035, USA}
\affiliation{Jacobs Engineering}

\author[0000-0001-7014-1771]{Dimitar Sasselov}
\affiliation{Center for Astrophysics $\vert$ Harvard \& Smithsonian 60 Garden St., Cambridge, MA 02138, USA}

\author{Shawn E. Seader}
\affiliation{Rincon Research Corporation, 101 N Wilmot Rd, Tucson, AZ 85711, USA}

\author[0000-0003-1179-3125]{Megan Shabram}
\affiliation{NASA Ames Research Center, Moffett Field, CA 94035, USA}

\author[0000-0002-1836-3120]{Avi Shporer}
\affiliation{Department of Physics and Kavli Institute for Astrophysics and Space Research, Massachusetts Institute of Technology, Cambridge, MA 02139, USA}

\author[0000-0002-6148-7903]{Jeffrey C. Smith}
\affiliation{NASA Ames Research Center, Moffett Field, CA 94035, USA}
\affiliation{SETI Institute, 189 Bernardo Ave, Suite 200, Mountain View, CA 94043, USA}

\author[0000-0003-2202-3847]{Jason H. Steffen}
\affiliation{University of Nevada, Las Vegas, 4505 S Maryland Pkwy, Las Vegas, NV 89154, USA}

\author{Martin Still}
\affiliation{Division of Astronomical Sciences, National Science Foundation, 2415 Eisenhower Ave, Alexandria, VA 22314, USA}

\author[0000-0002-5286-0251]{Guillermo Torres}
\affiliation{Center for Astrophysics $\vert$ Harvard \& Smithsonian 60 Garden St., Cambridge, MA 02138, USA}

\author{John Troeltzsch}
\affiliation{Ball Aerospace and Technologies Corp., Boulder, CO 80301, USA}

\author[0000-0002-6778-7552]{Joseph D. Twicken}
\affiliation{NASA Ames Research Center, Moffett Field, CA 94035, USA}
\affiliation{SETI Institute, 189 Bernardo Ave, Suite 200, Mountain View, CA 94043, USA}

\author{Akm Kamal Uddin}
\affiliation{Millennium Engineering \& Integration Services, Moffett Field, CA 94035, USA}

\author[0000-0002-4534-3969]{Jeffrey E. Van Cleve}
\affiliation{Ball Aerospace and Technologies Corp., Boulder, CO 80301, USA}

\author{Janice Voss}
\altaffiliation{deceased}
\affiliation{NASA Ames Research Center, Moffett Field, CA 94035, USA}

\author[0000-0002-3725-3058]{Lauren Weiss}
\affiliation{Institute for Astronomy, University of Hawai`i, 2680 Woodlawn Drive, Honolulu, HI 96822, USA}

\author[0000-0003-2381-5301]{William F. Welsh}
\affiliation{Department of Astronomy, San Diego State University, 5500 Campanile Drive, San Diego, CA 92182-1221 USA}

\author[0000-0002-5402-9613]{Bill Wohler}
\affiliation{NASA Ames Research Center, Moffett Field, CA 94035, USA}
\affiliation{SETI Institute, 189 Bernardo Ave, Suite 200, Mountain View, CA 94043, USA}

\author{Khadeejah A Zamudio}
\affiliation{NASA Ames Research Center, Moffett Field, CA 94035, USA}
\affiliation{KBRwyle, Houston, TX, USA}

\begin{abstract}
We present occurrence rates for rocky planets in the habitable zones (HZ) of main-sequence dwarf stars based on the Kepler DR25 planet candidate catalog and Gaia-based stellar properties.  We provide the first analysis in terms of star-dependent instellation flux, which allows us to track HZ planets.  We define $\eta_\oplus$ as the HZ occurrence of planets with radius between 0.5 and 1.5  $R_\oplus$ orbiting stars with effective temperatures between 4800~K and  6300~K.  We find that $\eta_\oplus$ for the conservative HZ is between $0.37^{+0.48}_{-0.21}$ (errors reflect 68\% credible intervals) and $0.60^{+0.90}_{-0.36}$ planets per star, while the optimistic HZ occurrence is between $0.58^{+0.73}_{-0.33}$ and $0.88^{+1.28}_{-0.51}$ planets per star.  These bounds reflect two extreme assumptions about the extrapolation of completeness beyond orbital periods where DR25 completeness data are available.  The large uncertainties are due to the small number of detected small HZ planets.  We find similar occurrence rates using both a Poisson likelihood Bayesian analysis and Approximate Bayesian Computation.  Our results are corrected for catalog completeness and reliability.  Both completeness and the planet occurrence rate are dependent on stellar effective temperature.  We also present occurrence rates for various stellar populations and planet size ranges.  We estimate with $95\%$ confidence that, on average, the nearest HZ planet around G and K dwarfs is $\sim$6~pc away, and there are $\sim 4$ HZ rocky planets around G and K dwarfs within 10~pc of the Sun.
 \end{abstract}



\section{Introduction} \label{section:introduction}
One of the primary goals of the \Kepler\ mission \citep{Borucki2010,Koch2010,Borucki2016} is to determine the frequency of occurrence of habitable-zone rocky planets around Sun-like stars, also known as ``$\eta_{\oplus}$''.  Habitable-zone rocky planets are broadly construed as any rocky planet in its star's {\it habitable zone} (HZ), roughly defined as being at the right distance from the star so that its surface temperature would permit liquid water (see \S\ref{section:habitability}). Measuring $\eta_{\oplus}$ informs theories of planet formation, helping us to understand why we are here, and is an important input to mission design for instruments designed to detect and characterize habitable-zone planets such as LUVOIR \citep{LUVOIR} and HabEX \citep{habex}. 

\Kepler's strategy to measure $\eta_\oplus$ was to continuously observe $>$150,000 Solar-like main-sequence dwarf stars (primarily F, G, and K) with a highly sensitive photometer in Solar orbit, identifying planets through the detection of transits. In the process, Kepler revolutionized our perspective of exoplanets in the Galaxy. The planet catalog in the final \Kepler\ data release 25 (DR25) contains 4034 planet candidates \citep[PCs;][]{Thompson2018}, leading to the confirmation or statistical validation of over 2,300 exoplanets\footnote{\url{https://exoplanetarchive.ipac.caltech.edu}\label{footnote:exoplanetArchive}} --- more than half of all exoplanets known today.

Identifying habitable zone rocky planets proved to be a greater challenge than anticipated. Based on the sensitivity of the \Kepler\ photometer and the expectation that Solar variability is typical of quiet main-sequence dwarfs, it was believed that four years of observation would detect a sufficient number of rocky habitable-zone planets to constrain their frequency of occurrence.  However, \Kepler\ observations showed that stellar variability was, on average, $\sim$50\% higher than Solar variability \citep{gilliland2011}, which suppressed the number of habitable-zone rocky planets that could be detected in four years.  In response, \Kepler's observational time was extended to eight years, but the failure of reaction wheels, required to maintain precise pointing, prevented continuation of high-precision observations in the original Kepler field after four years \citep{howell2014}. Furthermore, by definition, \Kepler\ planet candidates must have at least three observed transits.  The longest orbital period with three transits that can be observed in the four years of \Kepler\ data is 710 days (assuming fortuitous timing in when the transits occur). Given that the habitable zone of many F and late G stars require orbital periods longer than 710 days, \Kepler\ is not capable of detecting all habitable-zone planets around these stars.

The result is \Kepler\ data in which transiting rocky habitable zone planets are often near or beyond \Kepler's detection limit. Of the thousands of planets in the DR25 catalog, relatively few are unambiguously rocky and near their habitable zones: there are 56 such PCs with radius $\leq2.5~R_\oplus$, and 9 PCs with radius $\leq1.5~R_\oplus$ (using planet radii from \cite{Berger2020b}).  As described in \S\ref{section:habitability}, we expect many planets near the habitable zone larger than $1.5~R_\oplus$ to be non-rocky. These small numbers present challenges in the measurement of the frequency of occurrence of  habitable-zone planets.  

Converting a planet catalog into an underlying occurrence rate is also challenging due to the existence of selection effects and biases, with issues only exacerbated in the $\eta_{\oplus}$ regime. Planet candidate catalogs generally suffer from three types of error: 
\begin{itemize}
    \item The catalog is {\it incomplete}, missing real planets.
    \item The catalog is {\it unreliable}, being polluted with false positives (FPs).
    \item The catalog is {\it inaccurate}, with observational error leading to inaccurate planet properties.
\end{itemize}
Near the detection limit, both completeness and reliability can be low, requiring careful correction for the computation of occurrence rates.  The DR25 planet candidate catalog includes several products that facilitate the characterization of and correction for completeness and reliability \citep{Thompson2018}.  The data supporting completeness characterization, however, are only supplied for orbital periods of 500 days or less, requiring extrapolation of completeness for planets beyond these orbital periods. 

\begin{figure*}[ht]
  \centering
  \includegraphics[width=0.9\linewidth]{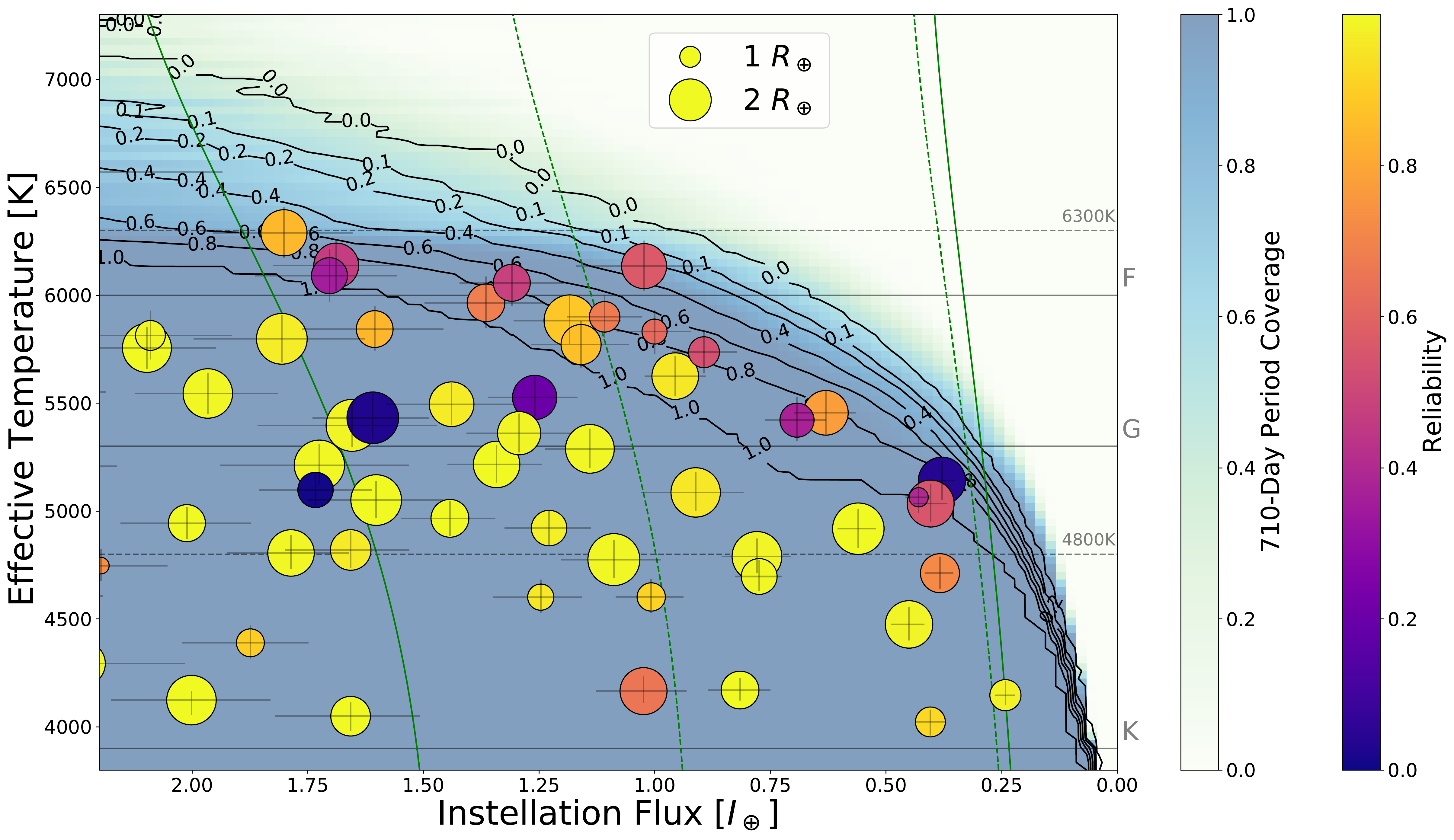} \\
  \includegraphics[width=0.91\linewidth]{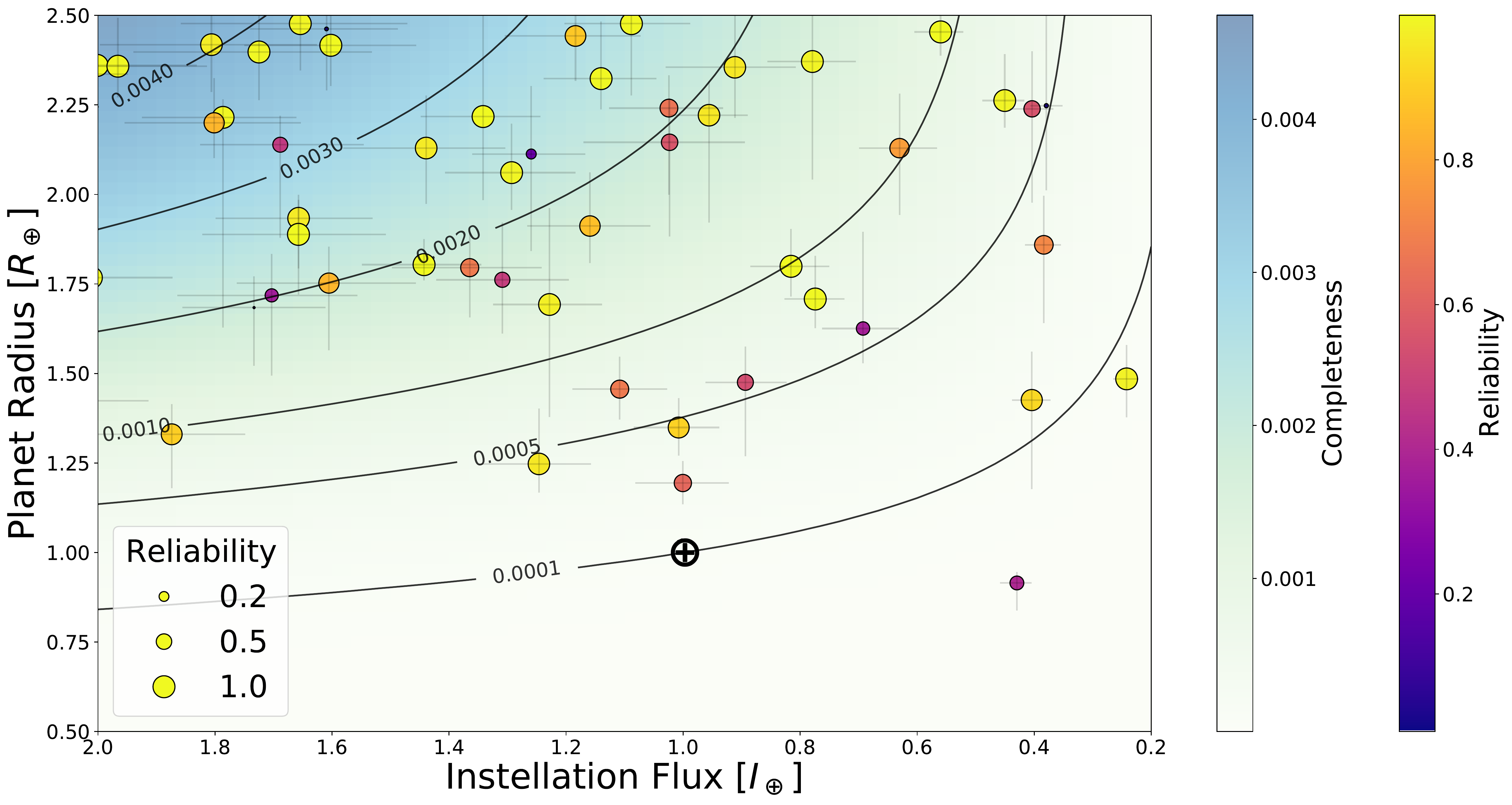} 
\caption{Two views of the DR25 PC population with radii smaller than $2.5~R_\oplus$ and instellation flux near their host star's habitable zone around main sequence dwarf stars. Top: Instellation flux vs. stellar effective temperature, showing the habitable zone and \Kepler\ observational coverage.  The background color map gives, at each point, the fraction of stars at that effective temperature and instellation flux that may have planets with with orbital periods of 710 days or less, so it is possible to observe three transits.  The contours show the fraction of planets with periods of 500 days or less, indicating available completeness measurements.  The solid green lines are the boundaries of the optimistic habitable zone, while the dashed green lines are the boundaries of the conservative habitable zone (see \S\ref{section:habitability}).  The planets are sized by their radius and colored by their reliability.  Bottom: Instellation flux vs. planet radius.  The color map and contours show the average completeness for the stellar population (\S\ref{section:completeness}).  The planets are sized and colored by reliability (\S\ref{section:reliability}), with radius and instellation flux error bars.  In the lower panel the $\oplus$ symbol shows the Earth.} \label{figure:populations}
\end{figure*}

These issues are summarized in Figure~\ref{figure:populations}, which show the DR25 PC population and its observational coverage, observational error, completeness, and reliability.  The details of these populations are given in Appendix~\ref{app:pcProperties}.

Our calculation of habitable zone occurrence will be in terms of planet radius and {\it instellation flux}, measuring the photon flux incident on the planet from its host star, which allows us to consider each star's habitable zone. We will proceed in two steps: 
\begin{enumerate}
    \item Develop a model describing the planet population in the neighborhood of the habitable zone (\S\ref{section:determineLambda}).  Because this is a statistical study, the model will be based on a large number of stars, using the observed DR25 planet candidate catalog and Gaia-based stellar properties, and will include corrections for catalog completeness and reliability.
    \item The derivation of the average number of rocky planets per star in each star's habitable zone from the planet population model (\S\ref{section:computeOccurrence}).  This will often be done in a subset of the parameter space used to compute the population model.
\end{enumerate}
When computing a quantity over a desired range of parameters such as radius and instellation flux, it is often the case that using data from a wider range will give better results.  For example, it is well known that polynomial fits to data have higher uncertainty near the boundaries of the data.   As explained in \S\ref{section:habitability}, we are primarily interested in rocky habitable zone planets, with planet radii between 0.5~$R_\oplus$ and 1.5~$R_\oplus$ and instellation flux within each star's estimated habitable zone, for stars with effective temperature between 4800~K and 6300~K.  To create our population model, we will use a larger domain with a planet radius range of $0.5~R_\oplus$ to $2.5~R_\oplus$, and an instellation range from 0.2 to 2.2 times Earth's insolation, which encloses the habitable zones of all the stars we consider.  We will focus on using two stellar populations: one exactly matching our desired effective temperature range of 4800~K to 6300~K and one with a larger range of 3900~K to 6300~K to investigate whether the larger range will improve our results.  Most of our results will be reported for both stellar populations because it is possible that including stars in the 3900~K to 4800~K range will bias our results.  We will have a population model for each choice of stellar population.

Once we have our population models, we will use them to compute our definition of $\eta_\oplus$, the average number of planets per star with radii between 0.5 and 1.5~$R_\oplus$, in the star's habitable zone, averaged over stars with effective temperature from 4800~K to 6300~K.  In the end we will find that the two stellar populations predict similar median values for $\eta_\oplus$, but the model using stars with effective temperatures from 3900~K to 6300~K yields significantly smaller (though still large) uncertainties.  While we are focused on our definition of $\eta_\oplus$, occurrence rates over other ranges of planet radius and stellar temperature are of interest.  We will use population model based on the 3900~K to 6300~K stellar population to compute the average number of habitable zone planets per star for various ranges of planet radius and stellar effective temperature.

\subsection{Previous \Kepler-based \texorpdfstring{$\eta_\oplus$}{eta Earth} Estimates} \label{section:previousWork}

Attempts to measure $\eta_{\oplus}$ and general occurrence rates with \Kepler\ have been made since the earliest \Kepler\ catalog releases \citep{Borucki2011}. \citet{Youdin2011} and \citet{Howard2012} were two influential early studies, in which planets found in only the first four months of data (Q0--Q2) were used to constrain \Kepler\ occurrence rates. \citet{Youdin2011} developed a maximum likelihood method to fit an underlying planetary distribution function, which later influenced the Poisson likelihood function method adopted by, e.g., \citet{Burke2015, Bryson2020}. \citet{Howard2012} took an alternative approach of estimating occurrence rates in bins of planets, defined over a 2D grid of planet radius and orbital period. In each bin, non-detections are corrected for by weighting each planet by the inverse of its detection efficiency. This inverse detection efficiency method (IDEM) is one of the most popular approaches in the literature. 

\citet{CatanzariteShao2011} and \citet{Traub2012} \citep[Q0--Q5,][]{Borucki2011} were among the first papers to focus on the $\eta_{\oplus}$ question specifically. Later $\eta_{\oplus}$ papers from \citet{DressingCharbonneau2013} \citep[Q1--Q6,][]{Batalha2013}, \citet{Kopparapu2013} (Q1--Q6), \citet{Burke2015} \citep[Q1--Q16,][]{Mullally2016}, and \citet{Silburt2015} (Q1--Q16) were able to take advantage of newer planet catalogs based on increased amounts of data. Other papers have used custom pipelines to search \Kepler\ light curves to estimate $\eta_{\oplus}$ with independently produced planet catalogs: namely, \citet{Petigura2013} (Q1--Q15), \citet{ForemanMackey2014} (Q1--Q15), \citet{Dressing2015} (Q1--Q16), and \citet{Kunimoto2020a} (Q1--Q17). Still more have been meta-analyses of results from the exoplanet community based on different \Kepler\ catalogs \citep[][]{Kopparapu2018, Garrett2018}.

Comparisons between these $\eta_{\oplus}$ studies are challenging due to the wide variety of catalogs used, some of which are based on only a fraction of the data as others. Characterization of completeness has also varied between authors, with some assuming a simple analytic model of detection efficiency \citep[e.g.,] []{Youdin2011, Howard2012}, some empirically estimating detection efficiency with transit injection/recovery tests \citep[e.g.,][]{Petigura2013, Burke2015}, and others simply assuming a catalog is complete beyond some threshold \citep[e.g.,][]{CatanzariteShao2011}. \citet{Borucki2011} provided a comprehensive analysis of completeness bias, reliability against astrophysical false positives and reliability against statistical false alarms based on manual vetting and simple noise estimates. Fully automated vetting was implemented via the Robovetter \citep{Coughlin2017} for the \Kepler\ DR24 \citep{Coughlin2016} and DR25 catalogs.  The final \Kepler\ data release (DR25), based on the full set of \Kepler\ observations and accompanied by comprehensive data products for characterizing completeness, has been essential for alleviating issues of completeness and reliability. The DR25 catalog is now the standard used by occurrence rate studies \citep[e.g.,][]{Mulders2018, Hsu2018, zinkChristiansen2019, Bryson2020}.

DR25 was the first catalog to include data products that allowed for the characterization of catalog reliability against false alarms due to noise and systematic instrumental artifacts, which are the most prevalent contaminants in the $\eta_{\oplus}$ regime.  Thus nearly all previous works did not incorporate reliability against false alarms in their estimates. \citet{Bryson2020} was the first to directly take into account reliability against both noise/systematics and astrophysical false positives, and in doing so found that occurrence rates for small planets in long-period orbits dropped significantly after reliability correction. \citet{Mulders2018} attempted to mitigate the impact of contamination by using a DR25 Disposition Score cut \citep[see \S7.3.4 of][]{Thompson2018} as an alternative to reliability correction. As shown in \citet{Bryson2020b}, while this approach does produce a higher reliability planet catalog, explicit accounting for reliability is still necessary for accurate occurrence rates.

Studies have also varied in stellar property catalogs used, and exoplanet occurrence rates have been shown to be sensitive to such choices. For instance, the discovery of a gap in the radius distribution of small planets, first uncovered in observations by \citet{Fulton2017}, was enabled by improvements in stellar radius measurements by the California Kepler Survey \citep[CKS;][]{Petigura2017,Johnson2017}. The use of Gaia DR2 parallaxes, which have resulted in a reduction in the median stellar radius uncertainty of \Kepler\ stars from $\approx27\%$ to $\approx4\%$ \citep{Berger2020a}, has been another significant improvement with important implications for $\eta_{\oplus}$. \citet{Bryson2020} showed that occurrence rates of planets near Earth's orbit and size can drop by a factor of 2 if one adopts planet radii based on Gaia stellar properties rather than pre-Gaia Kepler Input Catalog stellar properties.

\subsection{Our Work}\label{section:ourWork}

Measuring $\eta_{\oplus}$ requires a definition of what it actually means to be considered a rocky planet in the habitable zone. Different authors use different definitions, including regarding whether $\eta_{\oplus}$ refers to the number of rocky habitable zone planets per star, or the number of stars with rocky habitable zone planets. In this paper, for reasons detailed in \S\ref{section:habitability}, we define $\eta_\oplus$ as the average number of planets per star with planet radius between 0.5 and 1.5 Earth radii, in the star's habitable zone, where the average is taken over stars with effective temperatures between 4800~K and 6300~K. We compute $\eta_\oplus$ for both conservative and optimistic habitable zones, denoted respectively as $\eta_\oplus^\mathrm{C}$ and $\eta_\oplus^\mathrm{O}$.

\begin{figure}[ht]
  \centering
  \includegraphics[width=0.99\linewidth]{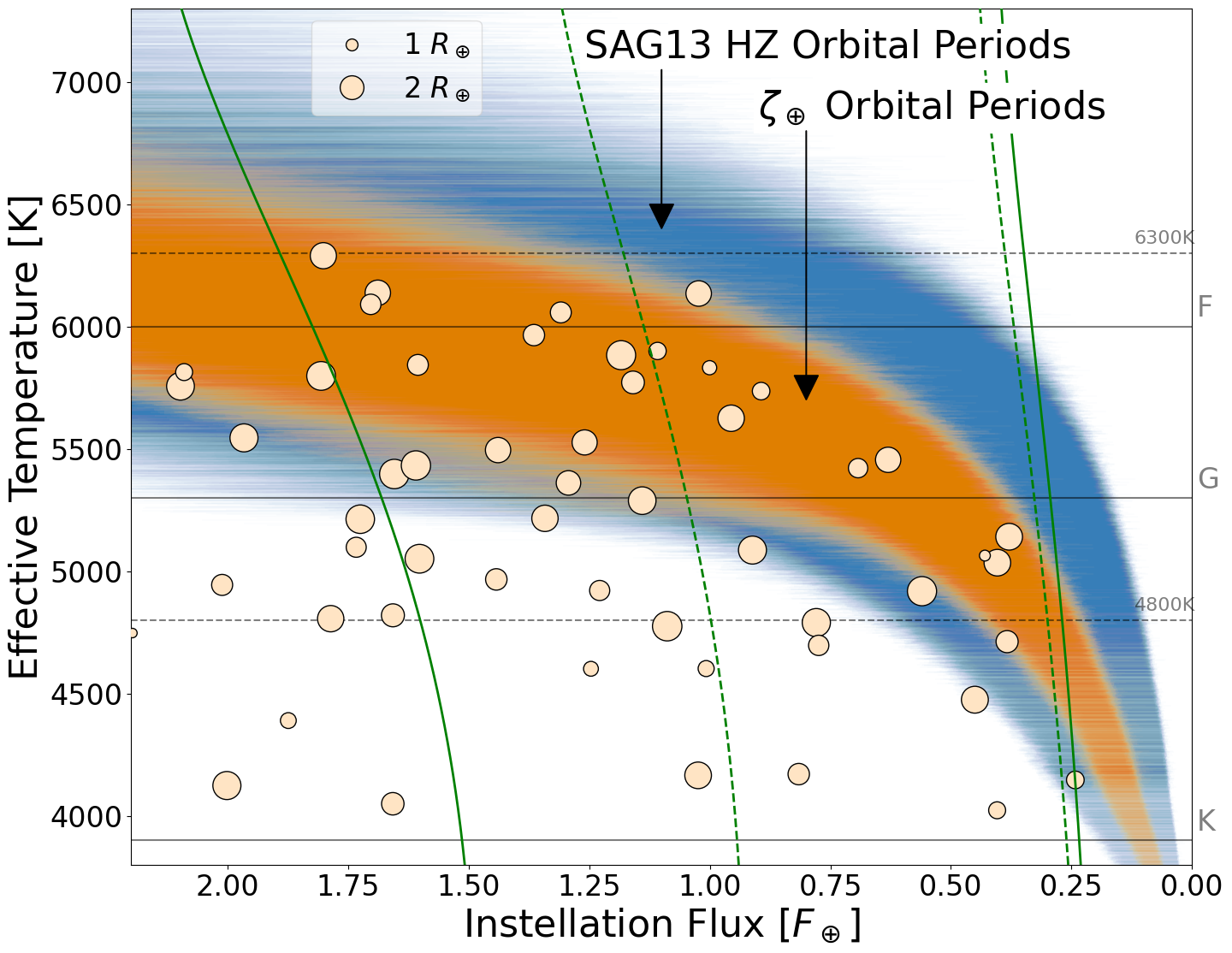} 
\caption{The habitable zone flux range compared with example orbital periods, previously used to estimate habitable zone occurrence for G, K, and F stars.  For each star in the stellar parent sample, we show the instellation flux range of each orbital period range, with the SAG13 instellation flux range shown as the blue region (comprised of a horizontal blue line for each star showing the flux range for that orbital period range) and $\zeta_\oplus$ shown as the orange region.  The solid green lines are the boundaries of the optimistic habitable zone, while the dashed green lines are the boundaries of the conservative habitable zone (see \S\ref{section:habitability}).  The planet population is the same as in Figure~\ref{figure:populations}, and are sized by their radius.} \label{figure:fluxPeriod}
\end{figure}

Most of the existing literature on habitable zone occurrence rates are in terms of orbital period, where a single period range is adopted to represent the bounds of the habitable zone for the entire stellar population considered. However, no single period range covers the habitable zone for a wide variety of stars. Figure~\ref{figure:fluxPeriod} shows two example period ranges used for habitable zone occurrence rate studies relative to the habitable zone of each star in our stellar parent sample.  The SAG13\footnote{\url{https://exoplanets.nasa.gov/exep/exopag/sag/\#sag13}\label{footnote:sag13}} habitable zone range of $237 \leq \mathrm{period} \leq 860$ days is shown in blue, and $\zeta_\oplus$, defined in \citet{Burke2015} as within 20\% of Earth's orbital period, is shown in orange.  While these period ranges cover much of the habitable zone for G stars, they miss significant portions of the habitable zones of K and F stars, and include regions outside the habitable zone even when restricted to G stars.  This will be true for any fixed choice of orbital period range for the range of stellar effective temperatures required for good statistical analysis.  Such coverage will not lead to accurate occurrence rates of planets in the habitable zone.  Given that the period ranges of many habitable zone definitions also extend beyond the detection limit of \Kepler, computing $\eta_{\oplus}$ requires extrapolation of a fitted population model to longer orbital periods.   \citet{Lopez2018} and \citet{Pascucci2019} present evidence and theoretical arguments that inferring the population of small rocky planets at low instellation from the population of larger planets at high instellation can introduce significant overestimates of $\eta_{\oplus}$.

For these reasons, we choose to work in terms of the {\it instellation flux}, measuring the photon flux incident on the planet from its host star, rather than orbital period. In \S\ref{section:methodology} we describe how we adopt existing occurrence rate methods and completeness characterizations to use instellation flux instead of orbital period. We address concerns with extrapolating completeness to long orbital periods by providing bounds on the impact of the limited coverage of completeness data (\S\ref{section:completenessAndReliability}). Following \citet{Howard2012,Youdin2011,Burke2015}, among others, we compute the number of planets per star $f$.  As in \citet{Youdin2011} and \citet{Burke2015}, we first compute a population model in terms of the differential rate $\lambda \equiv \mathrm{d}^2 f / \mathrm{d} r \, \mathrm{d} I $, where $r$ is the planet radius and $I$ is the instellation flux.  We consider several possible functional forms for $\lambda$, and will allow $\lambda$ to depend on the stellar host effective temperature.  We compute $\lambda$ over the radius range $0.5 \ R_\oplus \leq r \leq 2.5 \ R_\oplus$ and instellation flux range $0.2 \ I_\oplus \leq I \leq 2.2 \ I_\oplus$, averaged over the effective temperatures of the stellar population used for the computation (\S\ref{section:determineLambda}).  Occurrence rates will be computed by integrating $\lambda$ over the desired planet radius and instellation flux range, and averaging over the desired effective temperature range to give $f$, the average number of planets per star.  (\S\ref{section:computeOccurrence}).

By restricting our analysis to planets with $r \leq 2.5 \ R_\oplus$ in regions of instellation flux close to the habitable zone, we believe we are avoiding the biases pointed out by \citet{Lopez2018} and \citet{Pascucci2019} -- As seen in Figure~\ref{figure:populations}, there are more detected planets with $1.5 \ R_\oplus \leq r \leq 2.5 \ R_\oplus$ than planets with $0.5 \ R_\oplus \leq r \leq 1.5 \ R_\oplus$, so our results in \S\ref{section:results} will be driven by these larger planets, but all planets we consider are at similar low levels of instellation.  In Figure 2 of \citet{Lopez2018} we note that for instellation flux between 10 and 20 there is little change in the predicted relative sizes of the $1.5 \ R_\oplus \leq r \leq 2.5 \ R_\oplus$ and $0.5 \ R_\oplus \leq r \leq 1.5 \ R_\oplus$ planet populations.  Naively extrapolating this to instellation $<2$ in the HZ, we infer that the sizes of these larger and smaller planet populations in the HZ are similar. Therefore by working at low instellation flux we are likely less vulnerable to overestimating the HZ rocky planet population.

We use both Poisson likelihood-based inference with Markov-Chain Monte Carlo (MCMC) and likelihood-free inference with Approximate Bayesian Computation (ABC). The Poisson likelihood method is one of the most common approaches to calculating exoplanet occurrence rates \citep[e.g.][]{Burke2015, zinkChristiansen2019, Bryson2020}, while ABC was only applied for the first time in \citet{Hsu2018, Hsu2019}. As described in \S\ref{section:inferenceMethods}, these methods differ in their treatment of reliability and input uncertainty, and allow us to assess possible dependence of our result on the assumption of a Poisson likelihood function. We present our results in \S\ref{section:results}. Recognizing the importance of reliability correction in the $\eta_{\oplus}$ regime, and confirming the same impact of reliability as \citet{Bryson2020}, we choose to report only our reliability-incorporated results. We also present results both with and without incorporating uncertainties in planet radius and instellation flux and host star effective temperature in our analysis. Our final recommended population models and implications for $\eta_{\oplus}$, as well as how our results relate to previous estimates are discussed in \S\ref{section:discussion}.

All results reported in this paper were produced with Python code, mostly in the form of Python Jupyter notebooks, found at the paper GitHub site\footnote{\url{https://github.com/stevepur/DR25-occurrence-public/tree/master/insolation}}.
\subsection{Notation}\label{section:notation}
When summarizing a distribution, we use the notation $m^{+e1}_{-e2}$ to refer to the 68\% credible interval $[m-e2, m+e1]$, where $m$ is the median and ``$n\%$ credible interval'' means that the central $n\%$ of the values fall in that interval.  We caution our distributions are typically not Gaussian, so this interval should not be treated as ``$1 \sigma$''.  We will also supply 95\% and 99\% credible intervals for our most important results. 

We use the following notation throughout the paper:
\begin{enumerate}[\indent {}]
    \itemsep-0.3em 
    \item $r$: Planet radius in units of Earth radii $R_\oplus$.
    \item $I$: Planet instellation flux in units of Earth instellation $I_\oplus$.
    \item $T_\mathrm{eff}$: Stellar effective temperature in Kelvins.  When referring to a planet, this is the effective temperature of that planet's host star.
    \item $f$: The number of planets per star, typically a function of $r$, $I$ and $T_\mathrm{eff}$.
    \item $\lambda$: The differential rate population model $\equiv \mathrm{d}^2 f / \mathrm{d} r \, \mathrm{d} I $, typically a function of $r$, $I$ and $T_\mathrm{eff}$.  $\lambda$ is defined by several parameters, for example exponents when $\lambda$ is a power law.
    \item $\boldsymbol{\theta}$: the vector of parameters that define $\lambda$, whose contents depend on the particular form of $\lambda$.
\end{enumerate}
$\eta_\oplus$ without modification refers to the average number of habitable-zone planets per star with $0.5  \ R_\oplus \leq r \leq 1.5 \ R_\oplus$ and host star effective temperature between 3900~K and 6300~K, with the conservative or optimistic habitable zone specified in that context.  $\eta_\oplus^\mathrm{C}$ and  $\eta_\oplus^\mathrm{O}$, respectively, specifically refer to occurrence in the conservative and optimistic habitable zones.  Additional subscripts on $\eta_\oplus$ refer to different stellar populations.  For example $\eta_{\oplus, \mathrm{GK}}^\mathrm{C}$ is the occurrence of conservative habitable zone planets with $0.5  \ R_\oplus \leq r \leq 1.5 \ R_\oplus$ around GK host stars.


\section{Habitability} \label{section:habitability}


\subsection{Characterizing Rocky Planets in the Habitable Zone} \label{section:habCharacter}
A key aspect in computing habitable zone planet occurrence rates is the location and width of the Habitable Zone. Classically, it is defined as the region around a star in which a rocky-mass/size planet with an Earth-like atmospheric composition (CO$_{2}$, H$_{2}$O, and N$_{2}$) can sustain liquid water on its surface. The insistence on surface liquid water is important for the development of life as we know it, and the availability of water on the surface assumes that any biological activity on the surface alters the atmospheric composition of the planet, betraying the presence of life when observed with remote detection techniques.

 Various studies estimate the limits and the width of the HZ in the literature (see \citet{Kopparapu2018} and \citet{Kopparapu2019} for a review), and explored the effect  of physical processes such as tidal locking, rotation rate of the planet, combination of different greenhouse gases, planetary mass, obliquity, and eccentricity on HZ limits. These effects paint a more nuanced approach to identify habitability limits, and are particularly  useful to explore those environmental conditions where habitability could be maintained. However, for the purpose of calculating the occurrence rates of planets in the HZ, it is best to use a standard for HZ limits as a first attempt, such as Earth-like conditions. One reason is that it would become computationally expensive to estimate the occurrence rates of HZ planets considering all the various HZ limits arising from these planetary and stellar properties. Furthermore, future flagship mission concept studies like LUVOIR \citep{LUVOIR}, HabEX \citep{habex}, and OST \citep{ost} use the classical HZ limits as their standard case to estimate exoEarth mission yields and for identifying associated biosignature gases. Therefore, in this study we use the conservative and optimistic HZ estimates from \cite{Kopparapu2014}, where the conservative inner and outer edges of the HZ are defined by the `runaway greenhouse' and `maximum greenhouse' limits, and the optimistic inner and outer HZ boundaries are the `recent Venus' and `early Mars' limits. By using these HZ limits, we  (1) are able to make a consistent comparison with already published occurrence rates of HZ planets in the literature that have also used the same HZ limits, (2) provide a range of values for HZ planet occurrence, and (3)  obtain an `average' occurrence rate of planets in the HZ, as the conservative and optimistic HZ limits from \citet{Kopparapu2014} span the range of HZ limits from more complex models and processes.

We consider planets in the $0.5 - 1.5$ $R_\oplus$ size range to calculate rocky planet occurrence rates, as studies have suggested that planets that fall within these radius bounds are most likely to be rocky \citep{Rogers2015, Wolfgang2016, ChenKipping2017, Fulton2017}. While some studies have indicated that the rocky regime can extend to as high as 2.5 $R_\oplus$ \citep{Otegi2020}, many of these high radius-regime planets seem to be highly irradiated planets, receiving stellar fluxes much larger than the planets within the HZ. Nevertheless, we have also calculated occurrence rates of planets with radii up to 2.5 $R_\oplus$.  We note that \citet{Kane2016} also used Kopparapu et al. (2014) HZ estimates to identify HZ planet candidates using DR24 planet candidate catalog and DR25 stellar properties.

We also limit the host stellar spectral types to stars with $4800 \le$ $T_\mathrm{eff} \le 6300$~K, covering mid K to late F. The reason for limiting to $T_\mathrm{eff} > 4800$~K is two fold: (1) The inner working angle (IWA, the smallest angle  on the sky at which a direct imaging telescope can reach its designed ratio of planet to star flux) for the LUVOIR coronagraph instrument ECLIPS falls off below 48 milli arc sec at 1 micron (3$\lambda$/D) for a planet at 10~pc for $T_\mathrm{eff} \leq$ 4800~K, and (2) Planets are likely tidal-locked or synchronously rotating below 4800~K that could potentially alter the inner HZ limit significantly \citep{Yang2013, Yang2014b, Wolf2015a, Way2015, Godolt2015, Kopparapu2016, Kopparapu2017, Bin2018}. The upper limit of 6300~K is a result of planets in the HZs having longer orbital periods around early F-stars, where {\it Kepler} is not capable of detecting these planets, as described in \S\ref{section:introduction}.


\subsection{Effective Temperature Dependence of the Width of the Habitable Zone} \label{section:hzGeom}
The width of the HZ for hotter stars is larger than the width for cooler stars, implying that the habitable zone occurrence rate may be dependent on the host star effective temperature.  In this section we derive an approximate form for this effective temperature dependence, which we refer to as the ``geometric effect''.  

We compute the instellation flux $I$ on a planet orbiting a particular star as $I = R_*^2 T^4 / a^2 $, where $R_*$ is the stellar radius in Solar radii, $T = T_\mathrm{eff}/T_\odot$ is the effective temperature divided by the Solar effective temperature, and $a$ is the semi-major axis of the planet orbit in AU.  We assume the orbit is circular.  Then the size of the habitable zone $\Delta a$ is determined by the instellation flux at the inner and outer habitable zone boundaries, $I_\mathrm{inner}$ and $I_\mathrm{outer}$, as 
\begin{equation} \label{eqn:deltaA}
\begin{split}
    \Delta a 
    &=  a_\mathrm{outer} - a_\mathrm{inner} \\ 
    &=  R_* T^2 
    \left(\frac{1}{\sqrt{I_\mathrm{outer}}} - \frac{1}{\sqrt{I_\mathrm{inner}}} \right).
\end{split}
\end{equation}
The factor $\left(1 / \sqrt{I_\mathrm{outer}} - 1 / \sqrt{I_\mathrm{inner}} \right)$ has a weak $T_\mathrm{eff}$ dependence, ranging from 1.25 at 3900~K to 0.97 at 6300~K, which we crudely approximate as constant in this paragraph.  We also observe that, for the main-sequence dwarf stellar populations we use in our computations (described in \S\ref{section:stellarPopulation}), $R_*$ has an approximately linear dependence on $T$, which we write as $\left(\tau T + R_0 \right)$ ($\tau \approx 1.8$ and $R_0 \approx -0.74$).  Therefore 
\begin{equation} \label{eqn:deltaA2}
    \Delta a \propto  \left(\tau T + R_0 \right) T^2.
\end{equation}
So even if the differential occurrence rate $\lambda$ has no dependence on $a$, and therefore no dependence on $I$, the habitable zone occurrence rate may depend on $T_\mathrm{eff}$ simply because hotter stars have larger habitable zones.

Several studies, such as \citet{Burke2015} and \citet{Bryson2020},  studied planet occurrence in terms of the orbital period $p$ and have shown that $\mathrm{d} f / \mathrm{d} p$ is well-approximated by a power law $p^\alpha$.  In Appendix~\ref{app:InstellationDerivation} we show that this power law, combined with the relationship between instellation flux and period, implies that the instellation flux portion of the differential rate function $\lambda$, $\mathrm{d} f / \mathrm{d} I$, has the form
\begin{equation} \label{eqn:dfdI}
    \mathrm{d} f / \mathrm{d} I \approx  C I^\nu \left( \left(\tau T + R_0 \right) T^4 \right)^\delta
\end{equation}
where $\nu = -\frac{3}{4} \left( \alpha - \frac{7}{3} \right)$ and $\delta = -\nu - 1$.  This form incorporates the combined effects of the size of the habitable zone increasing with $T_\mathrm{eff}$, as well as dependence from the period power law $p^\alpha$.  The derivation in Appendix~\ref{app:InstellationDerivation} uses several crude approximations, so Equation (\ref{eqn:dfdI}) is qualitative rather quantitative. 

\begin{figure}[ht]
  \centering
  \includegraphics[width=0.95\linewidth]{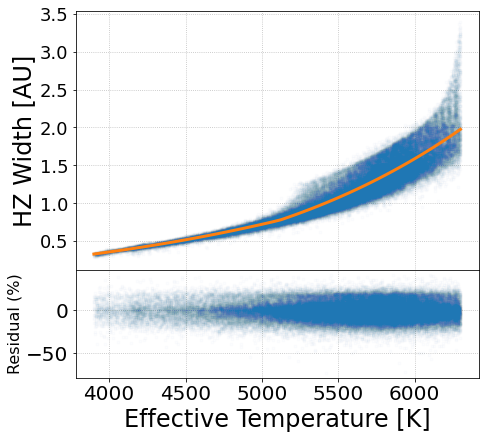} \\
\caption{The width of the optimistic habitable zone (outer HZ boundary minus inner HZ boundary) as a function of effective temperature for all the stars in our parent sample with effective temperature between 3900~K and 6300~K.  The line shows the broken power law fit in Equation (\ref{eqn:geomFunc}).  The percentage residual from the fit is shown in the lower panel.} \label{figure:hzWidthFit}
\end{figure}

In \S\ref{section:determineLambda} we consider forms of the population model $\lambda$ that separate the geometric effect in Equation~(\ref{eqn:deltaA2}) from a possible more physical dependence on $T_\mathrm{eff}$, and compare them with direct measurement of the $T_\mathrm{eff}$-dependence.  To separate the geometric effect, we incorporate a geometric factor $g(T_\mathrm{eff})$ inspired by Equation~(\ref{eqn:deltaA2}).  Because of the crude approximations used to derive Equations~(\ref{eqn:deltaA2}) and (\ref{eqn:dfdI}) we use an empirical fit to the habitable zone width $\Delta a$ for all stars in our stellar sample.
Because we will use this fit in models of the differential population population rate $\mathrm{d} f / \mathrm{d} I$ in \S\ref{section:models}, we perform the fit computing $\Delta a$ for each star using a fixed flux interval $\Delta I \in [0.25, 1.8]$.  Because $a = R_*^2 T^4 / \sqrt{I}$, $\Delta a$ is just a scaling of each star's luminance $R_*^2 T^4$ by the factor $1/\sqrt{0.25} - 1/\sqrt{1.8}$.  
As shown in Figure~\ref{figure:hzWidthFit}, $\Delta a$ is well-fit, with well-behaved residuals, by the broken power law
\begin{equation}
    g(T_\mathrm{eff}) = 
    \begin{cases} 
        10^{-11.84}~T_\mathrm{eff}^{3.16} & \text{if }T_\mathrm{eff}\leq 5117\mathrm{K}, \\ 
        10^{-16.77}~T_\mathrm{eff}^{4.49} & \text{otherwise}.
    \end{cases} \label{eqn:geomFunc}
\end{equation}

If the semi-major axes of planets are uniformly distributed in our stellar sample, then we expect that habitable zone planet occurrence would have a $T_\mathrm{eff}$ dependence due to Equation~\ref{eqn:geomFunc}.  In individual planetary systems, however, there is evidence of constant spacing in $\log(a)$ \citep{Weiss2018}, implying spacing proportional to $\Delta a/a$. In this case  there would be no impact of the larger habitable zones with increasing $T_\mathrm{eff}$: taking $a$ to be the average of the inner and outer semi-major axes, the star-dependent terms cancel, so $\Delta a/a$ is the same for all stars, independent of $T_\mathrm{eff}$.  This would imply that HZ occurrence has no $T_\mathrm{eff}$ dependence due to the increasing size of the HZ.

\section{Methodology} \label{section:methodology}
We base our occurrence rate of $f$ planets per star on a differential population rate model $\lambda(I, r, T_\mathrm{eff}) = \frac{\mathrm{d}^2 f(I, r, T_\mathrm{eff})}  {\mathrm{d} I \, \mathrm{d} r }$ that describes how $f$ varies as a function of  incident stellar flux $I$ and planet radius $r$.  We allow $\lambda(I, r, T_\mathrm{eff})$ (and therefore $f(I, r, T_\mathrm{eff})$) to depend on the host star effective temperature $T_\mathrm{eff}$.  In \S\ref{section:determineLambda} we use the DR25 planet candidate catalog to determine $\lambda$.  We cannot, however, simply take all the planet candidates in the DR25 catalog at face value.  We must statistically characterize and correct for errors in the catalog.

The DR25 planet candidate catalog contains 4034 planet candidates, identified through a uniform method of separating planet candidates from false positives and false alarms \citep{Thompson2018}.  This automated method is known to make mistakes, being both {\it incomplete} due to missing true transiting planets, and {\it unreliable} due to misidentifying various types of astrophysical false positives and instrumental false alarms as transiting planets.  Low completeness and low reliability are particularly acute near the \Kepler\ detection limit, which happens to coincide with the habitable zones of F, G, and K stars.  We characterize DR25 completeness and reliability using the synthetic data described in \citet{Thompson2018} with methods described in \citet{Bryson2020}.  We correct for completeness and reliability when determining the population rate $\lambda$ using the methods of \citet{Bryson2020} and \citet{Kunimoto2020b}.

The methods used in \citet{Bryson2020} and \citet{Kunimoto2020b} computed population models in orbital period and radius.  Generalizing these methods to instellation flux, radius, and effective temperature is relatively straightforward, with the treatment of completeness characterization presenting the largest challenge. In this section we summarize these methods, focusing on the changes required to operate in instellation flux rather than period and to allow for dependence on $T_\mathrm{eff}$.

\subsection{Stellar Populations} \label{section:stellarPopulation}
As in \citet{Bryson2020}, our stellar catalog uses the Gaia-based stellar properties from \citet{Berger2020a} combined with the DR25 stellar catalog at the NASA Exoplanet Archive\footref{footnote:exoplanetArchive}, with the cuts described in the baseline case of \citet{Bryson2020}. We summarize these cuts here for convenience.

We begin by merging the catalogs from \citet{Berger2020a}, the DR25 stellar catalog (with supplement), and \citet{berger18}, keeping only the 177,798 stars that are in all three catalogs.  We remove poorly characterized, binary and evolved stars, as well as stars whose observations were not well suited for long-period transit searches \citep{Burke2015, BurkeJCat2017} with the following cuts:
\begin{itemize}
    \item Remove stars with \citet{Berger2020a} goodness of fit ({\it iso\_gof}) $< 0.99$ and Gaia Renormalized Unit Weight Error \citep[RUWE;][]{gaiaRuwe2018}, as provided by \citet{Berger2020a}, $> 1.2$, leaving 162,219 stars.     \item Remove stars that, according to \citet{berger18}, are likely binaries, leaving 160,633 stars.
    \item Remove stars that have evolved off the main sequence, recomputing the Evol flag described in \citet{berger18} using the \citet{Berger2020a} stellar properties, leaving 105,118 stars.    
    \item Remove noisy targets identified in the KeplerPorts package\footnote{\url{https://github.com/nasa/KeplerPORTs/blob/master/DR25_DEModel_NoisyTargetList.txt}}, leaving 103,626 stars.
    \item Remove stars with NaN limb darkening coefficients, leaving 103,371 stars.
    \item Remove stars with NaN observation duty cycle, leaving 102,909 stars.
    \item Remove stars with a decrease in observation duty cycle $> 30\%$ due to data removal from other transits detected on these stars, leaving 98,672 stars.
    \item Remove stars with observation duty cycle $< 60\%$, leaving 95,335 stars.
    \item Remove stars with data span $< 1000$ days, leaving 87,765 stars.
    \item Remove stars with the DR25 stellar properties table {\it timeoutsumry} flag $\neq 1$, leaving 82,371 stars.
\end{itemize}
Selecting the FGK stars with effective temperature between 3900~K and 7300~K, which is a superset of the stellar populations we consider in this paper, we have 80,929 stars.

\begin{figure}[ht]
  \centering
  \includegraphics[width=0.95\linewidth]{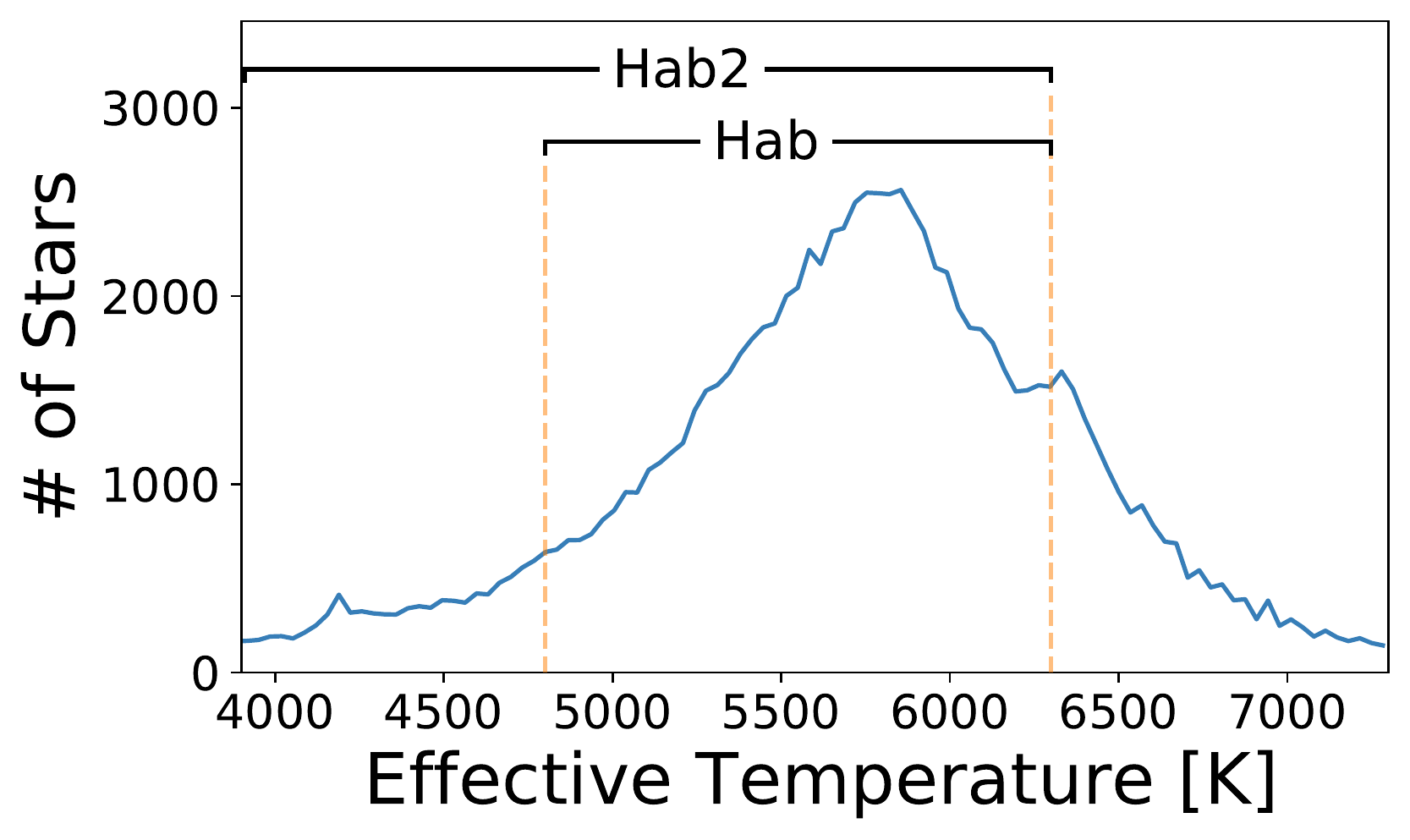} \\
\caption{The distribution of stellar effective temperature for the stellar populations used in this paper.  } \label{figure:stellarTeffDist}
\end{figure}

We are primarily interested in habitable zone occurrence rates for stars with effective temperatures hotter than 4800~K, and \Kepler\ observational coverage is very poor above 6300~K  (see \S\ref{section:habitability}).  We fit our population model using two stellar populations, and examine the solutions to determine which range of stellar temperature is best for computing the desired occurrence rate.  These stellar populations are:
\begin{itemize}
    \item {\bf Hab:} Stars with effective temperature between 4800~K and 6300~K (61,913 stars).
    \item {\bf Hab2:} Stars with effective temperature between 3900~K and 6300~K (68,885 stars).
\end{itemize}
The effective temperature distribution of the stars in these populations is shown in Figure~\ref{figure:stellarTeffDist}.  This distribution has considerably fewer cooler stars than we believe is the actual distribution of stars in the Galaxy.  Our analysis is weighted by the number of stars as a function of effective temperature.

There are two stellar population cuts recommended in \citet{BurkeJCat2017} that we do not apply.  The first is the requirement that stellar radii be less than 1.35 $R_\odot$ (1.25 $R_\odot$ in \citet{BurkeJCat2017}, but Burke now recommends 1.35 $R_\odot$ (private communication)). We do not impose this stellar radius cut, instead opting for the physically motivated selection based on the Evol flag.  After our cuts, 6.8\% of the the hab2 population contains stars that have radii larger than 1.35 $R_\odot$.  The completeness analysis for these stars is not expected to be as accurate as for smaller stars.

The second recommended cut that we do not apply is the requirement that the longest transit duration be less than 15 hours.  This cut is due to the fact planet search in the \Kepler\ pipeline does not examine transit durations longer than 15 hours \citep{Twicken2016}.  For the hab2 population, assuming circular orbits, transit durations of planets at the inner optimistic habitable zone boundary exceed 15 hours for 2.7\% of the stars.  Transit durations of planets at the outer optimistic habitable zone boundary exceed 15 hours for 35\% of the stars, with the duration being less than 25 hours for 98.7\% of the stars.  While transit durations longer than 15 hours will have an unknown impact on the completeness analysis of these stars, there is evidence that the impact is small.  KOI 5236.01, for example, has a transit duration of 14.54 hours, orbital period of 550.86 days and a transit signal to noise ratio (S/N) of 20.8.  KOI 5236.01 is correctly identified in the \Kepler\ pipeline in searches for transit durations of 3.5 to 15 hours.  KOI 7932.01, has a transit duration of 14.84 hours, orbital period of 502.256 days, and a transit S/N of 8.1, among the smallest transit S/N for planet candidates with period $> 450$ days.  KOI 7932.01 is correctly identified in searches using transit durations of 9 to 15 hours.  So even for low S/N transits, the transit can be identified in searches for transit durations 9/15 times the actual duration. If these examples are typical, we can expect that transit durations of up to 25 hours will be detected.  While these examples do not show that the impact of long transits on completeness is actually small, the bulk of these long durations occur in orbits beyond 500 days, so they are absorbed by the upper and lower bounds in completeness we describe in \S\ref{section:completenessExtrap}.  We are confident that long transit durations, as well as large stars, cause completeness to decrease, so their impact falls within these upper and lower bounds.  

\subsection{Planet Input Populations} \label{section:planetPopulation}
We use planet properties from the \Kepler\ DR25 Threshold Crossing Events (TCE) catalog \citep{Twicken2016}, with the Gaia-based planet radii and instellation flux from \citet{Berger2020b}.

Three DR25 small planet candidates that are near their host star's habitable zones (planet radius $\leq2.5~R_\oplus$ and instellation flux between 0.2 and 2.2 $I_\oplus$) are not included in our planet sample.  KOI 854.01 and KOI 4427.01 are orbiting host stars with effective temperatures $\leq 3900$~K, and KOI 7932.01's host star is cut from our stellar populations because it is marked ``Evolved'' \citep[see][]{Bryson2020}.

\subsection{Completeness and Reliability} \label{section:completenessAndReliability} 
\subsubsection{Detection and Vetting Completeness} \label{section:completeness}

The DR25 completeness products are based on {\it injected data} --- a ground-truth of transiting planets obtained by injecting artificial transit signals with known characteristics on all observed stars at the pixel level \citep{Christiansen2020}.  A large number of transits were also injected on a small number of target stars to measure the dependence of completeness on transit parameters and stellar properties \citep{BurkeJCat2017}. The data are then analyzed by the \Kepler\ detection pipeline \citep{Jenkins2010} to produce a catalog of detections at the injected ephemerides called {\it injected and recovered TCEs}, which are then sent through the same Robovetter used to identify planet candidates.  

{\bf Detection completeness} is defined as the fraction of injected transits that are recovered as TCEs by the \Kepler\ detection pipeline, regardless of whether or not those TCEs are subsequently identified as planet candidates.  We use the detection completeness of \citet{BurkeJCat2017}, which was computed for each target star as a function of period and simulated Multiple Event Statistic (MES), based on stellar noise properties measured in that star's \Kepler\ light curve.  MES is a measure of the signal-to-noise ratio (S/N) that is specific to the \Kepler\ pipeline \citep{Jenkins2010}.  The result is referred to as {\it completeness detection contours}.

{\bf Vetting completeness} is defined as the fraction of detected injected transits that were identified as planet candidates by the Robovetter \citep{Coughlin2017}.  We compute vetting completeness for a population of stars based on the simulated MES and orbital period of injected transits.  We use the method of \citet{Bryson2020}, which models vetting completeness as a binomial problem with a rate given by a product of rotated logistic functions of MES and orbital period.  We assume that vetting completeness and detection completeness are independent, so we can multiply them together to create combined completeness contours.

The product of vetting and detection completeness as a function of period and MES is converted to a function of period and planet radius for each star.  This product is further multiplied by the geometric transit probability for each star, which is a function of planet period and radius, given that star's radius.  The final result is a completeness contour for each star that includes detection and vetting completeness, and geometric transit probability.

We need to convert the completeness contours from radius and period to radius and instellation flux.  For each star, we first set the range of instellation flux to $0.2 \leq I \leq 2.2$, which contains the habitable zone for FGK stars.  We then interpolate the completeness contour from period to instellation flux via $I = R_*^2 T^4 / a^2 $, where $R_*$ is the stellar radius, $T = T_\mathrm{eff}/T_\odot$ is the effective temperature relative to the Sun, and $a$ is the semi-major axis of a circular orbit around this star with a given period.  

Once the completeness contours are interpolated onto radius and instellation flux for all stars, they are summed or averaged as required by the inference method used in \S\ref{section:inferenceMethods}.  

\begin{figure*}[ht]
  \centering
  \includegraphics[width=0.8\linewidth]{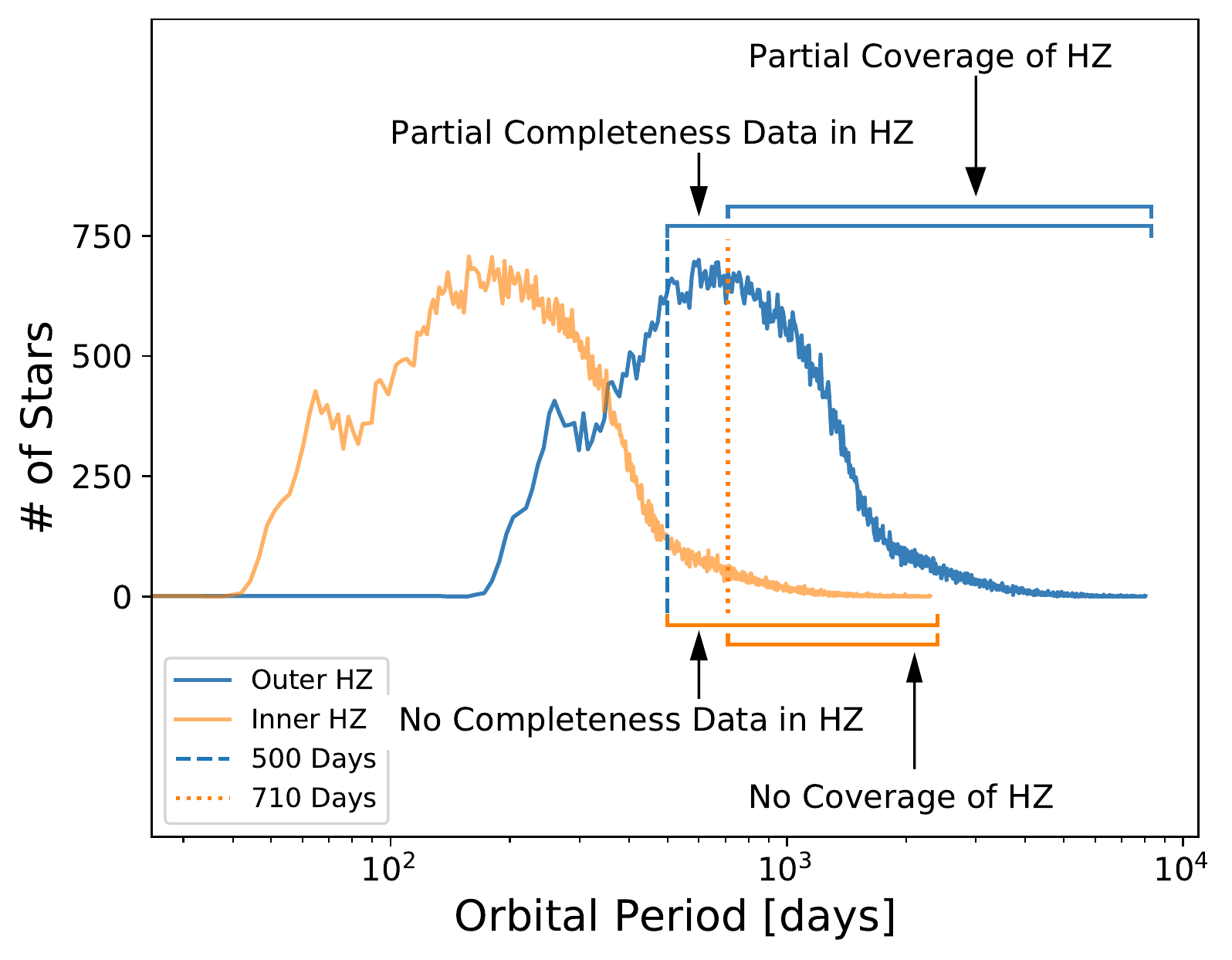} \\
\caption{The distribution orbital periods of the inner and outer optimistic habitable zone boundaries.  We show the orbital period distribution of circular orbits at the outer (blue) and inner (orange) boundaries of the optimistic habitable zone for our FGK stellar sample.  The blue vertical dashed line at 500 days indicates the limit of the completeness contours, beyond which there is no completeness data.  The orange vertical dotted line at 710 days shows the limit of \Kepler\ coverage, in the sense that beyond 710 days there is no possibility of three detected transits resulting in a planet detection.  Stars whose orbital period for the inner habitable zone boundary is beyond 500 days have no completeness data in their habitable zone, while stars whose outer habitable zone boundary orbital period is beyond 500 days require some completeness extrapolation.  Kepler cannot detect habitable zone planets for Stars whose inner habitable zone orbital period is beyond 710 days, while stars whose outer habitable zone orbital periods are beyond 710 days have only partial coverage, which will decrease completeness.  } \label{figure:starCoverage}
\end{figure*}

\subsubsection{Completeness Extrapolation} \label{section:completenessExtrap}
For most stars in our parent sample, there are regions of the habitable zone which require orbital periods beyond the 500-day limit of the period-radius completeness contours.  Figure~\ref{figure:starCoverage} shows the distribution of orbital periods at the inner and outer optimistic habitable zone boundaries for FGK stars in our stellar sample relative to the 500-day limit.  We see that a majority of these stars will require some completeness extrapolation to cover their habitable zones, and a small fraction of stars have no completeness information at all.  It is unknown precisely how the completeness contours will extrapolate out to longer period, but we believe that the possible completeness values can be bounded.

\begin{figure*}[ht]
  \centering
  \includegraphics[width=0.465\linewidth]{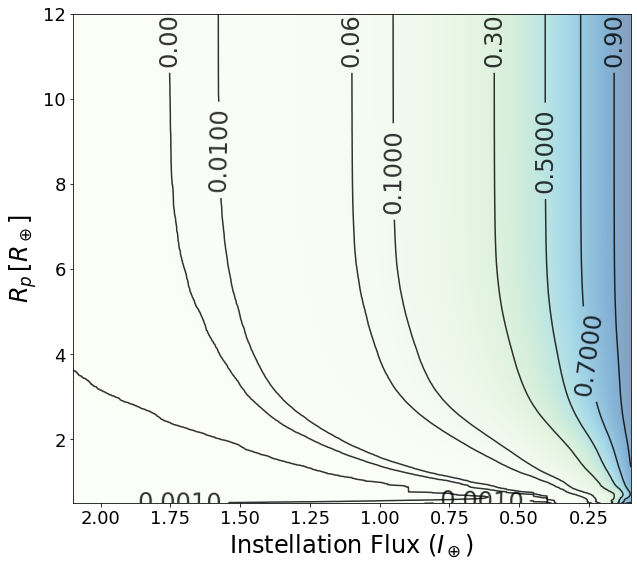}
  \includegraphics[width=0.48\linewidth]{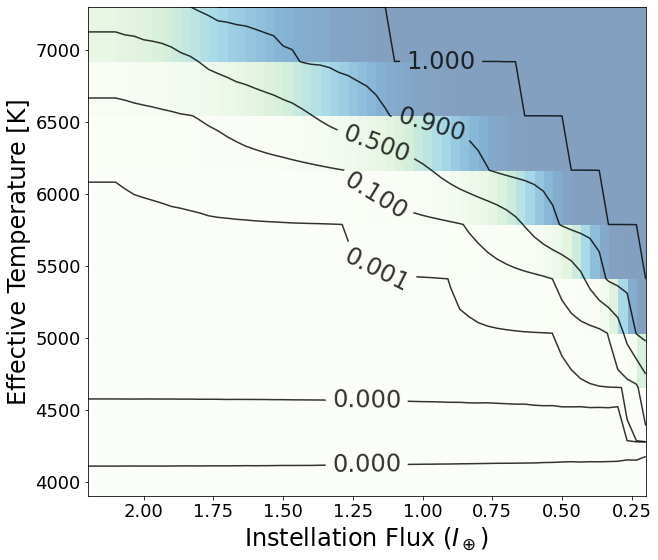} 
\caption{Left: The relative difference (difference divided by value) between the constant extrapolation and zero extrapolation completeness contours, summed over FGK stars, as a function of instellation flux and radius.  Right: the relative difference as a function of instellation flux and effective temperature} \label{figure:completenessDifference}
\end{figure*}

We assume that completeness is, on average, a decreasing function of orbital period.  Therefore, the completeness beyond 500 days will be less than the completeness at 500 days.  While this may not be a correct assumption for a small number of individual stars due to local completeness minima in period due to the window function \citep{BurkeJCat2017}, we have high confidence that this assumption is true on average.  We therefore bound the extrapolated completeness for each star by computing the two extreme extrapolation cases:
\begin{itemize}
    \item {\bf Constant completeness extrapolation,} where, for each radius bin, completeness for periods greater than 500 days is set to the completeness at 500 days.  This extrapolation will have higher completeness than reality, resulting in a smaller completeness correction and lower occurrence rates, which we take to be a lower bound.  In the tables below we refer to this lower bound as ``low'' values.  Here ``low'' refers to the resulting occurrence rates, and some population model parameters in the this case will have higher values.  
    \item {\bf Zero completeness extrapolation,}  where, for each radius bin, completeness for periods greater than 500 days is set to zero.  Zero completeness will have lower completeness than reality, resulting in a larger completeness correction and higher occurrence rates, which we take to be an upper bound.  In the tables below we refer to this upper bound as ``high'' values.  Here ``high'' refers to the resulting occurrence rates, and some population model parameters in  this case will have lower values.  
\end{itemize}
We solve for population models and compute occurrence rates for both extrapolation cases.  Figure~\ref{figure:completenessDifference} shows the relative difference in the completeness contours summed over all stars.  We see that for effective temperatures below $\sim$4500~K the difference between constant and zero completeness extrapolation is very close to zero, because these cooler stars are well-covered by completeness data (see Figure~\ref{figure:populations}).  We therefore expect the upper and lower occurrence rate bounds to converge for these stars.

\begin{figure*}[ht]
  \centering
  \includegraphics[width=0.495\linewidth]{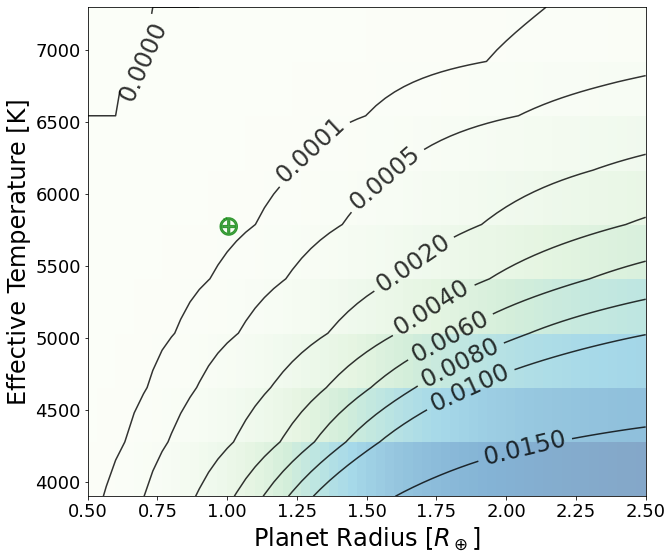}
  \includegraphics[width=0.48\linewidth]{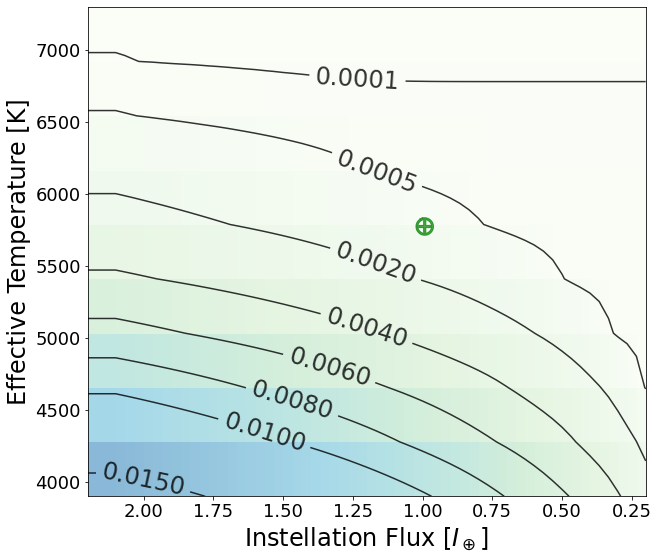} 
\caption{Example dependence of completeness on effective temperature, using the FGK stellar population and constant completeness extrapolation, which provides an upper completeness bound.  Left: Planet radius vs. effective temperature.  Right: Instellation flux vs. effective temperature.  
The location of the Earth-Sun system is shown with the $\oplus$ symbol.} \label{figure:teffCompleteness}
\end{figure*}

The Poisson likelihood we use requires the completeness contours summed over all stars, while the ABC method requires the completeness averaged over all stars.  We observe a significant dependence of summed completeness on effective temperature, shown in Figure~\ref{figure:teffCompleteness}.  We address this dependence of completeness on effective temperature by summing (for the Poisson likelihood) or averaging (for ABC) the completeness contours in effective temperature bins, as described in \S\ref{section:inferenceMethods}.

\subsubsection{Reliability}  \label{section:reliability}

We compute planet reliability as in \citet{Bryson2020}.  Because this is done as a function of multiple event statistic (MES) and period, there is no change from the methods of that paper.

\subsection{Computing the Population Model \texorpdfstring{$\lambda(I, r, T)$}{lambda}}  \label{section:determineLambda}
As described in \S\ref{section:ourWork}, we develop a planet population model using a parameterized differential rate function $\lambda$, and use Bayesian inference to find the model parameters that best explains the data.  To test the robustness of our results, we use both the Poisson-Likelihood MCMC method of \citet{Burke2015} and the Approximate Bayesian Computation method of \citet{Kunimoto2020a} to compute our population model. Both methods are modified to account for vetting completeness and reliability, with the Poisson-likelihood method described in \citet{Bryson2020} and the ABC method described in \citet{Kunimoto2020b} and \citet{Bryson2020b}. New to this work, we also take into account uncertainties in planet radius, instellation flux, and host star effective temperature, described in \S\ref{section:inferenceMethods}.

\subsubsection{Population Model Choices} \label{section:models}
We consider three population models for the differential rate function $\lambda(I, r, T)$.  These models are functions of instellation flux $I$, planet radius $r$ and stellar effective temperature $T_\mathrm{eff}$.  These models depend on possibly different sets of parameters, which we describe with the parameter vector $\boldsymbol{\theta}$.  For each model, we will solve for the $\boldsymbol{\theta}$ that best describes the planet candidate data.
\begin{equation} \label{eqn:models}
\begin{split}
    \lambda_1(I, r, T, \boldsymbol{\theta}) 
    &=  F_0 C_1 r^{\alpha} I^{\beta} T^{\gamma} g(T), 
    \ \boldsymbol{\theta} = \left(F_0, \alpha, \beta, \gamma \right) \\ 
    \lambda_2(I, r, T, \boldsymbol{\theta}) 
    &=  F_0 C_2 r^{\alpha} I^{\beta} T^{\gamma}, 
    \ \boldsymbol{\theta} = \left(F_0, \alpha, \beta, \gamma \right) \\ 
    \lambda_3(I, r, T, \boldsymbol{\theta}) 
    &=  F_0 C_3 r^{\alpha} I^{\beta} g(T), 
    \ \boldsymbol{\theta} = \left(F_0, \alpha, \beta \right) \\ 
\end{split}
\end{equation}
where $g(T)$ is given by Equation~(\ref{eqn:geomFunc}). The normalization constants $C_i$ in Equation~(\ref{eqn:models}) are chosen so that the integral of $\lambda$ from $r_{\min}$ to $r_{\max}$ and $I_{\min}$ to $I_{\max}$, averaged over $T_{\min}$ to $T_{\max}$, $= F_0$, so $F_0$ is the average number of planets per star in that radius, instellation flux and effective temperature range.

$\lambda_1$ allows for dependence on $T_\mathrm{eff}$ beyond the geometric dependence described in \S\ref{section:hzGeom}, breaking possible degeneracy between any intrinsic $T_\mathrm{eff}$ and the geometric dependence by fixing the geometric dependence as $g(T)$.  So, for example, if the planet population rate's dependence is entirely due to the larger HZ for hotter stars, captured in $\lambda_1$ by $g(T)$, then there is no additional $T_\mathrm{eff}$ dependence and $\gamma = 0$.  $\lambda_2$ does not separate out the geometric $T_\mathrm{eff}$ dependence.  $\lambda_3$ assumes that there is no $T_\mathrm{eff}$ dependence beyond the geometric effect.  

All models and inference calculations use uniform uninformative priors: $0 \leq F_0 \leq 50000$, $-5 \leq \alpha \leq 5$, $-5 \leq \beta \leq 5$, $-500 \leq \gamma \leq 50$.  The computations are initialized to a neighborhood of the maximum likelihood solution obtained with a standard non-linear solver.

\subsubsection{Inference Methods} \label{section:inferenceMethods}

Both the Poisson likelihood and ABC inference methods use the same stellar and planet populations, and the same characterization of completeness and reliability computed using the approach of \citet{Bryson2020}. These steps are as follows:
\begin{itemize}
    \item Select a target star population, which will be our parent population of stars that are searched for planets.  We apply various cuts intended to select well-behaved and well-observed stars.  We consider two such populations, defined by effective temperature range as described in \S\ref{section:stellarPopulation}, in order to explore the dependence of our results on the choice of stellar population.
    \item Use the injected data to characterize vetting completeness.
    \item Compute the detection completeness using a version of KeplerPorts\footnote{\url{https://github.com/nasa/KeplerPORTs}} modified for vetting completeness and insolation interpolation, incorporating vetting completeness and geometric probability for each star, and sum over the stars in effective temperature bins, as described in \S\ref{section:completeness}.
    \item Use observed, inverted, and scrambled data to characterize false alarm reliability, as described in \S\ref{section:reliability}.
    \item Assemble the collection of planet candidates, including computing the reliability of each candidate from the false alarm reliability and false positive probability.
    \item For each model in Equation~\ref{eqn:models}, use the Poisson likelihood or ABC methods to infer the model parameters $\boldsymbol{\theta}$ that are most consistent with the planet candidate data for the selected stellar population.
\end{itemize}
Because vetting completeness and reliability depend on the stellar population and the resulting planet catalog, all steps are computed for each choice of stellar population.  

A full study of the impact of the uncertainties in stellar and planet properties would include the impact of uncertainties on detection contours, and is beyond the scope of this paper.  However, we study the impact of the uncertainties in planet radius, instellation flux and host star effective temperature, shown Figure~\ref{figure:populations}, on our occurrence rates.  For both the Poisson likelihood and ABC methods we perform our inference computation both with and without uncertainties, which allows us to estimate the approximate contribution of input planet property uncertainties on the final occurrence rate uncertainties. In \citet{Bryson2020} it was shown that the uncertainties in reliability characterization have effectively no impact.  

For the Poisson likelihood inference of the parameters in Equation (\ref{eqn:models}) without input uncertainty, reliability is implemented by running the MCMC computation 100 times, with the planets removed with a probability given by their reliability.  The likelihood used in the Poisson method is Equation (\ref{equation:poisson5}) in Appendix~\ref{app:likelihoodDerivation}. For details see \citet{Bryson2020}.  

We treat input uncertainties similar to how we treat reliability: we run the Poisson MCMC inference 400 times, each time selecting the planet population according to reliability.  We then sample the planet instellation flux, radius and star effective temperature from the two-sided normal distribution with width given by the respective catalog uncertainties.  We perform this sampling prior to restricting to our period and instellation flux range of interest so planets whose median property values are outside the range may enter the range due to their uncertainties.  The posteriors from the 400 runs are concatenated together to give the posterior distribution of the parameters $\boldsymbol{\theta}$ for each model.  This approach to uncertainty does not recompute the underlying parent stellar population with re-sampled effective temperature uncertainties, because that would require re-computation of the completeness contours with each realization, which is beyond our computational resources.  \citet{Shabram2020} preforms a similar uncertainty study, properly re-sampling the underlying parent population and observes an impact of uncertainty similar to ours (see \S\ref{section:hzOccurrence}).  Our analysis of uncertainty should be considered an approximation.  While the result is not technically a sample from a posterior distribution, in \S\ref{section:results} we compare the resulting sample to the posterior sample from the model neglecting these uncertainties and find that the population parameter values and resulting occurrence rates change in a predictable way.   

The ABC-based inference of the parameters in Equation (\ref{eqn:models}) is computed using the approach of \citet{Kunimoto2020b}, with some modifications to accommodate temperature dependence and uncertainties on planet radius, instellation flux, and temperature.

In the ABC method, the underlying \Kepler\ population is simulated in each completeness effective temperature bin separately. $N_{p} = F_{0}N_{s}h(T)$ planets are drawn for each bin, where $N_{s}$ is the number of stars in the bin and $h(T)$ collects the model-dependent temperature terms from Equation (\ref{eqn:models}), averaged over the temperature range of the bin and normalized over the entire temperature range of the sample. Following the procedure of \citet{Mulders2018}, we assign each planet an instellation flux between 0.2 and 2.2 $I_{\oplus}$ from the cumulative distribution function of $I^{\beta}$, and a radius between 0.5 and 2.5 $R_{\oplus}$ from the cumulative distribution function of $r^{\alpha}$. The detectable planet sample is then simulated from this underlying population by drawing from a Bernoulli distribution with a detection probability averaged over the bin's stellar population. We compare the detected planets to the observed PC population using a distance function, which quantifies agreement between the flux distributions, radius distributions, and sample sizes of the catalogs. For the distances between the flux and radius distributions, we chose the two-sample Anderson-Darling (AD) statistic, which has been shown to be more powerful than the commonly used Kolmogorov-Smirnoff test \citep{Engmann2011}. The third distance is the modified Canberra distance from \citet{Hsu2019},

\begin{equation}\label{eqn:rho3}
\rho = \sum_{i=1}^{N} \frac{|n_{s,i} - n_{o,i}|}{\sqrt{n_{s,i} + n_{o,i}}},
\end{equation}

\noindent where $n_{s,i}$ and $n_{o,i}$ are the number of simulated and observed planets within the $i$th bin's temperature range, and the sum is over all $N$ bins.  For more details, see \citet{Bryson2020b}.  

These simulations are repeated within a Population Monte Carlo ABC algorithm to infer the parameters that give the closest match between simulated and observed catalogs. With each iteration of the ABC algorithm, model parameters are accepted when each resulting population's distance from the observed population is less than 75th quantile of the previous iteration's accepted distances. Following the guidance of \citet{prangle2017}, we confirmed that our algorithm converged by observing that the distances between simulated and observed catalogues approached zero with each iteration, and saw that the uncertainties on the model parameters flattened out to a noise floor.

This forward model is appropriate for estimating the average number of planets per star in a given flux, radius, and temperature range, similar to the Poisson likelihood method. However, rather than requiring many inferences on different catalogues to incorporate reliability or input uncertainty, we take a different approach. For reliability, we modify the distance function as described in \citet{Bryson2020b}. In summary, we replace the two-sample AD statistic with a generalized AD statistic developed in \citet{Trusina2020} that can accept a weight for each datapoint, and set each observed planet's weight equal to its reliability. We also alter the third distance (Equation~\ref{eqn:rho3}) so that a planet's contribution to the total number of planets in its bin is equal to its reliability. As demonstrated in \citet{Kunimoto2020b} and \citet{Bryson2020b}, this weighted distance approach gives results consistent with the Poisson likelihood function method with reliability. Meanwhile, to account for input uncertainty, the observed population is altered for every comparison with a simulated population by randomly assigning each observed planet a new radius, instellation flux, and host star effective temperature from the two-sided normal distribution with width given by their respective uncertainties.
\subsection{Computing Occurrence Rates} \label{section:computeOccurrence}
Once the population rate model $\lambda$ has been chosen and its parameters determined as described in \S\ref{section:determineLambda}, we can compute the number of habitable zone planets per star.  For planets with radius between $r_0$ and $r_1$ and instellation flux between $I_0$ and $I_1$, for a star with effective temperature $T_\mathrm{eff}$ the number of planets per star is
\begin{equation}
    f(T_\mathrm{eff}) = \int_{r_0}^{r_1}  \int_{I_0}^{I_1} \lambda(I, r, T_\mathrm{eff}, \boldsymbol{\theta})
        \, \mathrm{d} I \, \mathrm{d} r.
        \label{eqn:occRatePerStar}
\end{equation}
For a collection of stars with effective temperatures ranging from $T_0$ to $T_1$, we compute the average number of planets per star, assuming a uniform distribution of stars in that range, as 
\begin{equation}
    f = \frac{1}{T_1-T_0}\int_{T_0}^{T_1} f(T) \, \mathrm{d} T.
        \label{eqn:occRateAverage}
\end{equation}
We typically compute equation~(\ref{eqn:occRateAverage}) for every $\boldsymbol{\theta}$ in the posterior of our solution, giving a distribution of occurrence rates.  

The habitable zone is not a rectangular region in the $I$--$T_\mathrm{eff}$ plane (see Figure~\ref{figure:populations}), so to compute the occurrence in habitable zone for a given $T_\mathrm{eff}$, we integrate $I$ from the inner habitable zone flux $I_\mathrm{out}(T_\mathrm{eff})$ to the outer flux $I_\mathrm{in}(T_\mathrm{eff})$
\begin{equation}
    f_\mathrm{HZ}(T_\mathrm{eff}) = \int_{r_\mathrm{min}}^{r_\mathrm{max}}  \int_{I_\mathrm{out}(T_\mathrm{eff})}^{I_\mathrm{in}(T_\mathrm{eff})} \lambda(I, r, T_\mathrm{eff}, \boldsymbol{\theta})
        \, \mathrm{d} I \, \mathrm{d} r.
        \label{eqn:hzOccRatePerStar}
\end{equation}
The functions $I_\mathrm{out}(T)$ and $I_\mathrm{in}(T)$ are given in \citet{Kopparapu2014} and depend on the choice of the conservative vs. optimistic habitable zone.  $f_\mathrm{HZ}(T_\mathrm{eff})$ will be a distribution of occurrence rates if we use a distribution of $\boldsymbol{\theta}$.  For a collection of stars with effective temperatures ranging from $T_0$ to $T_1$, we compute $f_\mathrm{HZ}(T_\mathrm{eff})$ for a sampling of $T \in [T_0, T_1]$, and concatenate these distributions together to make a distribution of habitable zone occurrence rates $f_\mathrm{HZ}$ for that radius, flux and temperature range.  When we are computing $f_\mathrm{HZ}$ to determine the occurrence rate for a generic set of stars, we uniformly sample over $[T_0, T_1]$ (in practice we use all integer Kelvin values of $T \in [T_0, T_1]$).  The resulting distribution is our final result.

\begin{figure}[ht]
  \centering
  \includegraphics[width=0.95\linewidth]{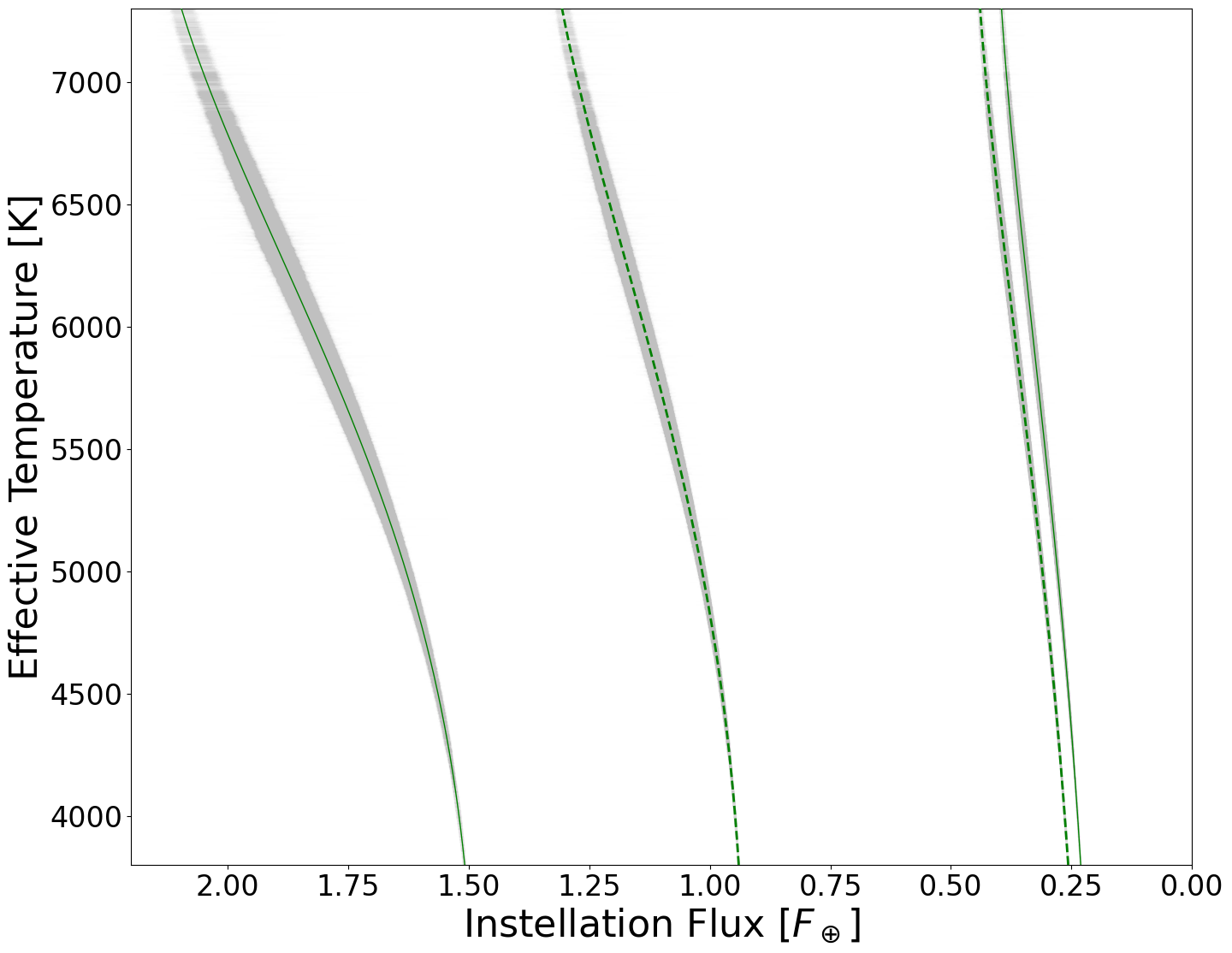} \\
\caption{The uncertainty in the habitable zone boundaries due to uncertainty in stellar effective temperature.  For every star, the inner and outer boundaries of the habitable zone is shown green, with the 68\% credible interval for each boundary shown in grey.  The solid green line is the optimistic habitable zone, and the dashed green line is the conservative habitable zone.} \label{figure:hzError}
\end{figure}

Figure~\ref{figure:hzError} shows the impact of uncertainty in stellar effective temperature on the habitable zone boundaries.  For each star we computed the uncertainty in the habitable zone boundaries with 100 realizations of that star's effective temperature with uncertainty, modeled as a two-sided Gaussian.  The grey regions in Figure~\ref{figure:hzError} show the 86\% credible intervals of the uncertainty of the habitable zone boundaries.  These intervals are small relative to the size of the habitable zone, and are well centered on the central value.  For example, consider the inner optimistic habitable zone boundary, which has the widest error distribution in Figure~\ref{figure:hzError}. The median of the difference between the median habitable zone uncertainty and the habitable zone boundary without uncertainty is less than 0.002\%, with a standard deviation less than 0.9\%.  Therefore, we do not believe that uncertainties in habitable zone boundaries resulting from stellar effective temperature uncertainties have a significant impact on occurrence rates.

\section{Results} \label{section:results}

\subsection{Inferring the Planet Population Model Parameters} \label{section:PopParams}

\begin{figure*}[ht]
  \centering
  \includegraphics[width=0.48\linewidth]{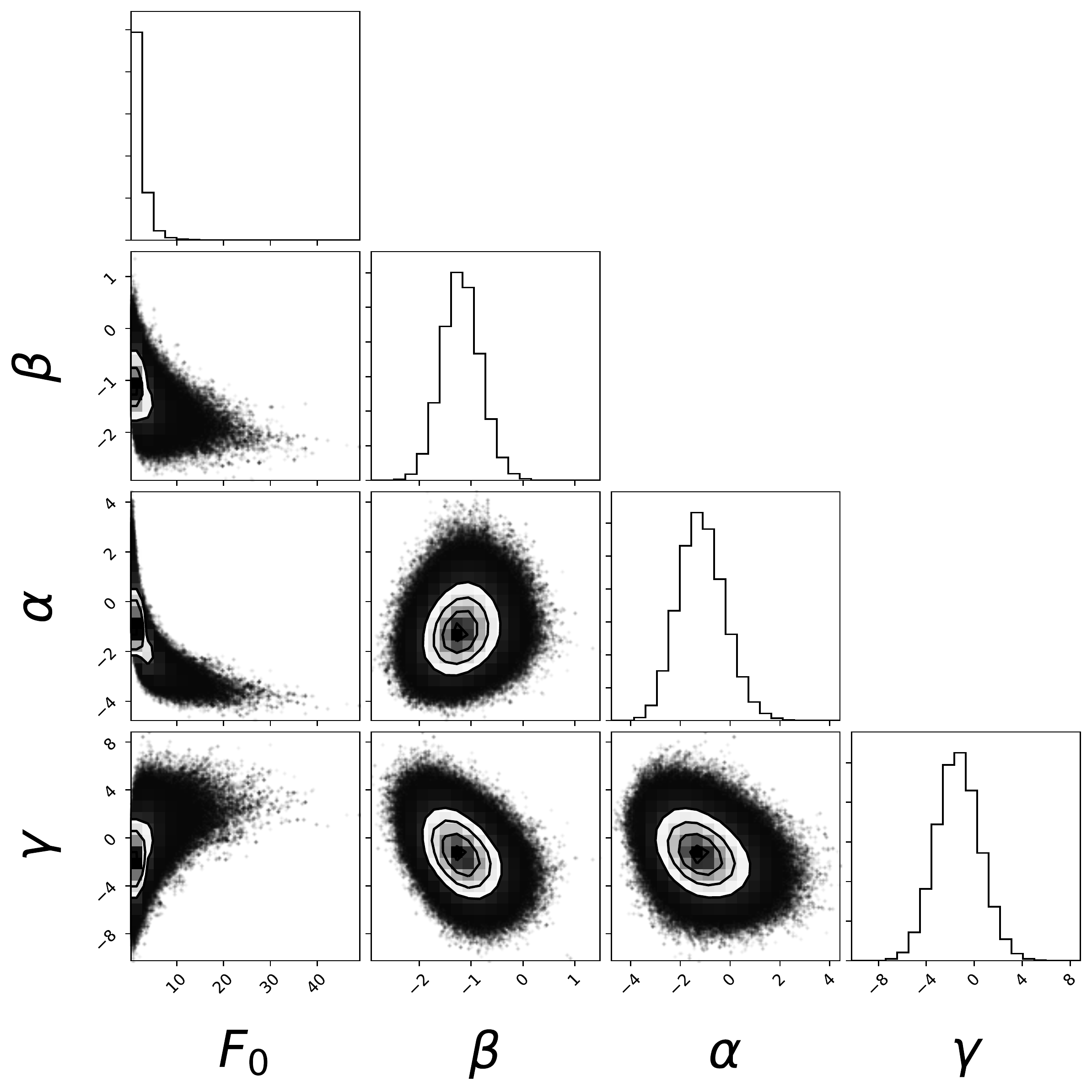} 
  \includegraphics[width=0.48\linewidth]{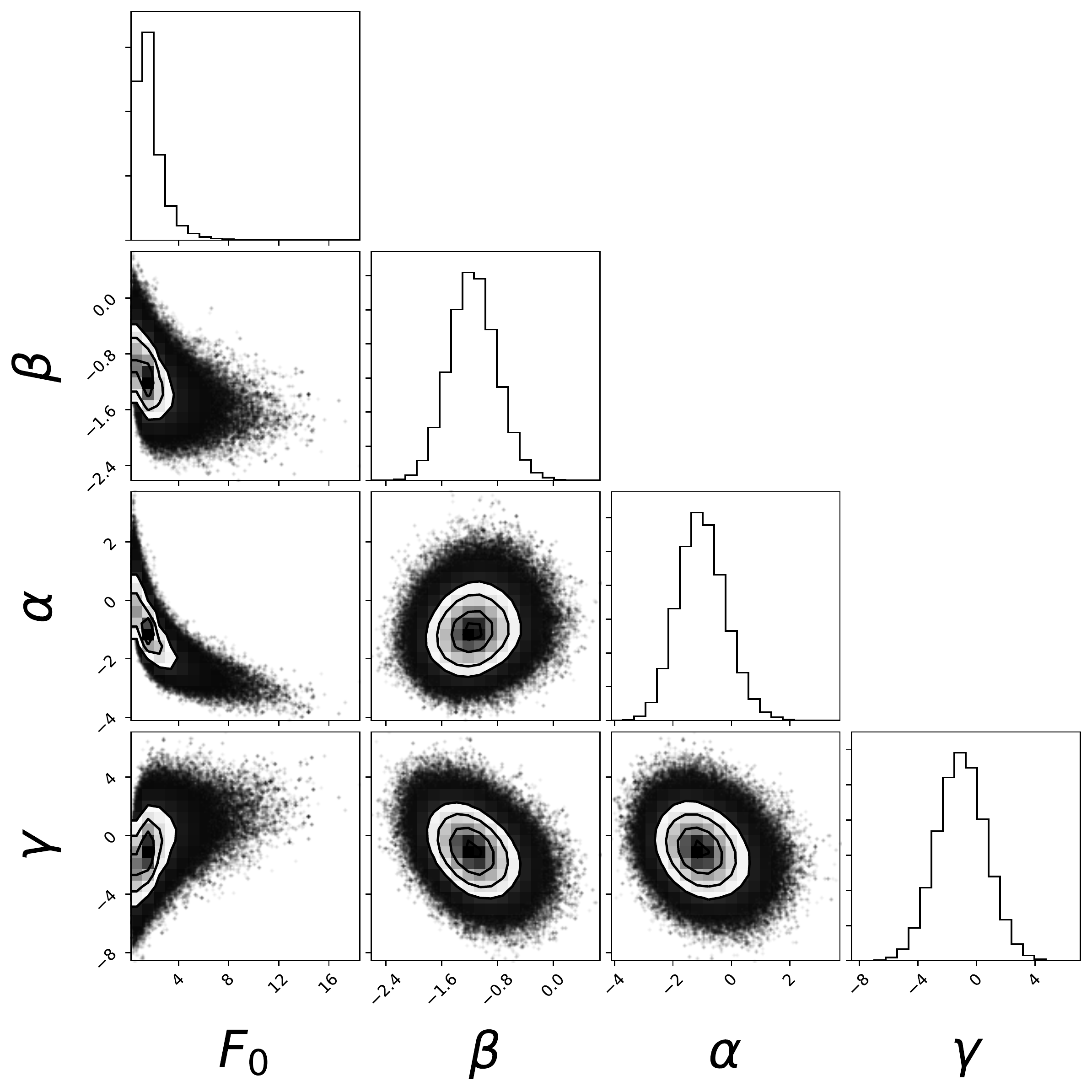} 
\caption{The posterior distributions of model 1 from Equation (\ref{eqn:models}) for the hab2 stellar population and zero-extrapolation completeness.  Left: with input uncertainty.  Right: without input uncertainty} \label{figure:occPost}
\end{figure*}

\begin{figure*}[ht]
  \centering
  \Large With Uncertainty \\
  \includegraphics[width=0.95\linewidth]{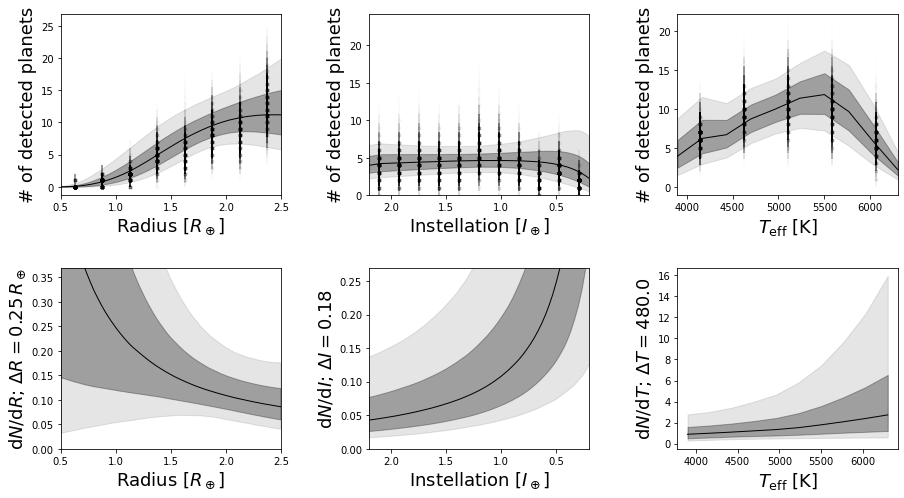} \\
  \Large Without Uncertainty \\
  \includegraphics[width=0.95\linewidth]{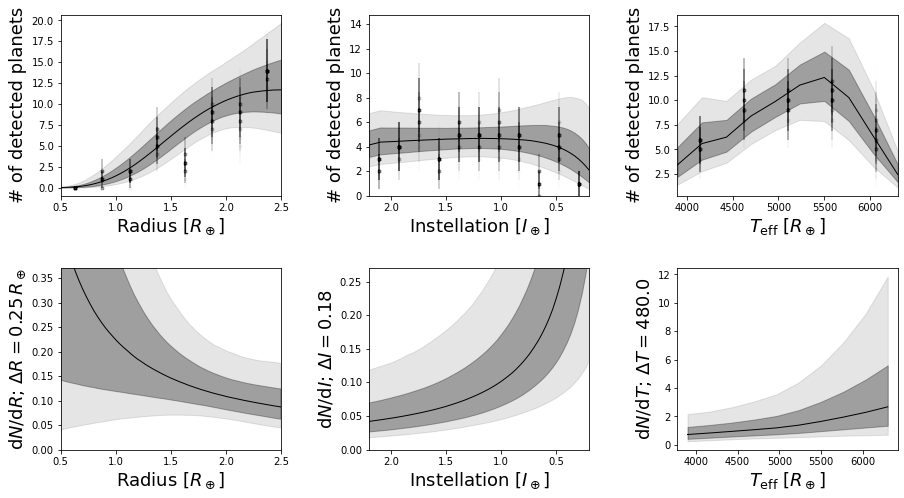}

\caption{The marginalized population rate of model 1 from Equation (\ref{eqn:models}) for the hab2 stellar population and zero-extrapolation completeness, with and without incorporating uncertainties on planet radius, instellation flux and host star effective temperature.  The top row for each case shows the completeness corrected population model compared with the observed planet population.  The bottom row for each case shows the underlying population model.  The dark grey regions are the 68\% credible intervals, and the light gray regions are the 95\% credible intervals.} \label{figure:occMarg}
\end{figure*}

\begin{figure*}[ht]
  \centering
  \includegraphics[width=0.98\linewidth]{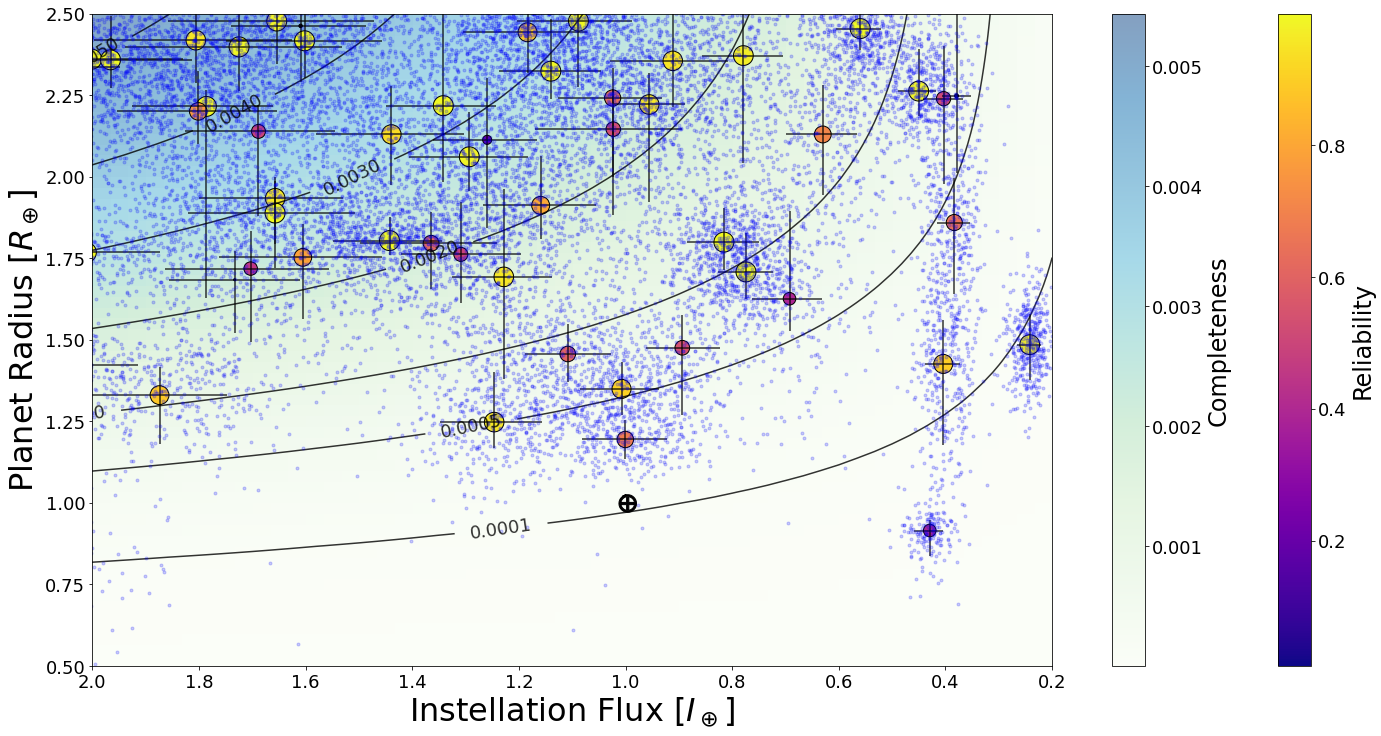}
\caption{The union of the planet instellation flux and radii for the 400 realizations in the uncertainty run shown by the blue dots superimposed on the lower panel of Figure~\ref{figure:populations}, for the hab2 stellar population and model 1.  See the caption of Figure~\ref{figure:populations} for explanation of the other elements of the figure.} \label{figure:uncertainty}
\end{figure*}

For each choice of population differential rate model from Equation~(\ref{eqn:models}) and stellar population from \S\ref{section:stellarPopulation}, we determine the parameter vector $\boldsymbol{\theta}$ with zero-extrapolated and constant extrapolated completeness, giving high and low bounds on occurrence rates. These solutions were computed over the radius range $0.5  \ R_\oplus \leq r \leq 2.5 \ R_\oplus$ and instellation flux range $0.2 \ I_\oplus \leq I \leq 2.2 \ I_\oplus$ using the hab and hab2 stellar populations described in \S\ref{section:stellarPopulation}.   We perform these calculations both without and with input uncertainties in the planet radius, instellation flux, and $T_\mathrm{eff}$ shown in Figure~\ref{figure:populations}.  Example of the resulting $\boldsymbol{\theta}$ posterior distributions are shown in Figure~\ref{figure:occPost} and the corresponding rate functions $\lambda$ are shown in Figure~\ref{figure:occMarg}.  The solutions for models 1--3 for the hab and hab2 stellar populations computed using the Poisson likelihood method are given in Table~\ref{table:allFits}, and the results computed using ABC are given in Table~\ref{table:allFitsABC}.  An example of the sampled planet population using input uncertainties is shown in Figure~\ref{figure:uncertainty}.  

We compared models 1--3 using the Akaike Information Criterion (AIC), but AIC did not indicate that one of the models was significantly more consistent with the data than another.  The resulting relative likelihoods from the AIC analysis relative to model 1 were 0.31 for model 2 and 2.07 for model 3.  Such small relative likelihoods ratios are not considered compelling.

The $F_0$ parameter, giving the average number of planets per star in the solution domain (see \S\ref{eqn:models}) indicates that solutions using zero completeness extrapolation (see \S\ref{section:completenessExtrap}), yield higher occurrence than constant completeness extrapolation. This is because zero completeness extrapolation induces larger completeness corrections.  The zero completeness extrapolation solution provides an upper bound on the habitable zone occurrence rate, and the constant extrapolation solution provides a lower bound.  Reality will be somewhere between, and is likely to be closer to the zero extrapolation case for hotter stars, which have lower completeness in their habitable zone.  

The hab stellar population is a subset of the hab2 population, so one may expect that they give similar solutions.  But the hab stellar population contain significant regions of extrapolated completeness and low-reliability planet candidates for the flux range considered in our solution (see Figure~\ref{figure:populations}).  While the hab2 population contains the same regions, hab2 also contains many cooler stars that host higher-reliability planet candidates.  These stars provide a better constraint on the power laws we use to describe the population.  The result is that the hab2 solution has significantly smaller uncertainties than the hab solution, as seen in Tables~\ref{table:allFits} and \ref{table:allFitsABC}. 


\renewcommand{\arraystretch}{1.25}
\setlength{\tabcolsep}{9pt}
\begin{table*}[ht]
\centering
\caption{Parameter fits with 68\% confidence limits for models 1--3 from Equation~\ref{eqn:models} for the hab and hab2 stellar populations from \S\ref{section:stellarPopulation}, computed with the Poisson likelihood method. }\label{table:allFits}
\begin{tabular}{ r c c | c c }
\hline
\hline
& \multicolumn{2}{c}{With Uncertainty} & \multicolumn{2}{c}{Without Uncertainty} \\
 & based on hab Stars
 & based on hab2 Stars
 & based on hab Stars
 & based on hab2 Stars
\\
& low bound -- high bound & low bound -- high bound & low bound -- high bound & low bound -- high bound \\
\hline
& \multicolumn{4}{c}{Model 1} \\
$F_0$
 & 
$+1.08^{+1.56}_{-0.57}$
 -- 
$+1.97^{+3.73}_{-1.17}$
 & 
$+1.11^{+0.88}_{-0.44}$
 -- 
$+1.59^{+1.56}_{-0.7}$
 & 
$+0.70^{+0.77}_{-0.32}$
 -- 
$+1.41^{+2.12}_{-0.77}$
 & 
$+1.02^{+0.66}_{-0.37}$
 -- 
$+1.46^{+1.18}_{-0.59}$
\\
$\alpha$
 & 
$-1.05^{+1.41}_{-1.2}$
 -- 
$-1.09^{+1.36}_{-1.18}$
 & 
$-1.08^{+0.94}_{-0.85}$
 -- 
$-1.18^{+0.96}_{-0.87}$
 & 
$-0.29^{+1.39}_{-1.18}$
 -- 
$-0.51^{+1.35}_{-1.15}$
 & 
$-0.96^{+0.83}_{-0.74}$
 -- 
$-1.03^{+0.83}_{-0.77}$
\\
$\beta$
 & 
$-0.56^{+0.48}_{-0.42}$
 -- 
$-1.18^{+0.6}_{-0.56}$
 & 
$-0.84^{+0.32}_{-0.3}$
 -- 
$-1.19^{+0.37}_{-0.36}$
 & 
$-0.43^{+0.46}_{-0.4}$
 -- 
$-1.13^{+0.54}_{-0.5}$
 & 
$-0.78^{+0.3}_{-0.28}$
 -- 
$-1.15^{+0.34}_{-0.33}$
\\
$\gamma$
 & 
$-1.84^{+3.33}_{-3.39}$
 -- 
$+0.91^{+3.87}_{-3.88}$
 & 
$-2.67^{+1.59}_{-1.57}$
 -- 
$-1.38^{+1.84}_{-1.78}$
 & 
$-2.13^{+3.06}_{-3.13}$
 -- 
$+0.25^{+3.39}_{-3.47}$
 & 
$-2.33^{+1.47}_{-1.46}$
 -- 
$-1.03^{+1.66}_{-1.64}$
\\
& \multicolumn{4}{c}{Model 2} \\
$F_0$
 & 
$+1.04^{+1.52}_{-0.55}$
 -- 
$+1.96^{+3.72}_{-1.18}$
 & 
$+1.13^{+0.92}_{-0.46}$
 -- 
$+1.50^{+1.43}_{-0.66}$
 & 
$+0.80^{+0.9}_{-0.38}$
 -- 
$+1.33^{+1.98}_{-0.71}$
 & 
$+1.00^{+0.7}_{-0.38}$
 -- 
$+1.43^{+1.21}_{-0.6}$
\\
$\alpha$
 & 
$-1.03^{+1.41}_{-1.23}$
 -- 
$-1.05^{+1.45}_{-1.18}$
 & 
$-1.13^{+0.97}_{-0.86}$
 -- 
$-1.12^{+0.94}_{-0.86}$
 & 
$-0.48^{+1.36}_{-1.17}$
 -- 
$-0.47^{+1.33}_{-1.16}$
 & 
$-0.92^{+0.88}_{-0.78}$
 -- 
$-1.01^{+0.88}_{-0.8}$
\\
$\beta$
 & 
$-0.56^{+0.48}_{-0.42}$
 -- 
$-1.20^{+0.6}_{-0.56}$
 & 
$-0.85^{+0.32}_{-0.3}$
 -- 
$-1.18^{+0.37}_{-0.35}$
 & 
$-0.51^{+0.45}_{-0.4}$
 -- 
$-1.09^{+0.55}_{-0.51}$
 & 
$-0.80^{+0.31}_{-0.28}$
 -- 
$-1.18^{+0.34}_{-0.33}$
\\
$\gamma$
 & 
$+2.60^{+3.56}_{-3.61}$
 -- 
$+5.33^{+3.94}_{-4.02}$
 & 
$+1.19^{+1.64}_{-1.63}$
 -- 
$+2.31^{+1.91}_{-1.85}$
 & 
$+2.03^{+3.1}_{-3.23}$
 -- 
$+4.66^{+3.43}_{-3.51}$
 & 
$+1.26^{+1.54}_{-1.54}$
 -- 
$+2.73^{+1.73}_{-1.71}$
\\
& \multicolumn{4}{c}{Model 3} \\
$F_0$
 & 
$+1.13^{+1.52}_{-0.58}$
 -- 
$+1.83^{+2.76}_{-1.0}$
 & 
$+1.41^{+1.14}_{-0.59}$
 -- 
$+1.89^{+1.51}_{-0.78}$
 & 
$+0.89^{+0.94}_{-0.4}$
 -- 
$+1.24^{+1.49}_{-0.6}$
 & 
$+1.25^{+0.93}_{-0.5}$
 -- 
$+1.75^{+1.26}_{-0.7}$
\\
$\alpha$
 & 
$-1.08^{+1.39}_{-1.18}$
 -- 
$-1.06^{+1.38}_{-1.18}$
 & 
$-1.37^{+0.91}_{-0.83}$
 -- 
$-1.28^{+0.9}_{-0.82}$
 & 
$-0.60^{+1.3}_{-1.12}$
 -- 
$-0.38^{+1.34}_{-1.18}$
 & 
$-1.21^{+0.88}_{-0.78}$
 -- 
$-1.17^{+0.85}_{-0.76}$
\\
$\beta$
 & 
$-0.56^{+0.48}_{-0.41}$
 -- 
$-1.16^{+0.55}_{-0.51}$
 & 
$-0.89^{+0.32}_{-0.29}$
 -- 
$-1.29^{+0.35}_{-0.32}$
 & 
$-0.49^{+0.45}_{-0.39}$
 -- 
$-1.06^{+0.53}_{-0.48}$
 & 
$-0.83^{+0.3}_{-0.28}$
 -- 
$-1.26^{+0.33}_{-0.31}$
\\
\end{tabular}
\tablecomments{The low and high bounds correspond to the constant and zero completeness extrapolation of \S\ref{section:completenessExtrap}.  ``With Uncertainty'' means planet candidate radius, instellation flux and host star effective temperature uncertainties were taken into account.}
\end{table*}

\renewcommand{\arraystretch}{1.25}
\setlength{\tabcolsep}{9pt}
\begin{table*}[ht]
\centering
\caption{Parameter fits with 68\% confidence limits for models 1--3 from Equation~\ref{eqn:models} for the four stellar populations from \S\ref{section:stellarPopulation}, computed with the ABC method. }\label{table:allFitsABC}
\begin{tabular}{ r c c | c c }
\hline
\hline
& \multicolumn{2}{c}{With Uncertainty} & \multicolumn{2}{c}{Without Uncertainty} \\
 & based on hab Stars
 & based on hab2 Stars
 & based on hab Stars
 & based on hab2 Stars
\\
& low bound -- high bound & low bound -- high bound & low bound -- high bound & low bound -- high bound \\
\hline
& \multicolumn{4}{c}{Model 1} \\
$F_0$
 & 
$+1.18^{+0.95}_{-0.56}$ -- $+2.04^{+1.44}_{-0.99}$
 & 
$+1.17^{+0.78}_{-0.44}$ -- $+1.61^{+1.05}_{-0.65}$
 & 
$+0.73^{+0.54}_{-0.29}$ -- $+1.37^{+1.08}_{-0.61}$
 & 
$+0.94^{+0.45}_{-0.32}$ -- $+1.41^{+0.99}_{-0.57}$
\\
$\alpha$
 & 
$-1.14^{+1.02}_{-0.89}$ -- $-0.95^{+0.99}_{-0.86}$
 & 
$-1.14^{+0.75}_{-0.77}$ -- $-1.18^{+0.72}_{-0.67}$
 & 
$-0.17^{+1.19}_{-0.97}$ -- $-0.11^{+1.17}_{-0.88}$
 & 
$-0.71^{+0.67}_{-0.68}$ -- $-0.83^{+0.77}_{-0.74}$
\\
$\beta$
 & 
$-0.69^{+0.41}_{-0.38}$ -- $-1.32^{+0.51}_{-0.44}$
 & 
$-0.90^{+0.31}_{-0.26}$ -- $-1.26^{+0.35}_{-0.31}$
 & 
$-0.67^{+0.38}_{-0.35}$ -- $-1.30^{+0.43}_{-0.41}$
 & 
$-0.89^{+0.25}_{-0.25}$ -- $-1.26^{+0.30}_{-0.32}$
\\
$\gamma$
 & 
$-0.84^{+3.81}_{-4.11}$ -- $+2.16^{+3.91}_{-3.68}$
& 
$-2.60^{+1.74}_{-1.84}$ -- $-1.14^{+2.15}_{-2.02}$
 & 
$-1.93^{+3.54}_{-3.37}$ -- $+1.82^{+3.94}_{-3.77}$
 & 
$-2.27^{+1.65}_{-1.71}$ -- $-0.78^{+2.11}_{-1.91}$
\\
& \multicolumn{4}{c}{Model 2} \\
$F_0$
 & 
$+1.06^{+0.90}_{-0.48}$ -- $+1.88^{+1.38}_{-0.87}$
& 
$+1.14^{+0.74}_{-0.43}$ -- $+1.66^{+1.26}_{-0.70}$
& 
$+0.70^{+0.44}_{-0.28}$ -- $+1.22^{+0.87}_{-0.57}$
 & 
$+0.96^{+0.45}_{-0.34}$ -- $+1.35^{+0.85}_{-0.51}$
\\
$\alpha$
 & 
$-1.07^{+1.11}_{-0.93}$ -- $-0.96^{+0.90}_{-0.90}$
& 
$-1.15^{+0.76}_{-0.76}$ -- $-1.25^{+0.80}_{-0.72}$
& 
$-0.06^{+1.02}_{-0.88}$ -- $-0.09^{+1.07}_{-0.98}$
 & 
$-0.75^{+0.70}_{-0.71}$ -- $-0.82^{+0.76}_{-0.66}$
\\
$\beta$
 & 
$-0.67^{+0.41}_{-0.35}$ -- $-1.30^{+0.45}_{-0.44}$
& 
$-0.90^{+0.28}_{-0.27}$ -- $-1.24^{+0.33}_{-0.36}$
& 
$-0.64^{+0.39}_{-0.36}$ -- $-1.19^{+0.43}_{-0.42}$
 & 
$-0.89^{+0.26}_{-0.23}$ -- $-1.27^{+0.32}_{-0.31}$
\\
$\gamma$
 & 
$+3.09^{+3.64}_{-3.42}$ -- $+5.68^{+2.60}_{-3.42}$
& 
$+1.34^{+1.95}_{-2.04}$ -- $+2.85^{+2.21}_{-2.24}$
& 
$+2.72^{+3.19}_{-3.78}$ -- $+5.03^{+2.90}_{-3.58}$
 & 
$+1.44^{+1.68}_{-1.64}$ -- $+3.04^{+2.01}_{-1.97}$
\\
& \multicolumn{4}{c}{Model 3} \\
$F_0$
 & 
$+1.20^{+0.93}_{-0.57}$ -- $+1.75^{+1.19}_{-0.77}$
 & 
$+1.57^{+0.93}_{-0.58}$ -- $+1.90^{+1.08}_{-0.69}$
& 
$+0.82^{+0.56}_{-0.31}$ -- $+1.21^{+0.82}_{-0.49}$
 & 
$+1.25^{+0.69}_{-0.47}$ -- $+1.62^{+0.81}_{-0.59}$
\\
$\alpha$
 & 
$-1.16^{+1.14}_{-0.88}$ -- $-0.93^{+0.98}_{-0.81}$
& 
$-1.54^{+0.68}_{-0.67}$ -- $-1.32^{+0.71}_{-0.65}$
& 
$-0.27^{+1.02}_{-0.98}$ -- $-0.09^{+1.06}_{-0.99}$
 & 
$-1.14^{+0.76}_{-0.70}$ -- $-0.95^{+0.70}_{-0.67}$
\\
$\beta$
 & 
$-0.71^{+0.41}_{-0.36}$ -- $-1.24^{+0.42}_{-0.41}$
& 
$-0.97^{+0.27}_{-0.23}$ -- $-1.33^{+0.26}_{-0.27}$
& 
$-0.66^{+0.32}_{-0.34}$ -- $-1.24^{+0.42}_{-0.39}$
 & 
$-0.96^{+0.25}_{-0.23}$ -- $-1.35^{+0.27}_{-0.26}$
\\
\end{tabular}
\tablecomments{The low and high bounds correspond to the constant and zero completeness extrapolation of \S\ref{section:completenessExtrap}.  ``With Uncertainty'' means planet candidate radius, instellation flux and host star effective temperature uncertainties were taken into account.  }
\end{table*}

\subsection{Habitable Zone Occurrence Rates} \label{section:hzOccurrence}
\begin{figure*}[ht]
  \centering
  \Large With Uncertainty \\
  \includegraphics[width=0.48\linewidth]{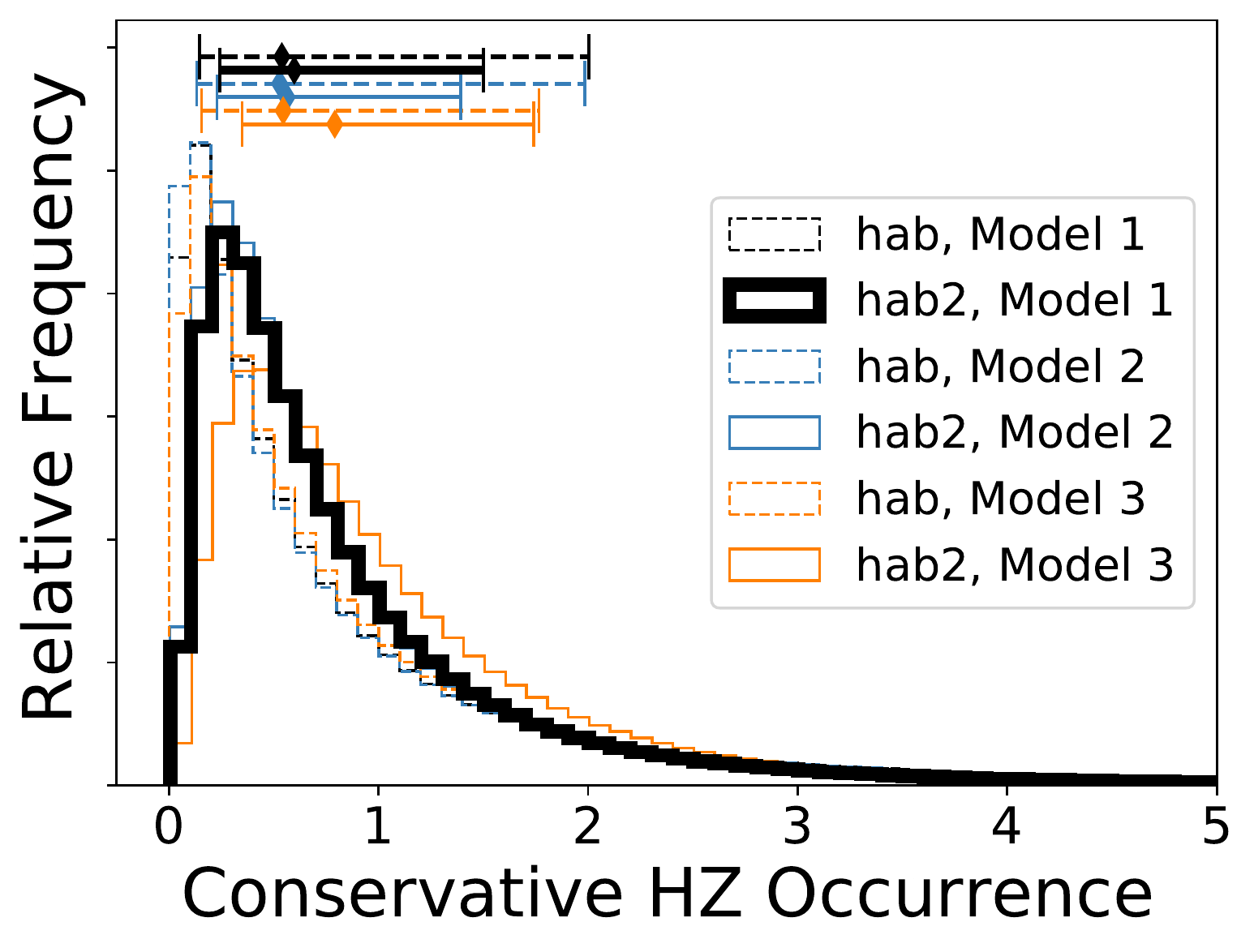}
  \includegraphics[width=0.48\linewidth]{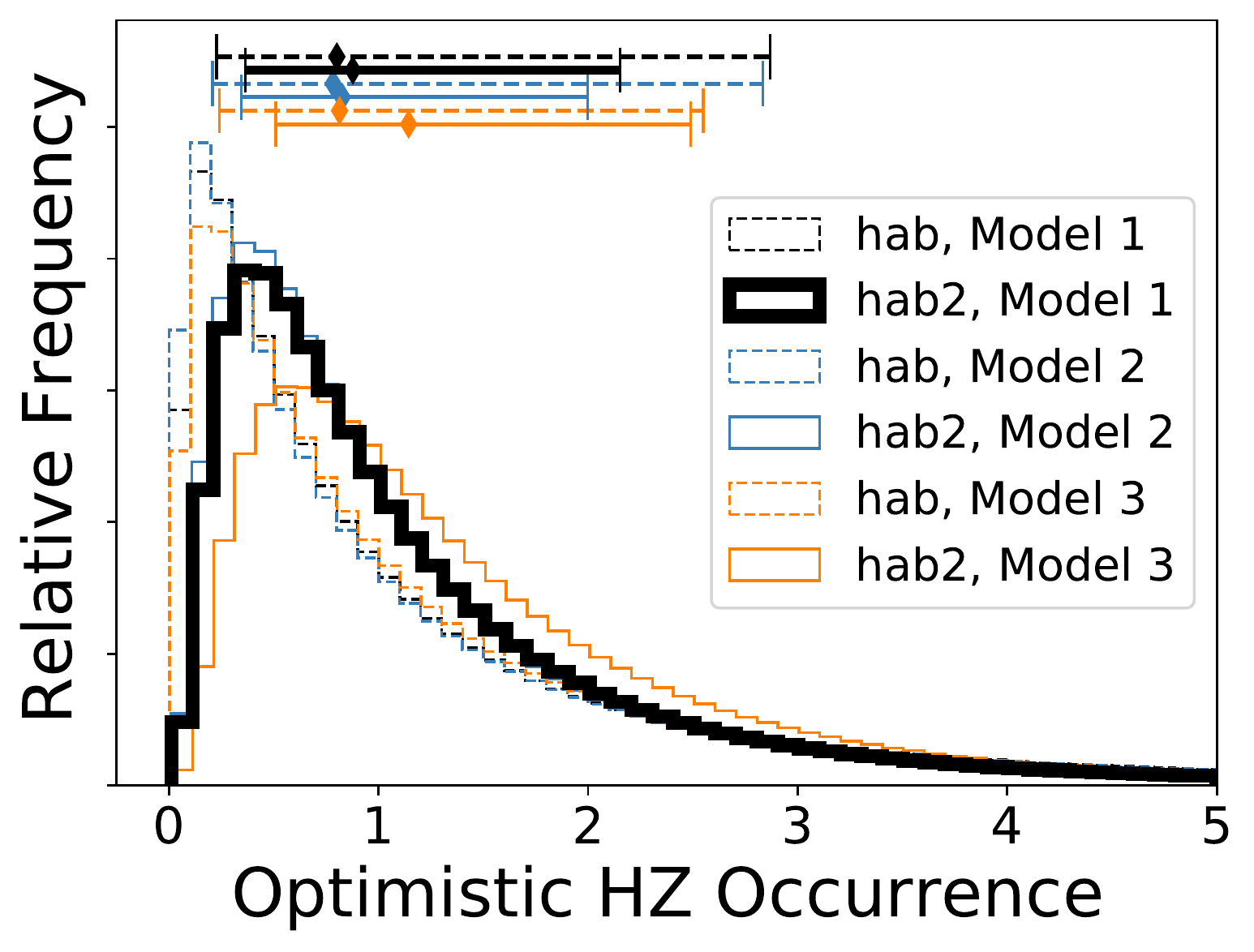} \\
  \Large No Uncertainty \\
  \includegraphics[width=0.48\linewidth]{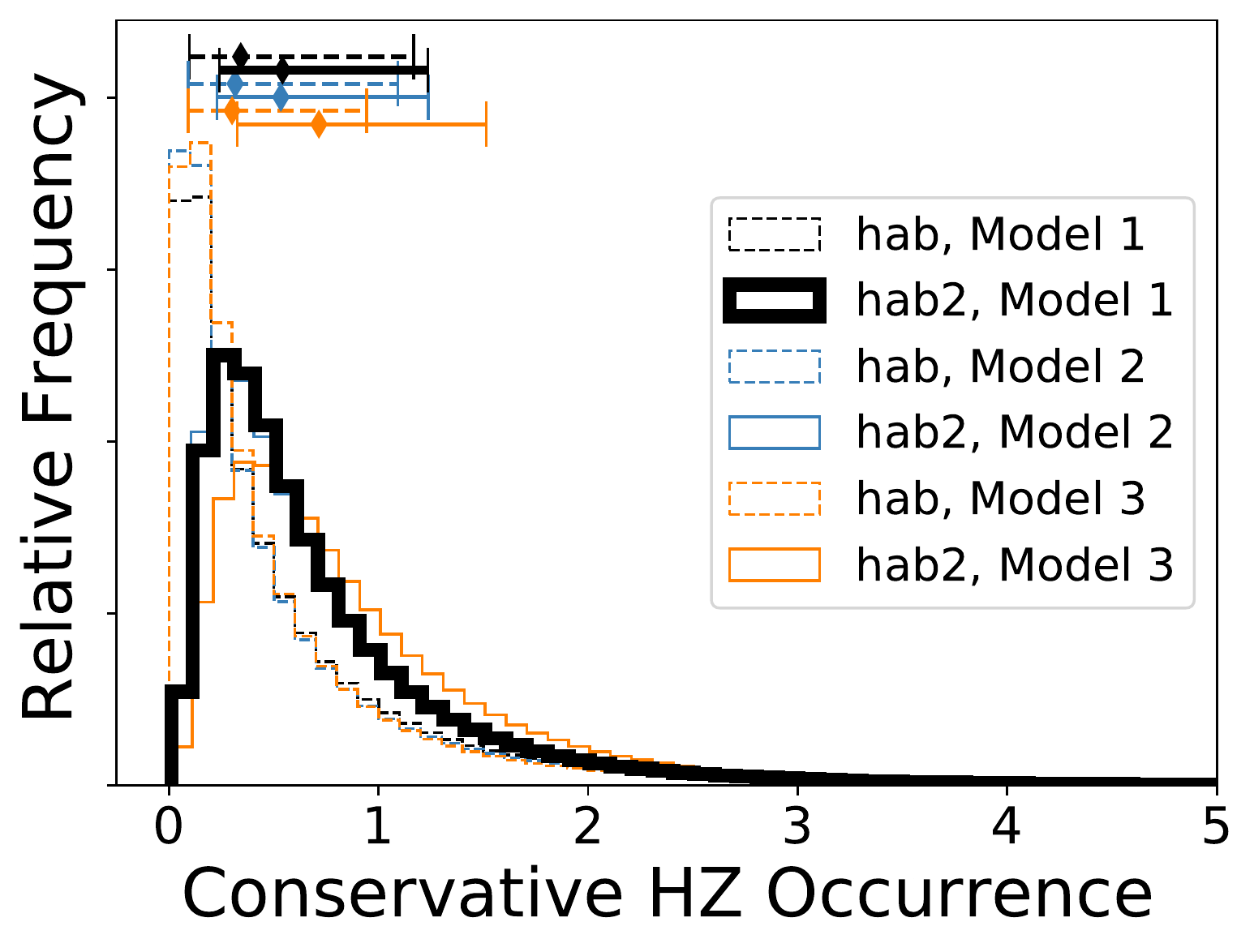}
  \includegraphics[width=0.48\linewidth]{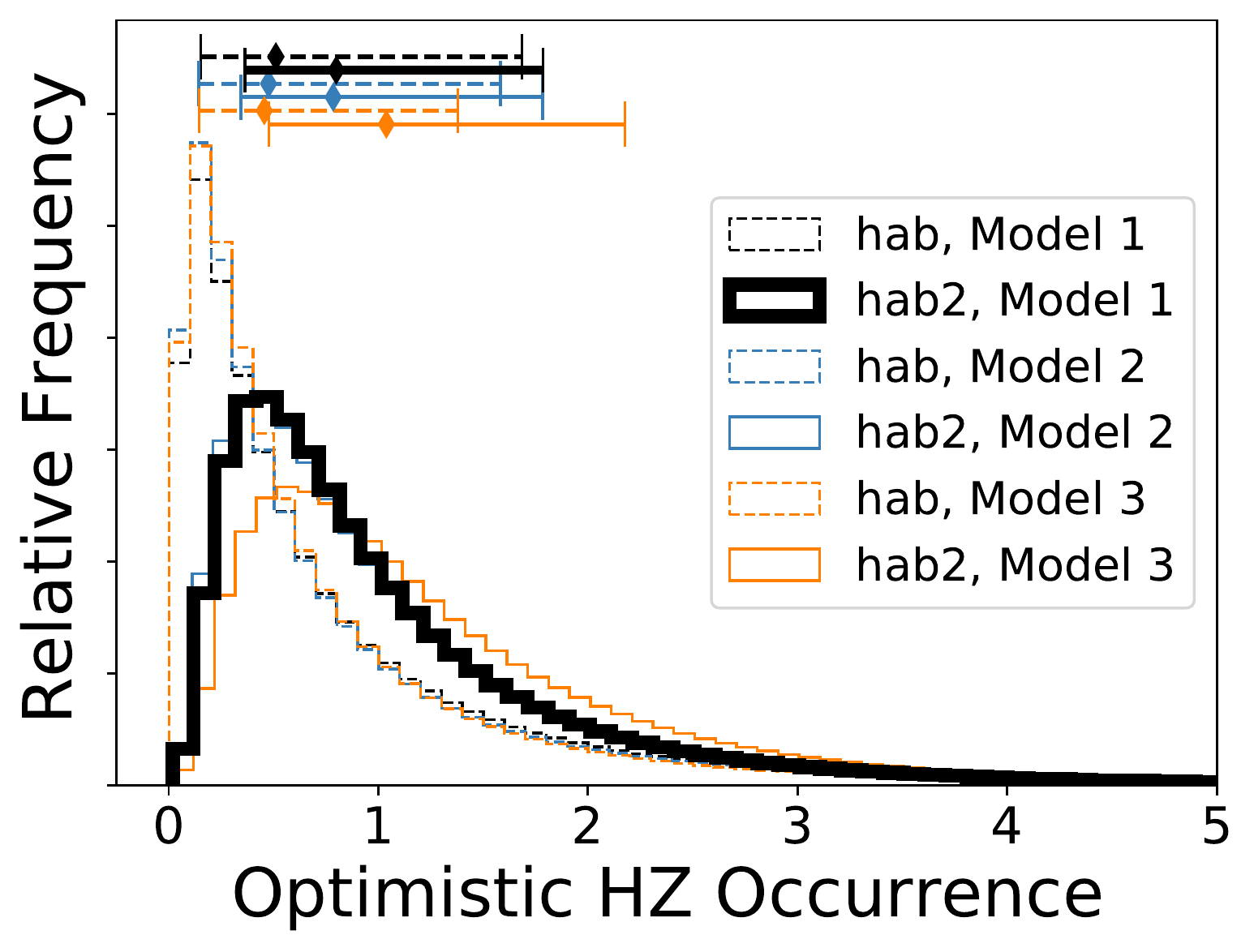} \\
\caption{The distribution of $\eta_\oplus$ for population models computed using the hab (dashed lines) and hab2 (solid lines) stellar populations, for the three models in Equation (\ref{eqn:models}), demonstrating that we get similar results  from models 1 and 2 for both stellar populations. Medians and 68\% credible intervals are shown above the distributions.  The result from the hab2 population including effective temperature dependence is shown with the thick black line.  Top: incorporating the uncertainty on planet radius, and stellar instellation and stellar effective temperature.   Bottom: without incorporating uncertainties.  Left: the conservative habitable zone.  Right: the optimistic habitable zone. } \label{figure:occDists}
\end{figure*}

Table~\ref{table:occStellarPop} gives $\eta_\oplus$, computed using the Poisson likelihood method for the optimistic and conservative habitable zones for the hab and hab2 stellar populations and models 1--3.  The low and high values correspond to the solutions using constant and zero completeness extrapolation, which bound the actual occurrence rates (see \S\ref{section:completenessExtrap}).  We see the expected behavior of zero completeness extrapolation leading to higher occurrence due to a larger completeness correction. Table~\ref{table:occStellarPopABC} gives the same occurrence rates computed using the ABC method.  The distributions of $\eta_\oplus$ using these models and the Poisson likelihood method are shown in Figure~\ref{figure:occDists}.  We see that for each model, when incorporating the input uncertainties, the hab and hab2 stellar populations yield consistent values of $\eta_\oplus$.  Without using the input uncertainties the hab population yields consistently lower values for $\eta_\oplus$, though the difference is still within the 68\% credible interval. Model 3 with hab2 gives generally larger occurrence rates than model 1.  We also see that the median occurrence rates are about $\sim 10\%$ higher when incorporating input uncertainties, qualitatively consistent with \citet{Shabram2020}, who also sees higher median occurrence rates when incorporating uncertainties.  It is not clear what is causing this increase in occurrence rates: on the one hand the sum of the inclusion probability, defined in Appendix~\ref{app:pcProperties}, for the planet candidates in Table~\ref{table:pcProperties} is 53.6, compared with 54 planet candidates in the analysis without uncertainty, indicating that, on average, more planets exit the analysis than enter the analysis when incorporating uncertainties.  On the other hand, the sum of the inclusion probability times the planet radius is 106.6, compared with 105.0 for the planet candidates in the analysis without uncertainty, indicating that, on average, larger planets are entering the analysis.  This may have an impact on the power law model, leading to higher occurrence rates.

\begin{figure*}[ht]
  \centering
  \includegraphics[width=0.48\linewidth]{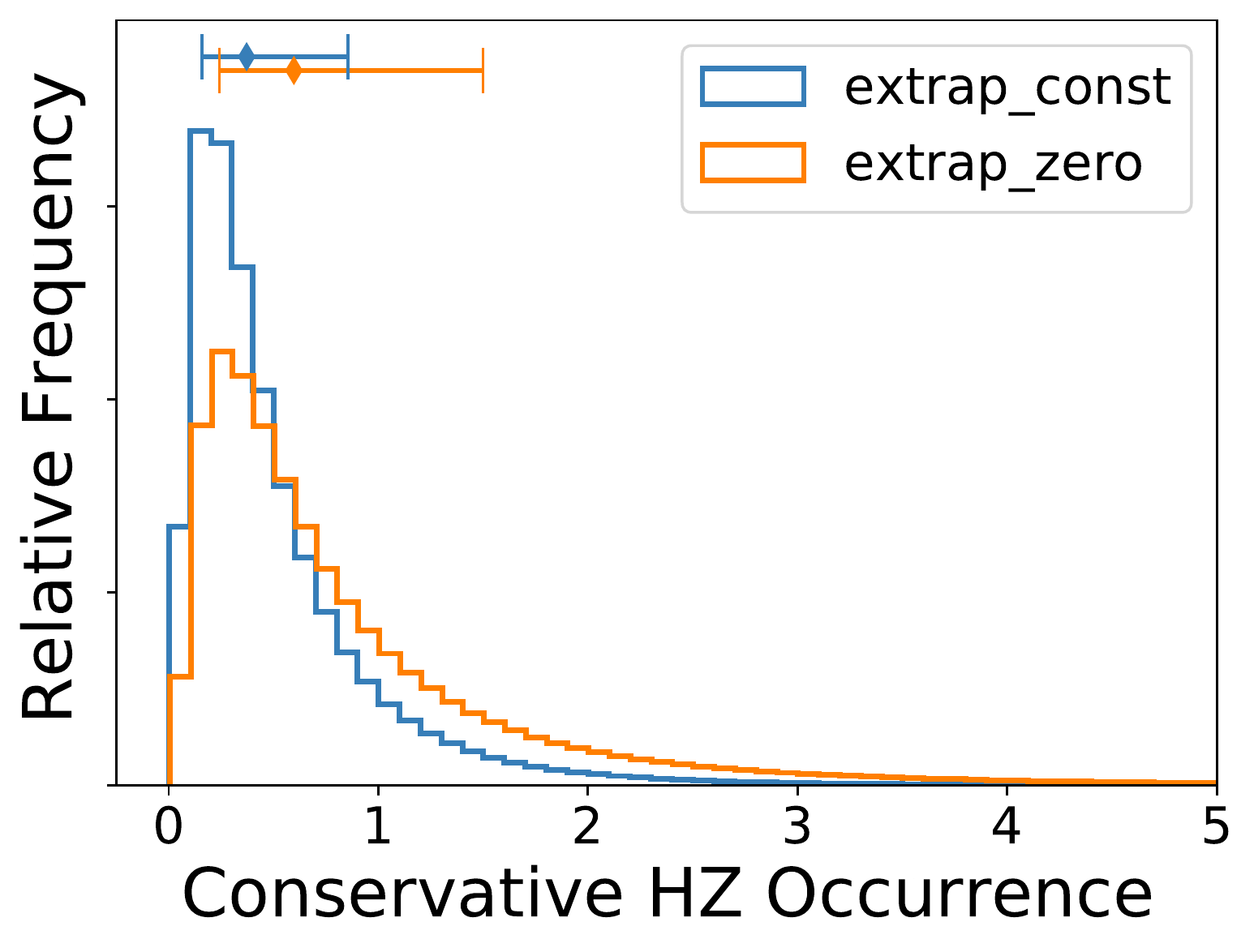}
  \includegraphics[width=0.48\linewidth]{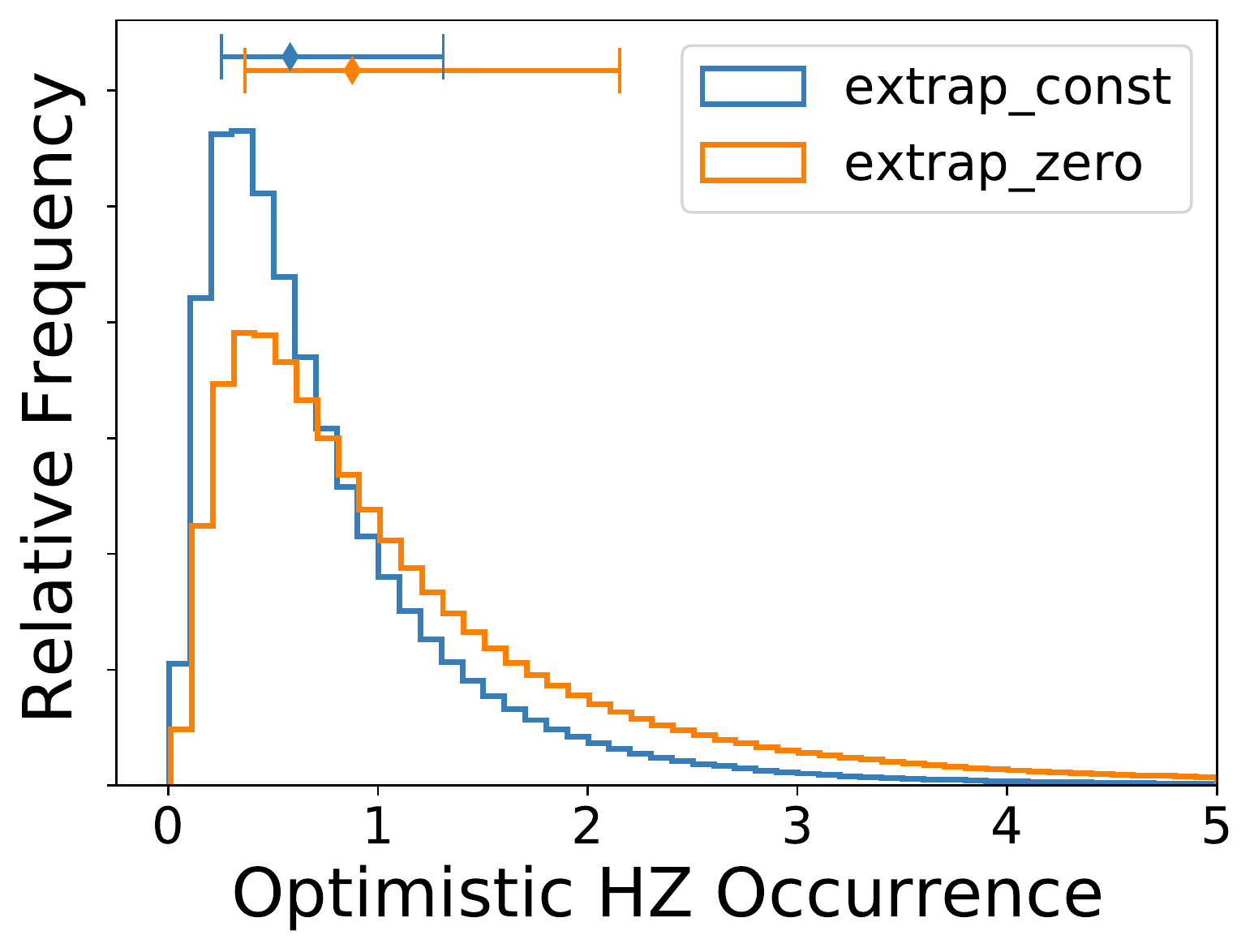}
\caption{The distribution of $\eta_\oplus$ for the two bounding extrapolation cases, computed with model 1 and hab2 with input uncertainties.  Left: the conservative habitable zone.  Right: the optimistic habitable zone. } \label{figure:occExtrap}
\end{figure*}

Table~\ref{table:hab2AllOcc} gives occurrence rates for a variety of planet radius and host star effective temperature ranges, computed using the hab2 stellar population and models 1--3.  We see that the uncertainties for the 1.5 -- 2.5 $R_\oplus$ planets are significantly smaller than for the 0.5 -- 1.5 $R_\oplus$ planets, indicating that the large uncertainties in $\eta_\oplus$ are due to the small number of observed planets in the 0.5 -- 1.5 $R_\oplus$ range.  The distributions of occurrence for the two bounding extrapolation types is shown in Figure~\ref{figure:occExtrap}.  The difference between these two bounding cases is smaller than the uncertainties.  Table~\ref{table:occConfIntervals} gives the 95\% and 99\% intervals for the 0.5 -- 1.5 $R_\oplus$ planets using model 1 computed with hab2.  

\begin{figure*}[ht]
  \centering
  \includegraphics[width=0.48\linewidth]{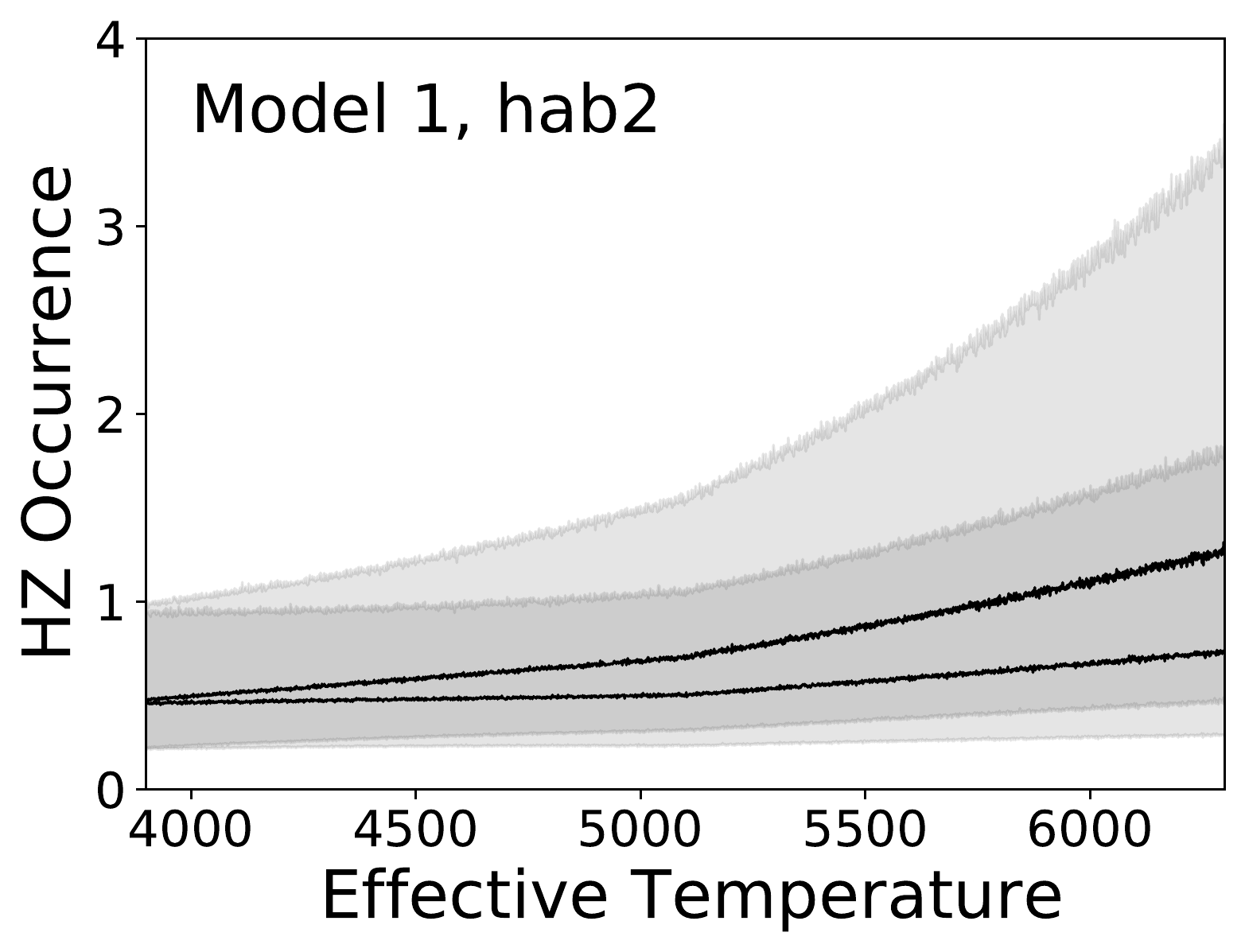} 
  \includegraphics[width=0.48\linewidth]{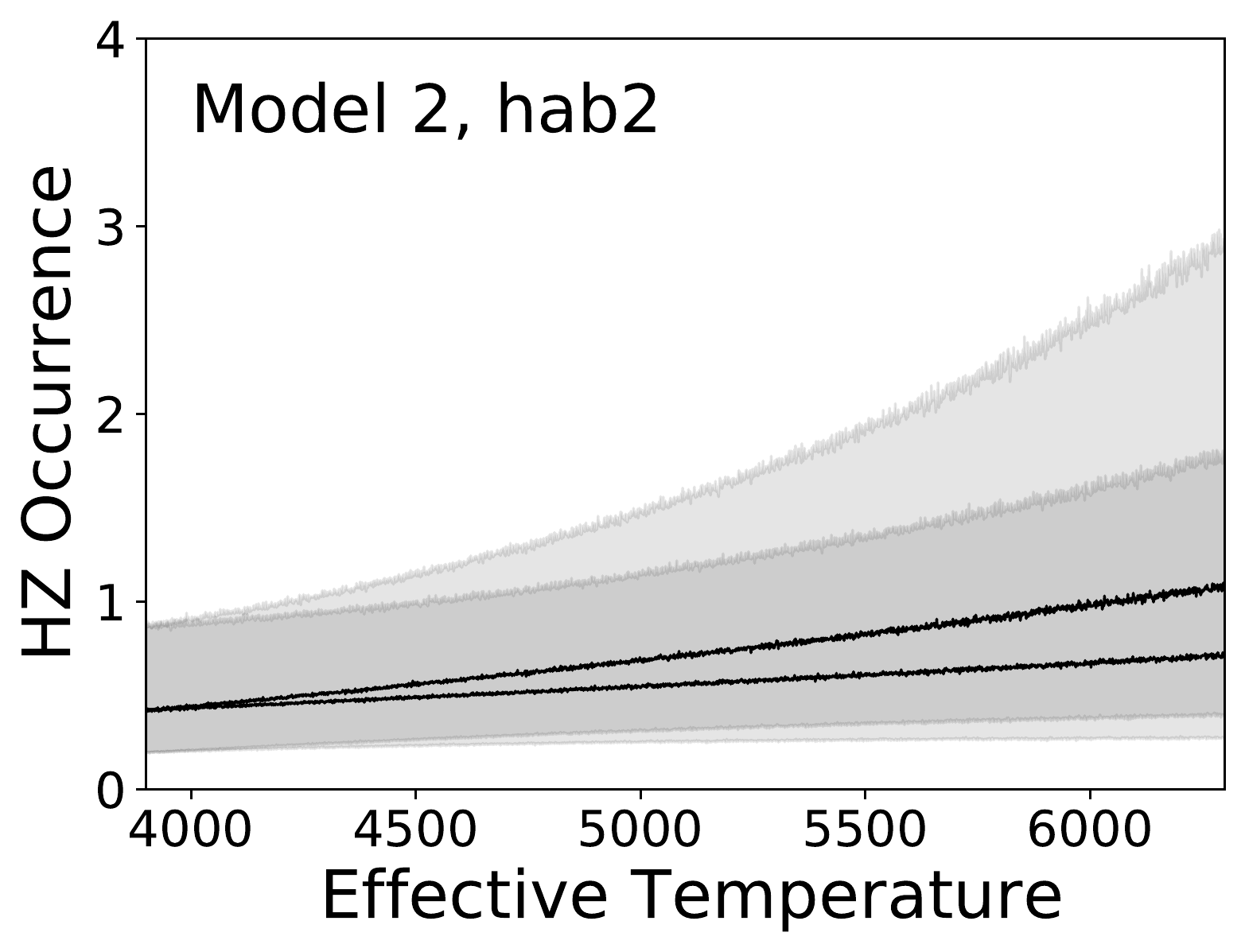} \\
  \includegraphics[width=0.48\linewidth]{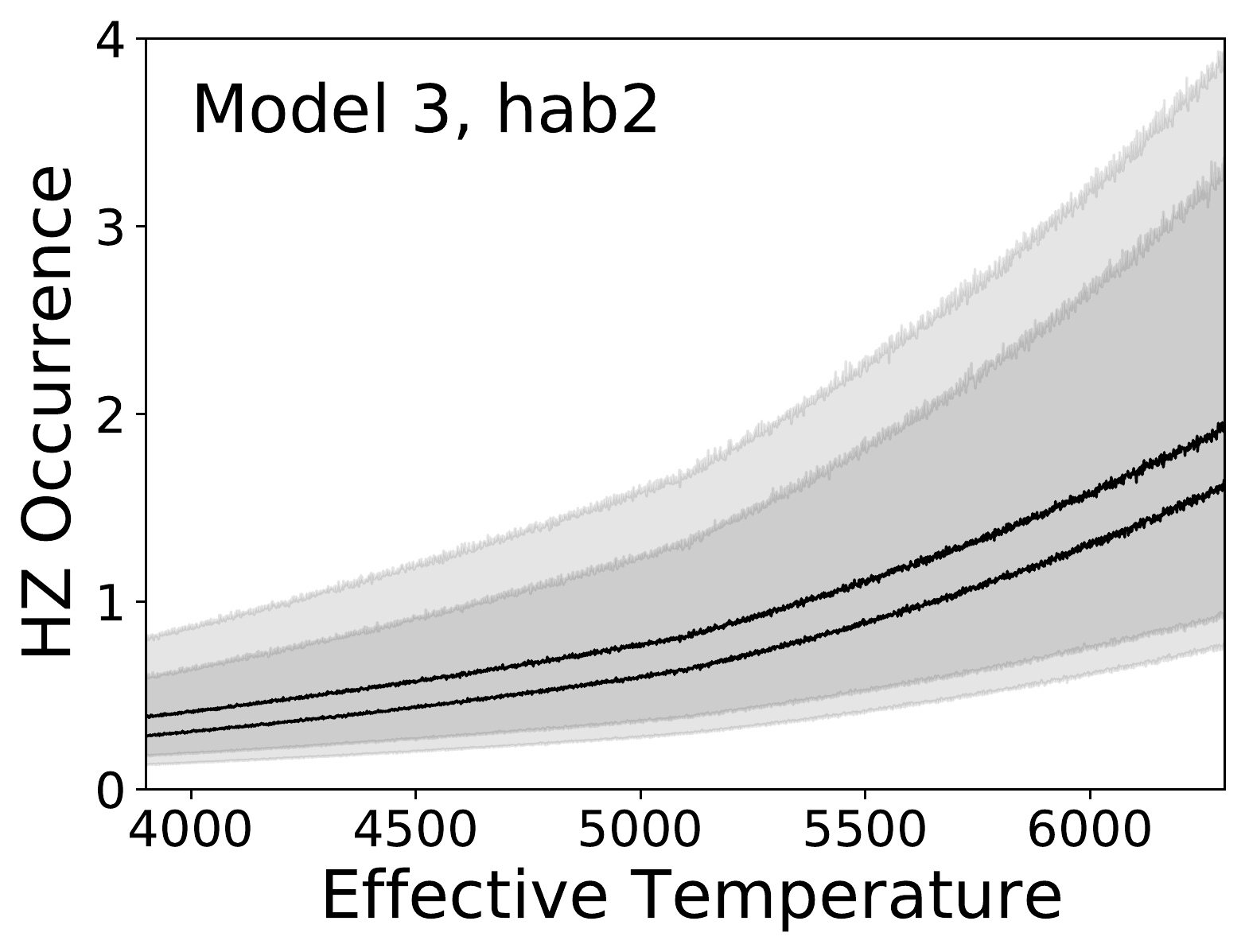} 
  \includegraphics[width=0.48\linewidth]{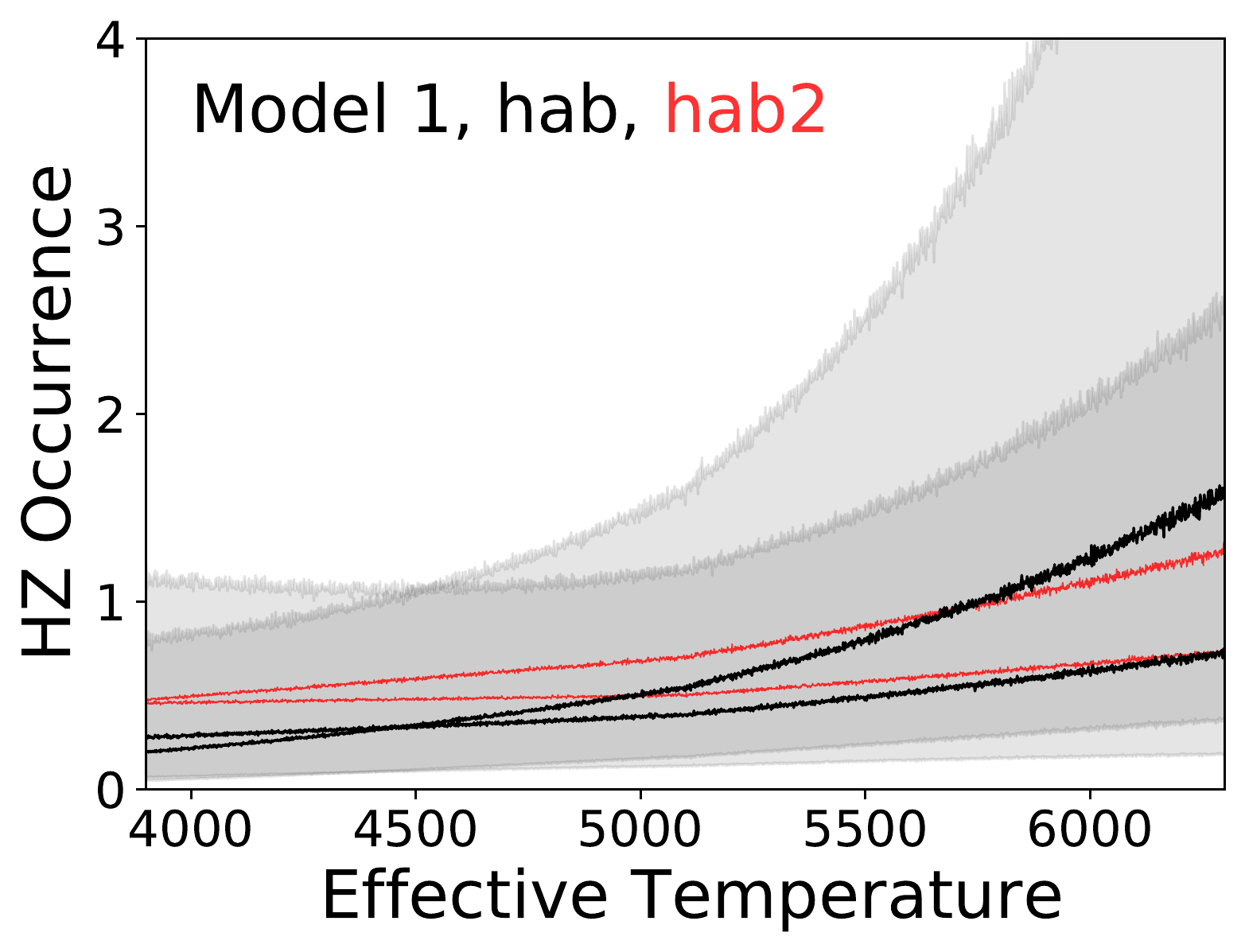} \\
\caption{Optimistic habitable zone rate occurrence for planets with radii between 0.5 and 1.5 $R_\oplus$ as a function of host star effective temperature.  $\eta_\oplus$ is the average over the temperature range $4800$~K  $\leq T_\mathrm{eff} \leq 6300$~K.  The black lines show the median occurrence rate when using zero completeness extrapolation (upper line) and constant completeness extrapolation (lower line).  The grey areas show the 68\% confidence limits for the two completeness extrapolation cases, and the darker grey areas are the overlap of the 68\% confidence regions. Upper Left: Model 1 based on hab2. Right: Model 2 based on hab2. Bottom Left: Model 3 based on hab2. Bottom Right: Model 1 based on hab, with the medians from model 1 based on hab2 in red. } \label{figure:occVsTeff}
\end{figure*}

Figure~\ref{figure:occVsTeff} shows the dependence of the habitable zone occurrence rate on effective temperature for models 1--3 based on the hab2 stellar population, and model 1 for the hab stellar population.  For each model, the occurrence using zero and constant extrapolation is shown.  Models 1 and 2 show a weak increase in occurrence with increasing effective temperature. Model 3 shows a stronger increase occurrence with effective temperature, consistent with model 3's assumption that the only temperature dependence is the geometric effect described in \S\ref{section:hzGeom}.  However, as shown in Figure~\ref{figure:completenessDifference}, the difference between constant and zero extrapolated completeness is near zero for $T_\mathrm{eff} \leq 4500$~K, so we would expect the difference in occurrence rates to be close to zero in that temperature range.  This is true for models 1 and 2, but not true for model 3.  We take this as evidence that models 1 and 2 are correctly measuring a $T_\mathrm{eff}$ dependence beyond the geometric effect.  We recognize that the statistical evidence for this $T_\mathrm{eff}$ dependence is not compelling, since the overlapping 68\% credible intervals for the two completeness extrapolations would allow an occurrence rate independent of $T_\mathrm{eff}$.

An issue that arises with zero completeness extrapolation (= high bound) is that, strictly speaking, PCs with orbital periods $>500$ days are in a region of zero completeness around their host star and would not contribute to the Poisson likelihood (see Equation~(\ref{equation:poisson4}) in Appendix~\ref{app:likelihoodDerivation}). There is one such PC in the hab2 stellar sample with reliability = 0.67.  Performing our Poisson inference, removing planets with period $> 500$ days, for model 1 and incorporating input uncertainties yields an optimistic $\eta_\oplus$ = $0.70^{+1.01}_{-0.41}$, compared with $0.88^{+1.27}_{-0.51}$ (from Table~\ref{table:occStellarPop}) when including the planet.  While well within the 68\% credible interval of the result with this planet included, removing this planet has a noticeable impact on the upper bound for the optimistic $\eta_\oplus$.  However this planet was in fact detected, implying that the completeness is not zero for periods $> 500$ days, at least for this planet's host star.  If the actual completeness is very close to zero, a planet detection implies a large population.  We therefore leave this planet in the analysis, thinking of ``zero completeness'' as a limit of the completeness going to zero when the habitable zone includes orbital periods $>500$ days, summed or averaged over the stellar population for our computations.

\renewcommand{\arraystretch}{1.25}
\setlength{\tabcolsep}{7pt}
\begin{table*}[ht]
\centering
\caption{$\eta_\oplus$, computed using population models based on the hab and hab2 stellar populations, with and without uncertainties for models 1--3 and using the Poisson method. }\label{table:occStellarPop}
\begin{tabular}{ r c c | c c}
\hline
\hline
& \multicolumn{2}{c}{With Uncertainty} & \multicolumn{2}{c}{Without Uncertainty} \\
 & based on hab Stars
 & based on hab2 Stars
 & based on hab Stars
 & based on hab2 Stars
\\
& low bound -- high bound & low bound -- high bound & low bound -- high bound & low bound -- high bound \\
\hline
& \multicolumn{4}{c}{{\bf Conservative Habitable Zone}} \\
Model 1
 & $0.30^{+0.69}_{-0.21}$ -- $0.54^{+1.46}_{-0.39}$ 
 & $0.37^{+0.48}_{-0.21}$ -- $0.60^{+0.90}_{-0.36}$ 
 & $0.15^{+0.32}_{-0.11}$ -- $0.34^{+0.83}_{-0.25}$ 
 & $0.34^{+0.37}_{-0.18}$ -- $0.54^{+0.69}_{-0.30}$ 
\\
Model 2
 & $0.28^{+0.66}_{-0.20}$ -- $0.53^{+1.46}_{-0.39}$ 
 & $0.39^{+0.51}_{-0.23}$ -- $0.56^{+0.83}_{-0.33}$ 
 & $0.19^{+0.39}_{-0.13}$ -- $0.32^{+0.78}_{-0.23}$ 
 & $0.33^{+0.38}_{-0.18}$ -- $0.53^{+0.70}_{-0.30}$ 
\\
Model 3
 & $0.31^{+0.69}_{-0.22}$ -- $0.55^{+1.22}_{-0.39}$ 
 & $0.59^{+0.74}_{-0.34}$ -- $0.79^{+0.95}_{-0.44}$ 
 & $0.21^{+0.42}_{-0.15}$ -- $0.30^{+0.64}_{-0.21}$ 
 & $0.50^{+0.60}_{-0.28}$ -- $0.72^{+0.80}_{-0.39}$ 
\\
& \multicolumn{4}{c}{{\bf Optimistic Habitable Zone}} \\
Model 1
 & $0.50^{+1.09}_{-0.35}$ -- $0.80^{+2.07}_{-0.57}$ 
 & $0.58^{+0.73}_{-0.33}$ -- $0.88^{+1.27}_{-0.51}$ 
 & $0.26^{+0.52}_{-0.18}$ -- $0.51^{+1.17}_{-0.36}$ 
 & $0.54^{+0.56}_{-0.28}$ -- $0.80^{+0.99}_{-0.44}$ 
\\
Model 2
 & $0.48^{+1.06}_{-0.33}$ -- $0.78^{+2.05}_{-0.58}$ 
 & $0.61^{+0.77}_{-0.35}$ -- $0.83^{+1.17}_{-0.48}$ 
 & $0.32^{+0.62}_{-0.22}$ -- $0.48^{+1.11}_{-0.33}$ 
 & $0.52^{+0.58}_{-0.28}$ -- $0.78^{+1.00}_{-0.44}$ 
\\
Model 3
 & $0.53^{+1.10}_{-0.37}$ -- $0.81^{+1.73}_{-0.57}$ 
 & $0.92^{+1.12}_{-0.52}$ -- $1.14^{+1.35}_{-0.63}$ 
 & $0.36^{+0.68}_{-0.25}$ -- $0.46^{+0.92}_{-0.31}$ 
 & $0.79^{+0.92}_{-0.44}$ -- $1.04^{+1.14}_{-0.56}$ 
\\
\end{tabular}
\tablecomments{The low and high bounds correspond to the constant and zero completeness extrapolation of \S\ref{section:completenessExtrap}.  ``With Uncertainty'' means planet candidate radius, instellation flux and host star effective temperature uncertainties were taken into account.}
\end{table*}

\renewcommand{\arraystretch}{1.25}
\setlength{\tabcolsep}{7pt}
\begin{table*}[ht]
\centering
\caption{$\eta_\oplus$, computed using population models based on the hab and hab2 stellar populations, with and without uncertainties for models 1--3 and using the ABC method. }\label{table:occStellarPopABC}
\begin{tabular}{ r c c | c c}
\hline
\hline
 & \multicolumn{2}{c}{With Uncertainty} & \multicolumn{2}{c}{Without Uncertainty} \\
 & based on hab Stars
 & based on hab2 Stars
 & based on hab Stars
 & based on hab2 Stars
\\
& low bound -- high bound & low bound -- high bound & low bound -- high bound & low bound -- high bound \\
\hline
& \multicolumn{4}{c}{{\bf Conservative Habitable Zone}} \\
Model 1
 & $0.33^{+0.46}_{-0.20}$ -- $0.50^{+0.69}_{-0.31}$
 & $0.40^{+0.45}_{-0.21}$ -- $0.61^{+0.63}_{-0.33}$
 & $0.16^{+0.24}_{-0.10}$ -- $0.26^{+0.44}_{-0.16}$ 
 & $0.29^{+0.28}_{-0.15}$ -- $0.49^{+0.61}_{-0.27}$
\\
Model 2
 & $0.30^{+0.41}_{-0.19}$ -- $0.52^{+0.63}_{-0.31}$ 
 & $0.40^{+0.43}_{-0.21}$ -- $0.64^{+0.73}_{-0.35}$ 
 & $0.14^{+0.19}_{-0.09}$ -- $0.25^{+0.36}_{-0.15}$
 & $0.31^{+0.28}_{-0.15}$ -- $0.47^{+0.50}_{-0.24}$ 
\\
Model 3
 & $0.35^{+0.47}_{-0.22}$ -- $0.50^{+0.56}_{-0.29}$
 & $0.69^{+0.64}_{-0.35}$ -- $0.81^{+0.69}_{-0.40}$
 & $0.18^{+0.27}_{-0.11}$ -- $0.27^{+0.36}_{-0.16}$ 
 & $0.50^{+0.48}_{-0.26}$ -- $0.62^{+0.55}_{-0.31}$ 
\\
& \multicolumn{4}{c}{{\bf Optimistic Habitable Zone}} \\
Model 1
 & $0.54^{+0.72}_{-0.33}$ -- $0.73^{+0.98}_{-0.45}$ 
 & $0.62^{+0.66}_{-0.32}$ -- $0.89^{+0.89}_{-0.47}$ 
 & $0.26^{+0.37}_{-0.16}$ -- $0.39^{+0.62}_{-0.23}$ 
 & $0.45^{+0.42}_{-0.23}$ -- $0.71^{+0.86}_{-0.38}$ 
\\
Model 2
 & $0.48^{+0.66}_{-0.30}$ -- $0.75^{+0.90}_{-0.44}$
 & $0.62^{+0.63}_{-0.32}$ -- $0.92^{+1.02}_{-0.49}$
 & $0.24^{+0.30}_{-0.14}$ -- $0.37^{+0.52}_{-0.22}$
 & $0.47^{+0.41}_{-0.23}$ -- $0.68^{+0.70}_{-0.34}$ 
\\
Model 3
 & $0.56^{+0.72}_{-0.36}$ -- $0.73^{+0.81}_{-0.41}$ 
 & $1.05^{+0.96}_{-0.52}$ -- $1.15^{+0.99}_{-0.57}$ 
 & $0.30^{+0.43}_{-0.18}$ -- $0.39^{+0.51}_{-0.23}$
 & $0.75^{+0.72}_{-0.39}$ -- $0.89^{+0.78}_{-0.44}$
\\
\end{tabular}
\tablecomments{The low and high bounds correspond to the constant and zero completeness extrapolation of \S\ref{section:completenessExtrap}.  ``With Uncertainty'' means planet candidate radius, instellation flux and host star effective temperature uncertainties were taken into account.}
\end{table*}

\renewcommand{\arraystretch}{1.25}
\setlength{\tabcolsep}{5pt}
\begin{table*}[ht]
\centering
\caption{Number of planets per star for various ranges of planet radii and host star effective temperature, computed using the population model based on the hab2 stellar population with the Poisson likelihood method and incorporating uncertainties in planet radius, instellation flux and host star effective temperature.  }\label{table:hab2AllOcc}
\begin{tabular}{ r c c c c}
\hline
\hline
Planet Radius 
 & 4800~K -- 6300K
 & 3900~K -- 6300K
 & 3900~K -- 5300~K (K)
 & 5300~K -- 6000~K (G)
\\
\hline
& low bound -- high bound & low bound -- high bound & low bound -- high bound & low bound -- high bound \\
\hline
& \multicolumn{4}{c}{{\bf Conservative Habitable Zone}} \\
& \multicolumn{4}{c}{Model 1} \\
$0.5$ -- $1.5 R_\oplus$
& $\boldsymbol{0.37^{+0.48}_{-0.21}}$
-- $\boldsymbol{0.60^{+0.90}_{-0.36}}$
& $0.35^{+0.43}_{-0.19}$
-- $0.50^{+0.73}_{-0.29}$
& $0.32^{+0.35}_{-0.17}$
-- $0.42^{+0.50}_{-0.23}$
& $0.38^{+0.50}_{-0.22}$
-- $0.63^{+0.94}_{-0.38}$
\\
$1.5$ -- $2.5 R_\oplus$
& $0.16^{+0.07}_{-0.05}$
-- $0.24^{+0.14}_{-0.08}$
& $0.15^{+0.06}_{-0.05}$
-- $0.20^{+0.12}_{-0.07}$
& $0.14^{+0.05}_{-0.04}$
-- $0.17^{+0.07}_{-0.06}$
& $0.17^{+0.06}_{-0.05}$
-- $0.26^{+0.13}_{-0.09}$
\\
$0.5$ -- $2.5 R_\oplus$
& $0.54^{+0.52}_{-0.24}$
-- $0.85^{+0.99}_{-0.42}$
& $0.51^{+0.46}_{-0.22}$
-- $0.71^{+0.80}_{-0.34}$
& $0.46^{+0.37}_{-0.19}$
-- $0.60^{+0.52}_{-0.26}$
& $0.56^{+0.53}_{-0.25}$
-- $0.90^{+1.01}_{-0.45}$
\\
& \multicolumn{4}{c}{Model 2} \\
$0.5$ -- $1.5 R_\oplus$
& $\boldsymbol{0.39^{+0.51}_{-0.23}}$
-- $\boldsymbol{0.56^{+0.83}_{-0.33}}$
& $0.36^{+0.44}_{-0.20}$
-- $0.47^{+0.67}_{-0.27}$
& $0.33^{+0.36}_{-0.18}$
-- $0.40^{+0.47}_{-0.22}$
& $0.40^{+0.53}_{-0.23}$
-- $0.59^{+0.86}_{-0.35}$
\\
$1.5$ -- $2.5 R_\oplus$
& $0.16^{+0.06}_{-0.05}$
-- $0.24^{+0.13}_{-0.08}$
& $0.15^{+0.06}_{-0.05}$
-- $0.20^{+0.11}_{-0.07}$
& $0.14^{+0.05}_{-0.04}$
-- $0.17^{+0.07}_{-0.06}$
& $0.17^{+0.06}_{-0.05}$
-- $0.25^{+0.12}_{-0.08}$
\\
$0.5$ -- $2.5 R_\oplus$
& $0.56^{+0.54}_{-0.26}$
-- $0.81^{+0.91}_{-0.39}$
& $0.51^{+0.47}_{-0.23}$
-- $0.68^{+0.75}_{-0.32}$
& $0.47^{+0.38}_{-0.20}$
-- $0.58^{+0.50}_{-0.25}$
& $0.57^{+0.56}_{-0.27}$
-- $0.85^{+0.94}_{-0.42}$
\\
& \multicolumn{4}{c}{Model 3} \\
$0.5$ -- $1.5 R_\oplus$
& $\boldsymbol{0.59^{+0.74}_{-0.34}}$
-- $\boldsymbol{0.79^{+0.95}_{-0.44}}$
& $0.43^{+0.63}_{-0.26}$
-- $0.59^{+0.82}_{-0.35}$
& $0.31^{+0.37}_{-0.17}$
-- $0.43^{+0.51}_{-0.24}$
& $0.64^{+0.72}_{-0.35}$
-- $0.85^{+0.93}_{-0.46}$
\\
$1.5$ -- $2.5 R_\oplus$
& $0.20^{+0.11}_{-0.07}$
-- $0.29^{+0.15}_{-0.10}$
& $0.15^{+0.13}_{-0.06}$
-- $0.22^{+0.17}_{-0.09}$
& $0.11^{+0.05}_{-0.04}$
-- $0.16^{+0.07}_{-0.05}$
& $0.22^{+0.07}_{-0.06}$
-- $0.32^{+0.11}_{-0.09}$
\\
$0.5$ -- $2.5 R_\oplus$
& $0.81^{+0.79}_{-0.39}$
-- $1.11^{+1.02}_{-0.52}$
& $0.60^{+0.71}_{-0.32}$
-- $0.84^{+0.92}_{-0.44}$
& $0.42^{+0.39}_{-0.20}$
-- $0.60^{+0.54}_{-0.28}$
& $0.87^{+0.75}_{-0.38}$
-- $1.18^{+0.97}_{-0.51}$
\\
& \multicolumn{4}{c}{{\bf Optimistic Habitable Zone}} \\
& \multicolumn{4}{c}{Model 1} \\
$0.5$ -- $1.5 R_\oplus$
& $\boldsymbol{0.58^{+0.73}_{-0.33}}$
-- $\boldsymbol{0.88^{+1.28}_{-0.51}}$
& $0.54^{+0.64}_{-0.29}$
-- $0.73^{+1.02}_{-0.41}$
& $0.49^{+0.51}_{-0.26}$
-- $0.60^{+0.69}_{-0.33}$
& $0.60^{+0.75}_{-0.34}$
-- $0.93^{+1.32}_{-0.55}$
\\
$1.5$ -- $2.5 R_\oplus$
& $0.25^{+0.09}_{-0.06}$
-- $0.35^{+0.19}_{-0.11}$
& $0.23^{+0.09}_{-0.06}$
-- $0.29^{+0.17}_{-0.10}$
& $0.21^{+0.07}_{-0.06}$
-- $0.25^{+0.09}_{-0.08}$
& $0.26^{+0.09}_{-0.07}$
-- $0.37^{+0.17}_{-0.11}$
\\
$0.5$ -- $2.5 R_\oplus$
& $0.84^{+0.78}_{-0.37}$
-- $1.25^{+1.39}_{-0.59}$
& $0.78^{+0.68}_{-0.33}$
-- $1.04^{+1.13}_{-0.47}$
& $0.71^{+0.53}_{-0.28}$
-- $0.86^{+0.72}_{-0.36}$
& $0.87^{+0.79}_{-0.38}$
-- $1.33^{+1.42}_{-0.63}$
\\
& \multicolumn{4}{c}{Model 2} \\
$0.5$ -- $1.5 R_\oplus$
& $\boldsymbol{0.61^{+0.77}_{-0.35}}$
-- $\boldsymbol{0.83^{+1.17}_{-0.48}}$
& $0.55^{+0.66}_{-0.31}$
-- $0.68^{+0.95}_{-0.39}$
& $0.50^{+0.54}_{-0.27}$
-- $0.57^{+0.66}_{-0.31}$
& $0.63^{+0.79}_{-0.36}$
-- $0.87^{+1.22}_{-0.51}$
\\
$1.5$ -- $2.5 R_\oplus$
& $0.25^{+0.09}_{-0.06}$
-- $0.34^{+0.17}_{-0.10}$
& $0.23^{+0.09}_{-0.07}$
-- $0.29^{+0.16}_{-0.10}$
& $0.21^{+0.07}_{-0.06}$
-- $0.25^{+0.09}_{-0.08}$
& $0.26^{+0.08}_{-0.07}$
-- $0.37^{+0.16}_{-0.11}$
\\
$0.5$ -- $2.5 R_\oplus$
& $0.87^{+0.81}_{-0.39}$
-- $1.19^{+1.28}_{-0.56}$
& $0.79^{+0.70}_{-0.34}$
-- $0.99^{+1.05}_{-0.45}$
& $0.72^{+0.55}_{-0.29}$
-- $0.84^{+0.69}_{-0.35}$
& $0.90^{+0.83}_{-0.40}$
-- $1.25^{+1.32}_{-0.59}$
\\
& \multicolumn{4}{c}{Model 3} \\
$0.5$ -- $1.5 R_\oplus$
& $\boldsymbol{0.92^{+1.12}_{-0.52}}$
-- $\boldsymbol{1.14^{+1.35}_{-0.63}}$
& $0.67^{+0.96}_{-0.40}$
-- $0.85^{+1.17}_{-0.50}$
& $0.46^{+0.55}_{-0.26}$
-- $0.61^{+0.70}_{-0.33}$
& $1.00^{+1.08}_{-0.54}$
-- $1.23^{+1.32}_{-0.65}$
\\
$1.5$ -- $2.5 R_\oplus$
& $0.31^{+0.16}_{-0.11}$
-- $0.42^{+0.21}_{-0.15}$
& $0.22^{+0.20}_{-0.10}$
-- $0.31^{+0.25}_{-0.13}$
& $0.16^{+0.07}_{-0.05}$
-- $0.23^{+0.10}_{-0.07}$
& $0.34^{+0.10}_{-0.08}$
-- $0.46^{+0.14}_{-0.12}$
\\
$0.5$ -- $2.5 R_\oplus$
& $1.26^{+1.19}_{-0.59}$
-- $1.60^{+1.43}_{-0.73}$
& $0.92^{+1.07}_{-0.49}$
-- $1.20^{+1.30}_{-0.62}$
& $0.63^{+0.58}_{-0.29}$
-- $0.85^{+0.75}_{-0.38}$
& $1.35^{+1.11}_{-0.57}$
-- $1.71^{+1.36}_{-0.71}$
\\
\end{tabular}
\tablecomments{ $\eta_\oplus$ values for model 1 are shown in boldface.  The low and high bounds correspond to the constant and zero completeness extrapolation of \S\ref{section:completenessExtrap}.  As explained in \S\ref{section:discussSelection} we recommend model 1 as the baseline model. Results from other models are included for comparison.}
\end{table*}

\renewcommand{\arraystretch}{1.25}
\setlength{\tabcolsep}{9pt}
\begin{table}[ht]
\centering
\caption{Credible intervals of the upper and lower bounds on habitable zone occurrence for model 1 computed using the population model based on the hab2 stellar population (see \S\ref{figure:populations}) and accounting for input uncertainty}\label{table:occConfIntervals}
\begin{tabular}{ r c c c}
\hline
\hline
& low & high & total \\
\hline
\multicolumn{4}{c}{95\% Credible Interval} \\
$\eta_\oplus^\mathrm{C}$
 & [0.07, 1.91]& [0.10, 3.77] & [0.07, 3.77] \\
$\eta_\oplus^\mathrm{O}$
 & [0.11, 2.88]& [0.16, 5.29] & [0.11, 5.29] \\
$\eta_{\oplus, \mathrm{G}}^\mathrm{C}$
 & [0.07, 1.92]& [0.10, 3.76] & [0.07, 3.76] \\
$\eta_{\oplus, \mathrm{G}}^\mathrm{O}$
 & [0.11, 2.90]& [0.16, 5.26] & [0.11, 5.26] \\
$\eta_{\oplus, \mathrm{K}}^\mathrm{C}$
 & [0.07, 1.34]& [0.09, 1.92] & [0.07, 1.92] \\
$\eta_{\oplus, \mathrm{K}}^\mathrm{O}$
 & [0.11, 1.96]& [0.13, 2.66] & [0.11, 2.66] \\
\multicolumn{4}{c}{99\% Credible Interval} \\
$\eta_\oplus^\mathrm{C}$
 & [0.04, 3.19]& [0.06, 6.91] & [0.04, 6.91] \\
$\eta_\oplus^\mathrm{O}$
 & [0.06, 4.76]& [0.09, 9.58] & [0.06, 9.58] \\
$\eta_{\oplus, \mathrm{G}}^\mathrm{C}$
 & [0.04, 3.13]& [0.06, 6.57] & [0.04, 6.57] \\
$\eta_{\oplus, \mathrm{G}}^\mathrm{O}$
 & [0.06, 4.65]& [0.10, 9.09] & [0.06, 9.09] \\
$\eta_{\oplus, \mathrm{K}}^\mathrm{C}$
 & [0.04, 2.06]& [0.05, 3.06] & [0.04, 3.06] \\
$\eta_{\oplus, \mathrm{K}}^\mathrm{O}$
 & [0.07, 2.97]& [0.08, 4.20] & [0.07, 4.20] \\
\end{tabular}
\tablecomments{See \S\ref{section:notation} for the definitions of the different types of $\eta_\oplus$.}
\end{table}

\section{Discussion} \label{section:discussion}

\subsection{Instellation Flux vs. Orbital Period}
We choose to compute our occurrence rates as a function of instellation flux for two major reasons: this allows a more direct characterization of each star's habitable zone, and it allows a more constrained extrapolation to longer orbital periods than working directly in orbital period.  

By considering instellation flux, we can measure habitable zone occurrence by including observed planets from the habitable zone of their host stars, which is not possible across a wide range of stellar temperatures when using orbital period (see Figure~\ref{figure:fluxPeriod}).  Instellation flux also allows a direct measurement of the impact of uncertainties in stellar effective temperature and planetary instellation flux.  

The habitable zone of most G and F stars includes orbital periods that are beyond those periods well-covered by \Kepler\ observations (see Figures~\ref{figure:populations} and \ref{figure:fluxPeriod}), requiring significant extrapolation of orbital-period based planet population models to long orbital periods.  Such extrapolation is poorly known or constrained, leading to possible significant and unbounded inaccuracies.  In instellation flux, however, there is planet data throughout those regions of our domain of analysis that have reasonable completeness (see Figure~\ref{figure:populations}) so no extrapolation in instellation flux is required.  In this sense, replacing orbital period with instellation flux (determined by orbital period for each star) moves the problem of extrapolating the population model to longer orbital period to extrapolating the completeness data to lower instellation flux.  In \S\ref{section:completenessExtrap} we argue that completeness, on average, decreases monotonically with decreasing instellation flux.  This allows us to bound the extrapolated completeness between no decrease at all (constant extrapolation) and zero completeness for instellation flux for orbital periods beyond the 500-day limit where completeness was measured.  We then perform our analysis for the two extrapolation cases, and find that their difference in habitable zone occurrence rates is small relative to our uncertainties.  In this way we provide a bounded estimate of habitable zone occurrence rates using instellation flux, rather than the unbounded extrapolation resulting from using orbital period.

\subsection{Comparing the Stellar Population and Rate Function Models} \label{section:discussSelection}
Our approach to measuring $\eta_\oplus$ is to compute the planet population rate model $\lambda(r, I, T_\mathrm{eff}) \equiv \mathrm{d}^2 f(r, I, T_\mathrm{eff}) / \mathrm{d} r \, \mathrm{d} I $, integrate over $r$ and $I$ and average over $T_\mathrm{eff}$.  We compute the population model $\lambda$ using the hab and hab2 stellar populations (\S\ref{section:stellarPopulation}) to measure the sensitivity of our results to stellar type, and we consider several possible functional forms for $\lambda$ (Equation~\ref{eqn:models}).

\subsubsection{Comparing Population Rate Function Models}
We believe that we have detected a weak dependence of habitable zone occurrence on host star effective temperature $T_\mathrm{eff}$, with hotter stars having slightly higher habitable zone occurrence.  Model 2 ($F_0 r^{\alpha} I^{\beta} T_\mathrm{eff}^{\gamma}$), which directly measures $T_\mathrm{eff}$ dependence as a power law with exponent $\gamma$, indicates a weak $T_\mathrm{eff}$ dependence for the zero completeness extrapolation case, though in the constant extrapolation case $\gamma$ includes 0 in the 68\% credible interval (see Table~\ref{table:allFits}).  This is somewhat remarkable given that, as discussed in \S\ref{section:hzGeom}, the size of the habitable zone grows as at least $T_\mathrm{eff}^3$.  This is consistent with model 1 ($F_0 r^{\alpha} I^{\beta} T_\mathrm{eff}^{\gamma} g(T_\mathrm{eff})$), which includes a fixed $g(T_\mathrm{eff})$ term from Equation~\ref{eqn:geomFunc}, reflecting the increase in size of the habitable zone with increasing temperature, and an additional $T_\mathrm{eff}^\gamma$ power law to capture any additional $T_\mathrm{eff}$.  In Table~\ref{table:allFits} we see that model 1 yields a very weak or negative value for $\gamma$, consistent with the weak direct detection of $T_\mathrm{eff}$ dependence in model 2.  The consistency between models 1 and 2 is further indicated by the fact that they yield very similar occurrence rates, as shown in Tables~\ref{table:occStellarPop}, \ref{table:occStellarPopABC} and \ref{table:hab2AllOcc}, as well as Figure~\ref{figure:occVsTeff}.  

Model 3 ($F_0 r^{\alpha} I^{\beta} g(T_\mathrm{eff}))$) assumes that the $T_\mathrm{eff}$ dependence of habitable zone occurrence is entirely due to the increase in size of the habitable zone with increasing $T_\mathrm{eff}$.  When averaged over our $\eta_\oplus$ effective temperature range of 4800~K -- 6300~K, model 3 yields somewhat higher occurrence rates than models 2 and 3 (see Tables~\ref{table:occStellarPop} and \ref{table:occStellarPopABC}, and Figure~\ref{figure:occDists}).  

Models 1 and 2 have the expected behavior of the high and low bounds converging for cooler stars (see Figure~\ref{figure:occVsTeff}), consistent with the extrapolation options coinciding for these stars (see Figure~\ref{figure:completenessDifference}). Model 3 ($\lambda_3 =  F_0 C_3 r^{\alpha} I^{\beta} g(T))$ does not have this behavior but model 3's fixed $T_\mathrm{eff}$ dependence does not allow such a convergence.

Because models 1 and 2 detect a weaker $T_\mathrm{eff}$ dependence than would be expected due to the larger HZ for hotter stars (Equation~(\ref{eqn:geomFunc}) if planets were uniformly distributed, we don't believe that model 3 is the best model for the data.

Models 1 and 2 yield essentially the same habitable zone occurrence results, but model 1 separates the geometric effect from intrinsic $T_\mathrm{eff}$ dependence.  We therefore emphasize model 1, but model 2 provides a direct measure of the total $T_\mathrm{eff}$ dependence.

\subsubsection{Comparing the hab and hab2 Stellar Populations} 
Without input uncertainties, the hab and hab2 stellar populations yield interestingly different values for $\eta_\oplus$.  However, Tables~\ref{table:allFits} and \ref{table:allFitsABC} shows that the $F_0$ parameter fits for hab have significantly larger relative uncertainties than hab2, with hab having $\approx 200\%$ positive uncertainties compared with the $\approx 100\%$ positive uncertainties for the hab2 fits. In addition, the effective temperature exponent $\gamma$ has larger absolute uncertainties for hab than hab2.  These larger uncertainties propagate to larger relative uncertainties in the occurrence rates in Tables~\ref{table:occStellarPop}, \ref{table:occStellarPopABC}, and \ref{table:hab2AllOcc}. This can also be seen by comparing hab and hab2 for model 1 in Figure~\ref{figure:occVsTeff}.  These large uncertainties result in the differences between hab and hab2 being less than the 68\% credible interval. With input uncertainties, the results for the hab and hab2 stellar populations are more consistent, being well inside the 68\% credible interval, but hab still has larger relative uncertainties.

We believe the larger uncertainties for hab relative to hab2 is due to the hab2 population being less well covered by the \Kepler\ observations than hab.   A larger fraction of planet candidates for stars in the hab effective temperature range of 4800~K -- 6300~K are in a region of lower completeness and reliability (Figure~\ref{figure:teffCompleteness}), and have poorer observational coverage (Figure~\ref{figure:populations}).  The hab2 population, with an effective temperature range of 3900~K -- 6300~K includes regions with better observational coverage and more reliable planet detections.  

Basing our occurrence estimates on hab2 covers the full range of K stars without extrapolation, allowing us to produce, for example, GK or G or K habitable zone occurrence rates using the same population model. This avoids possible ambiguities that may be due to different population models.  Finally, when considering $T_\mathrm{eff}$ uncertainties, there are several planets close to the lower hab boundary at 4800~K, which lead to larger impact of $T_\mathrm{eff}$ uncertainties on the population model because those planets will be in some uncertainty realizations and not in others (see Figure~\ref{figure:populations}).  In contrast, the lower $T_\mathrm{eff}$ boundary for hab2 is outside the 68\% credible interval for all detected planets.   Therefore, although the hab population exactly matches our effective temperature range for $\eta_\oplus$, we prefer models computed using the hab2 population. 

To summarize, we adopt model 1 based on hab2 for our primary reported result, but we also provide the results for models 1--3 and the hab stellar populations.

\subsection{Computing \texorpdfstring{$\eta_{\oplus}$}{eta\_Earth}} \label{section:computingEtaEarth}
We find reasonable consistency in $\eta_{\oplus}$ across models for both the hab and hab2 stellar population as shown in Tables~\ref{table:occStellarPop} and \ref{table:occStellarPopABC}.  Table~\ref{table:hab2AllOcc} gives occurrence rates for several planet radius and stellar effective temperature ranges, using the population model from the hab2 population.  The uncertainties reflecting our 68\% credible intervals for our $\eta_{\oplus}$, counting HZ planets with radius $0.5$ -- $1.5~R_\oplus$, are large, with positive uncertainties being nearly 150\% of the value when using input uncertainties. Comparing occurrence rate uncertainties with and without input uncertainties in Tables~\ref{table:occStellarPop} and \ref{table:occStellarPopABC}, we see that the bulk of the uncertainties occur without input uncertainties, while using input uncertainties increases the output uncertainty by nearly 20\%.  The much smaller uncertainties for the larger planets ($1.5  \ R_\oplus \leq r \leq 2.5 \ R_\oplus$) in Table~\ref{table:hab2AllOcc} suggest the large uncertainties in $\eta_{\oplus}$ are driven by the very small number of detections in the $\eta_{\oplus}$ range, combined with the very low completeness (see Figure~\ref{figure:populations}).  Low completeness will cause large completeness corrections, which will magnify the Poisson uncertainty due few planet detections.  There are more planets with larger radius, which are in a regime of higher completeness, resulting in lower uncertainties for larger planet occurrence rates.

\subsection{Implications of \texorpdfstring{$\eta_{\oplus}$}{eta\_Earth}} \label{section:etaEarthImplications}

Estimates of $\eta_{\oplus}$ are useful in calculating  exoEarth yields from direct imaging missions, such as the flagship concept studies like LUVOIR and HabEX. These mission studies assumed an occurrence rate of $\eta_{\oplus} = 0.24^{+0.46}_{-0.16}$ for Sun-like stars based on the NASA ExoPAG SAG13 meta-analysis of Kepler data \citep{Kopparapu2018a}. The expected exoEarth candidate yields from the mission study reports are $54^{+61}_{-34}$ for LUVOIR-A (15m), $28^{+30}_{-17}$ for LUVOIR-B (8m), and $8$ for HabEx 4m baseline configuration which combines a more traditional
coronagraphic starlight suppression system with
a formation-flying starshade occulter. Table~\ref{table:hab2AllOcc} provides $\eta_{\oplus}$ values based on three models for G (Sun-like) and K-dwarfs.  If we assume the range of $\eta_{\oplus, \mathrm{G}}$ from conservative to optimistic HZ from Table~\ref{table:hab2AllOcc} for planets in the $0.5 - 1.5$ $R_{\oplus}$ from the ``low'' end, say for Model 1, $\eta_{\oplus, \mathrm{G}}$ would be between $0.38^{+0.50}_{-0.22}$ and $0.60^{+0.75}_{-0.34}$. {\it Added after acceptance:} While these $\eta_{\oplus}$ values appear to be larger than the $0.24^{+0.46}_{-0.16}$ occurrence rate assumed by the mission studies, it should be noted that these studies adopted radius range of 0.82 to 1.4 $R_{\oplus}$, and a lower limit of $0.8*a^{-0.5}$, where $a$ is the HZ-corrected semi-major axis. This is slightly a smaller HZ region and lower radius than the one used in our study. As a result, it is possible that we might be agreeing with their assumed $\eta_{\oplus}$ value  if we use the same bounding boxes.  Computing the conservative habitable zone as described in \S\ref{section:computeOccurrence} but replacing the planet radius range with $0.82 \leq r \leq 1.4~R_{\oplus}$ gives a lower bound of $0.18^{+0.16}_{-0.09}$ and an upper bound of $0.28^{+0.30}_{-0.14}$, nicely bracketing the value assumed in mission studies.

$\eta_{\oplus}$ can also be used to estimate, on average, the nearest HZ planet around a G and K-dwarf assuming the planets are distributed randomly. Within the Solar neighborhood, the stellar number density ranges from 0.0033 to 0.0038~pc$^{-3}$ for G-dwarfs, and 0.0105 to 0.0153~pc$^{-3}$ for K-dwarfs \citep{Mamajek2008, kirk2012}\footnote{\url{http://www.pas.rochester.edu/~emamajek/memo_star_dens.html}}. For G-dwarfs, multiplying with the conservative ($0.38^{+0.50}_{-0.22}$) Model 1 ``low'' end of the $\eta_{\oplus, \mathrm{G}}$ values (i.e, the number of planets per star), we  get between $0.0013^{+0.0016}_{-0.0007}$ and $0.0014^{+0.0019}_{-0.0008}$ HZ planets pc$^{-3}$. The nearest HZ planet around a G-dwarf would then be expected to be at a distance of $d = (3/(4 \pi \times n_{p}))^{1/3}$, where $n_{p}$ is the planet number density in pc$^{-3}$. Substituting, we get $d$ between $5.9^{+1.76}_{-1.26}$~pc and $5.5^{+1.83}_{-1.34}$~pc, or essentially around $\sim$6~pc away. A similar calculation for K-dwarfs assuming Model 1 conservative HZ $\eta_{\oplus, \mathrm{K}}$ values from Table~\ref{table:hab2AllOcc} indicates that, on average, the nearest HZ planet could be between $4.1^{+1.19}_{-0.90}$~pc and $3.6^{+1.05}_{-0.79}$~pc, or around $\sim$4~pc away.  

An additional speculative calculation one could do is to take the  number of G-dwarfs in the Solar neighborhood within 10~pc --- 19 from RECONS\footnote{\url{http://www.recons.org/census.posted.htm}} --- and multiply it with the ``low''  conservative $\eta_{\oplus, \mathrm{G}}$ value from Model 1,  $0.38^{+0.50}_{-0.22}$. We then get $7.2^{+9.5}_{-4.2}$ HZ planets around G-dwarfs (of all sub-spectral types) within 10~pc. A similar calculation for K-dwarfs from the same RECONS data with 44 stars, and Model 1 ``low'' value, conservative HZ $\eta_{\oplus, \mathrm{K}} = 0.32^{+0.35}_{-0.17}$ indicates that there are $14^{+15}_{-7.5}$ HZ planets around K-dwarfs within 10~pc.  It should be noted that the numbers for the nearest HZ planet and the number of HZ planets in the solar neighborhood used the ``low'' end of the rocky planet (0.5--1.5~$R_{\oplus}$) occurrence rate values from Table~\ref{table:hab2AllOcc}. As such, these represent the lower bound estimates. In other words, there may potentially be a HZ rocky planet around a G or a K-dwarf closer, and may be more HZ planets, than the values quoted above. 

This can be quantified from the numbers shown in  Table \ref{table:occConfIntervals}, which provides the 95$\%$ and $99\%$ credible intervals of the upper and lower bounds on habitable zone occurrence for model 1 computed with hab2 and accounting for input uncertainty. If we use only the ``low'' and the lower end of the conservative HZ occurrence values from this table (0.07 for 95$\%$, 0.04 for $99\%$ credible intervals), then the nearest HZ planet around a G or K-dwarf star is within $\sim$6~pc away with 95$\%$ confidence, and within $\sim7.5$~pc away with $99\%$ confidence. Similarly, there could be $\sim$4 HZ planets within 10~pc with $95\%$ confidence, and $\sim$3 HZ planets with $99\%$ confidence.

    We again caution that these are only estimates and do not necessarily indicate actual number of planets that could be detectable or exist. The numbers provided in this section are first order estimates to simply show a meaningful application of $\eta_{\oplus}$, given its uncertainties. \cite{Mulders2018} and \cite{He2020} have shown that there is strong evidence for multiplicity and clustering of planets within a system.  This implies that the nearest such planets would be farther than if there were not clustering. 

\subsection{Comparison with previous estimates of  \texorpdfstring{$\eta_{\oplus}$}{eta\_Earth}} \label{section:gammaEarthEstimate}
Our work is the first to compute habitable zone occurrence rates using the incident stellar flux for habitable zones around subsets of FGK stars based on the DR25 catalog, Gaia-based stellar properties, and using the habitable zone definition of \citet{Kopparapu2014}. Other works in the literature have produced occurrence rates for orbital periods related to FGK habitable zones, but as discussed in \S\ref{section:ourWork} and \S\ref{section:computeOccurrence}, we are measuring occurrence between habitable zone boundaries as a function of $T_\mathrm{eff}$.  This is a different quantity than occurrence rates based on orbital period.  The few occurrence rate estimates in the literature based on instellation flux, such as \citet{Petigura2013}, used a rectangular region in radius and instellation flux, and only approximated the habitable zone.  Therefore, we do not directly compare our occurrence rates with those in previous literature.

We provide a formal computation of $\Gamma_\oplus$, which is commonly used to compare $\eta_{\oplus}$ estimates \citep[see, for example, Figure 14 of][]{Kunimoto2020a}.  For a planet of period $p$ and radius $r$, we define $\Gamma \equiv \mathrm{d}^2 f / \mathrm{d} \log p \, \mathrm{d} \log r = p  \, r  \, \mathrm{d}^2 f / \mathrm{d} p \, \mathrm{d} r$, and $\Gamma_\oplus$ is $\Gamma$ evaluated at Earth's period and radius.  We need to express $\Gamma$ in terms of instellation flux $I$, which we will do using $\mathrm{d}^2 f / \mathrm{d} p \, \mathrm{d} r = \left( \mathrm{d}^2 f / \mathrm{d} I \, \mathrm{d} r \right) \, \left(\mathrm{d} I / \mathrm{d} p\right)$.

For a particular star, the instellation on a planet at period $p$ in years is given by $I=R_*^2 T^4 M_*^{-2/3} p^{-4/3} $, where $R_*$ is the stellar radius in Solar radii, $M_*$ is the stellar mass in Solar masses and $T = T_\mathrm{eff}/T_\odot$ is the star's effective temperature divided by the Solar effective temperature.  Then 
\begin{equation} \label{eqn:gammaEarth}
\begin{split}
    \Gamma
    & = p  \, r  \, \frac{\mathrm{d}^2 f} {\mathrm{d} I \, \mathrm{d} r} \left(\frac {\mathrm{d} I}{\mathrm{d} p}\right) \\
    & = -\frac{4}{3}  \, \frac{R_*^2 T^4}{M_*^\frac{2}{3}} r  \, p^{ -\frac{4}{3}}  \, \lambda(I, r, T, \boldsymbol{\theta}) 
\end{split}
\end{equation}
because $\mathrm{d}^2 f / \mathrm{d} I \, \mathrm{d} r = \lambda(I, r, T, \boldsymbol{\theta})$, one of the differential population rate models from Equation~(\ref{eqn:models}). To compute $\Gamma_\oplus$ for a particular star, we evaluate Equation~\ref{eqn:gammaEarth} at $r = 1 ~ R_\oplus$, $p = 1$ year, and $I=R_*^2 T^4 M_*^{-2/3}$, the instellation a planet with a one-year orbital period would have from that star.  The result is the $\Gamma_\oplus$ in radius and period implied by our differential population rate function in radius and instellation for that star, and may be compared directly with $\Gamma_\oplus$ from period-based occurrence studies.

We compute $\Gamma_\oplus$ using model 1 from Equation~(\ref{eqn:models}) with input uncertainty on the hab2 stellar population.  For each star in hab2, we evaluate Equation~\ref{eqn:gammaEarth} using the posterior $\boldsymbol{\theta}$ distribution, and concatenate the resulting $\Gamma_\oplus$ distributions from all the stars.  We do this for both completeness extrapolations in \S\ref{section:completenessExtrap}, giving low and high bounds.  This results in a $\Gamma_\oplus$ between $0.45^{+0.46}_{-0.24}$ and $0.50^{+0.46}_{-0.26}$.  While this is a formal mathematical exercise that has not been demonstrated to be truly equivalent to $\Gamma_\oplus$ defined in period space, the match between this value and our conservative habitable zone $\eta_\oplus^\mathrm{C} = 0.37^{+0.48}_{-0.21}$ -- $0.60^{+0.90}_{-0.36}$ for the model 1, hab2 with input uncertainty in Table~\ref{table:occStellarPop} is remarkable.

Our value of $\Gamma_\oplus$ is somewhat higher than values using post-Gaia stellar and planet data (see, for example, figure 14 of \citet{Kunimoto2020a}), but not significantly so. For example, \citet{Bryson2020} found $\Gamma_\oplus = 0.09^{+0.07}_{-0.04}$ when correcting for reliability in period-radius space.  Using the same population model \citet{Bryson2020} found a SAG13 $\eta_\oplus$ value of $0.13^{+0.10}_{-0.06}$.  It is not clear how much we should expect $\Gamma_\oplus$ to correspond with $\eta_\oplus$.

\subsection{Effective Temperature Dependence}
As described in \S\ref{section:hzOccurrence} and Figure~\ref{figure:occVsTeff}, our results indicate a weak, but not compelling, increase in HZ planet occurrence with increasing stellar effective temperature.  This $T_\mathrm{eff}$ dependence is weaker than would be expected if planet occurrence were uniformly spaced in semi-major axis (see \S\ref{section:hzGeom}) because hotter stars have larger HZs.  This can be seen quantitatively in the median differential population rates for models 1 and 2 using the hab2 population in Tables~\ref{table:allFits} and \ref{table:allFitsABC}. In model 2 we observe a median $T_\mathrm{eff}$ exponent $\gamma < 3$, compared with the prediction of $\gamma \approx 3$ to 4.5 due to the larger HZ for hotter stars from Equation~(\ref{eqn:geomFunc}).  This is reflected in model 1, which includes the correction for the larger HZ so if $T_\mathrm{eff}$ dependence were due only to the larger HZ then $\gamma$ would equal 0.  The high bound of model 1 finds a median $\gamma < -1$ indicating that we see fewer HZ planets than expected in the larger HZ of hotter stars if the planets were uniformly spaced.  However the upper limits of $\gamma$'s 68\% credible interval in Tables~\ref{table:allFits} and \ref{table:allFitsABC} are consistent with the prediction of uniform planet spacing in larger HZs for hotter stars.  For example, the posterior of $\gamma$ for model 1 (hab 2 population, high bound) has $\gamma \geq  0$ 22.3\% of the time. 

Our detection of a weaker $T_\mathrm{eff}$ dependence than expected from larger HZs is qualitatively consistent with the increasing planet occurrence with lower $T_\mathrm{eff}$ found in \citet{Garrett2018} and \citet{Mulders2015}.  But our uncertainties do not allow us to make definitive conclusions about this $T_\mathrm{eff}$ dependence.

\subsection{Dependence on the Planet Sample}
To study the dependence of our result on the planet sample, we performed a bootstrap analysis.  We ran the Poisson likelihood inference using hab2 and model 1 with zero extrapolation (high bound) 400 times, re-sampling the planet candidate population with replacement.  Each re-sampled run removed planets according to their reliability as described in \S\ref{section:inferenceMethods}, but did not consider input uncertainty.  The concatenated posterior of these re-sampled runs gives $F_0=1.404^{+1.768}_{-0.680}$, $\alpha=-0.920^{+1.236}_{-1.072}$, $\beta=-1.175^{+0.465}_{-0.444}$ and $\gamma=-1.090^{+2.446}_{-2.217}$.  These parameters yield $\eta_\oplus^\mathrm{C} = 0.483^{+0.997}_{-0.324}$ and  $\eta_\oplus^\mathrm{O} = 0.716^{+1.413}_{-0.472}$.  Comparing with the hab2 model 1 high value without uncertainty in Tables~\ref{table:allFits} and \ref{table:occStellarPop}, we see that the central values from the bootstrap study are well within the 68\% credible interval of our results, and the uncertainties are as much as 50\% higher. 

A similar study of the dependence on the stellar sample is not feasible because each re-sampled stellar population would require a full re-computation of detection and vetting completeness and reliability.  Performing hundreds of these computations is beyond our available resources.

\subsection{Impact of Catalog Reliability Correction}
\begin{figure}[ht]
  \centering
  \includegraphics[width=0.98\linewidth]{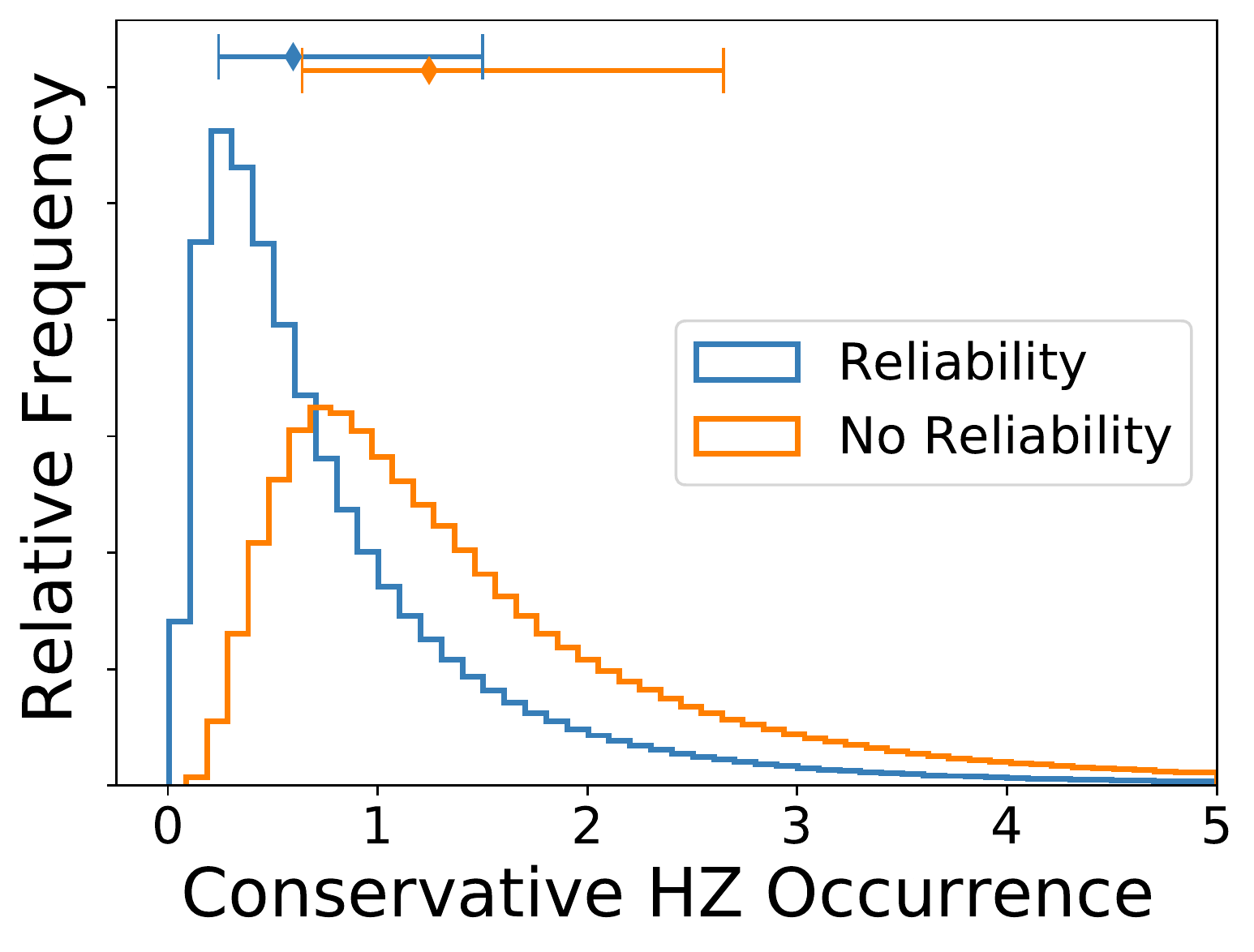}
\caption{A comparison of the distributions, with and without reliability correction, of the conservative habitable zone $\eta_\oplus^\mathrm{C}$ computed with model 1, zero completeness extrapolation (high value), accounting for input uncertainty and using the hab2 stellar population.} \label{figure:reliabilityExample}
\end{figure}

All results presented in this paper are computed with corrections for planet catalog completeness and reliability (see \S\ref{section:completenessAndReliability}).  Figure~\ref{figure:reliabilityExample}  shows an example of what happens when there is no correction for catalog reliability.  We compute $\eta_\oplus^\mathrm{C}$, occurrence in the conservative habitable zone, with model 1, zero completeness extrapolation (high value), accounting for input uncertainty and using the hab2 stellar population.  With reliability correction, we have $\eta_\oplus^\mathrm{C} = 0.60^{+0.90}_{-0.36}$ and without reliability correction we have $\eta_\oplus^\mathrm{C} = 1.25^{+1.40}_{-0.60}$.  In this typical case reliability has a factor-of-two impact, consistent with \citet{Bryson2020}, though because of the large uncertainties the difference is less than the 68\% credible interval.

\subsection{\texorpdfstring{$\eta_{\oplus}$}{eta\_Earth} based on the GK and FGK Stellar Populations}
Our definition of $\eta_\oplus$, restricted to stars with effective temperatures between 4800~K and 6300~K, varies somewhat from the literature.  To connect with other occurrence rate studies we repeat our analysis using the GK ($3900$~K $\leq T_\mathrm{eff} \leq 6000$~K) and FGK ($3900$~K $\leq T_\mathrm{eff} \leq 7300$~K) stellar populations to compute planet population models, with results in Table~\ref{table:allFitsFGK}.  We provide our $\eta_\oplus$ derived from these stellar populations as well as habitable zone occurrence for the GK and FGK $T_\mathrm{eff}$ ranges.  The values for our definition of $\eta_\oplus$ are consistent with the values in Tables~\ref{table:occStellarPop} and \ref{table:occStellarPopABC}.  We caution, however, that the FGK population extends well into stellar effective temperatures where there are no planet detections and very low or zero completeness, so an FGK result necessarily involves extrapolation from cooler stars. 

\renewcommand{\arraystretch}{1.25}
\setlength{\tabcolsep}{7pt}
\begin{table*}[ht]
\centering
\caption{Parameter fits and $\eta_\oplus$ with 68\% confidence limits for model 1 from Equation~\ref{eqn:models} computed using the population model from the Poisson likelihood method applied to the GK and FGK stellar populations.}\label{table:allFitsFGK}
\begin{tabular}{ r c c | c c }
\hline
\hline
& \multicolumn{2}{c}{With Uncertainty} & \multicolumn{2}{c}{Without Uncertainty} \\
 & based on GK Stars
 & based on FGK Stars
 & based on GK Stars
 & based on FGK Stars
\\
& low bound -- high bound & low bound -- high bound & low bound -- high bound & low bound -- high bound \\
\hline
& \multicolumn{4}{c}{Model 1} \\
$F_0$
 & 
$+1.01^{+0.82}_{-0.4}$
 -- 
$+1.23^{+1.12}_{-0.52}$
 & 
$+1.26^{+1.2}_{-0.55}$
 -- 
$+2.36^{+2.94}_{-1.2}$
 & 
$+0.89^{+0.64}_{-0.33}$
 -- 
$+1.12^{+0.83}_{-0.44}$
 & 
$+1.24^{+1.03}_{-0.51}$
 -- 
$+2.05^{+2.14}_{-0.98}$
\\
$\alpha$
 & 
$-1.00^{+1.0}_{-0.9}$
 -- 
$-1.05^{+1.0}_{-0.9}$
 & 
$-1.03^{+0.95}_{-0.87}$
 -- 
$-1.16^{+0.94}_{-0.85}$
 & 
$-0.80^{+0.93}_{-0.83}$
 -- 
$-0.90^{+0.87}_{-0.78}$
 & 
$-1.00^{+0.84}_{-0.77}$
 -- 
$-0.96^{+0.87}_{-0.78}$
\\
$\beta$
 & 
$-0.90^{+0.34}_{-0.32}$
 -- 
$-1.13^{+0.37}_{-0.36}$
 & 
$-0.82^{+0.32}_{-0.3}$
 -- 
$-1.23^{+0.36}_{-0.34}$
 & 
$-0.84^{+0.33}_{-0.3}$
 -- 
$-1.08^{+0.36}_{-0.34}$
 & 
$-0.82^{+0.3}_{-0.27}$
 -- 
$-1.21^{+0.34}_{-0.32}$
\\
$\gamma$
 & 
$-3.00^{+1.75}_{-1.72}$
 -- 
$-2.25^{+1.9}_{-1.86}$
 & 
$-2.67^{+1.57}_{-1.58}$
 -- 
$-1.07^{+1.8}_{-1.77}$
 & 
$-2.76^{+1.61}_{-1.62}$
 -- 
$-1.98^{+1.75}_{-1.69}$
 & 
$-2.67^{+1.43}_{-1.43}$
 -- 
$-1.15^{+1.64}_{-1.66}$
\\
$\eta_\oplus^\mathrm{C}$
& $0.34^{+0.48}_{-0.20}$ -- $0.46^{+0.71}_{-0.28}$ 
& $0.36^{+0.46}_{-0.20}$ -- $0.63^{+0.92}_{-0.37}$ 
& $0.29^{+0.37}_{-0.16}$ -- $0.41^{+0.53}_{-0.23}$ 
& $0.35^{+0.40}_{-0.19}$ -- $0.53^{+0.68}_{-0.30}$ 
\\
$\eta_{\oplus,\mathrm{GK}}^\mathrm{C}$
& $0.32^{+0.41}_{-0.18}$ -- $0.40^{+0.54}_{-0.23}$ 
& $0.32^{+0.38}_{-0.18}$ -- $0.48^{+0.64}_{-0.27}$ 
& $0.26^{+0.31}_{-0.15}$ -- $0.35^{+0.40}_{-0.19}$ 
& $0.32^{+0.33}_{-0.16}$ -- $0.41^{+0.49}_{-0.22}$ 
\\
$\eta_{\oplus,\mathrm{FGK}}^\mathrm{C}$
& $0.35^{+0.54}_{-0.21}$ -- $0.47^{+0.81}_{-0.29}$ 
& $0.37^{+0.53}_{-0.21}$ -- $0.63^{+1.18}_{-0.39}$ 
& $0.30^{+0.43}_{-0.17}$ -- $0.42^{+0.64}_{-0.24}$ 
& $0.36^{+0.46}_{-0.20}$ -- $0.53^{+0.88}_{-0.31}$ 
\\
$\eta_\oplus^\mathrm{O}$
& $0.52^{+0.72}_{-0.30}$ -- $0.68^{+1.01}_{-0.41}$ 
& $0.56^{+0.70}_{-0.31}$ -- $0.92^{+1.29}_{-0.54}$ 
& $0.45^{+0.56}_{-0.25}$ -- $0.61^{+0.77}_{-0.34}$ 
& $0.55^{+0.60}_{-0.29}$ -- $0.77^{+0.96}_{-0.43}$ 
\\
$\eta_{\oplus,\mathrm{GK}}^\mathrm{O}$
& $0.48^{+0.59}_{-0.27}$ -- $0.59^{+0.76}_{-0.33}$ 
& $0.50^{+0.56}_{-0.27}$ -- $0.70^{+0.90}_{-0.39}$ 
& $0.41^{+0.46}_{-0.22}$ -- $0.51^{+0.57}_{-0.27}$ 
& $0.49^{+0.49}_{-0.25}$ -- $0.59^{+0.68}_{-0.32}$ 
\\
$\eta_{\oplus,\mathrm{FGK}}^\mathrm{O}$
& $0.54^{+0.81}_{-0.32}$ -- $0.69^{+1.17}_{-0.42}$ 
& $0.57^{+0.81}_{-0.33}$ -- $0.91^{+1.70}_{-0.55}$ 
& $0.46^{+0.64}_{-0.27}$ -- $0.62^{+0.93}_{-0.35}$ 
& $0.57^{+0.70}_{-0.30}$ -- $0.77^{+1.28}_{-0.45}$ 
\\
\end{tabular}
\tablecomments{The low and high bounds correspond to the constant and zero completeness extrapolation of \S\ref{section:completenessExtrap}. The superscripts C and O on $\eta_\oplus$ refer to the conservative and optimistic habitable zones.  $\eta_{\oplus,\mathrm{GK}}$ is the HZ occurrence for GK ($3900$~K $\leq T_\mathrm{eff} \leq 6000$~K) and $\eta_{\oplus,\mathrm{FGK}}$ is HZ occurrence for FGK ($3900$~K $\leq T_\mathrm{eff} \leq 7300$~K) stars.}
\end{table*}

\subsection{Caveats} \label{section:unresolvedIssues}
While this study takes care to incorporate detection and vetting completeness, and importantly both reliability and observational uncertainty, there are still unresolved issues. We summarize these issues here, each of which can motivate future improvements to our occurrence rate model and methodology.

\textbf{Power Law Assumption}: Products of power laws in radius and period are commonly adopted for planet population models in occurrence rate studies, but there is growing evidence that calls their suitability into question. For instance, improvements to stellar radius measurements have revealed that the radius distribution for small, close-in planets is bi-modal, rather than a smooth or broken power law \citep{Fulton2017}, which has also been observed in K2 data \citep{Hardegree-Ullman2020}. Power laws are not capable of describing such non-monotonic populations.  Looking at the bottom panels of Figure~\ref{figure:occMarg} (without uncertainty), some data points in the radius and instellation flux distributions do not lie along our inferred power laws. However, using input uncertainties (top panels of Figure ~\ref{figure:occMarg}) washes out this structure, making it more difficult to discern a failure or success of a power law model as a descriptor of the data.  There is also strong evidence that populations are not well described by products of power laws in radius and period \citep{petigura2018,Lopez2018} for orbital periods $< 100$ days.  Therefore a product of power laws such as Equation~(\ref{eqn:models}) in radius and instellation flux is unlikely to be a good description of planet populations at high instellation.  At the low instellation of the habitable zone, however, the observed PC population does not indicate any obvious structure (see Figure~\ref{figure:populations}) Given that our domain of analysis is plagued by few detections, low completeness, and low reliability, more observations are likely needed to determine more appropriate population models.  Therefore, because most of our planet population have radii larger than $1.5 R_\oplus$, those larger planets are likely driving the population model, and may be causing bias in the model in the smaller planet regime due to our simple product power laws in Equation~(\ref{eqn:models}). 

\textbf{Planetary Multiplicity}: \citet{zinkChristiansen2019} point out that when short-period planets are detected in the \Kepler\ pipeline, data near their transits are removed for subsequent searches, which can suppress the detection of longer period planets around the same star.  They find that for planets with periods greater than 200 days detection completeness can be suppressed by over 15\% on average.  Our stellar population has removed stars for which more than 30\% of the data has been removed due to transit detection via the {\it dutycycle\_post} stellar property from the DR25 stellar properties table (for details see \citet{Bryson2020}).  We have not attempted to quantify the extent to which this stellar cut mitigates the impact identified in \citet{zinkChristiansen2019}, nor have we accounted for this effect in our analysis.

\textbf{Stellar Multiplicity Contamination}: 
Several authors \citep{Ciardi2015,Furlan2017a,Furlan2017b,Furlan2020} have shown that undetected stellar multiplicity can impact occurrence rate studies in at least two ways.  Stellar multiplicity can reveal planet candidates to be false positives, reducing the planet population, and valid planet candidates in the presence of unknown stellar multiplicity will have incorrect planet radii due to flux dilution.  They estimate that these effects can have an overall impact at the 20\% level.  Stellar multiplicity can also bias the parent stellar sample because unaccounted for flux dilution will bias the completeness estimates.  Our analysis does not take possible stellar multiplicity into account.  However stellar multiplicity has been shown to be associated with poor quality metrics, specifically the BIN flag of \citet{berger18} and the GAIA RUWE metric \citep{gaiaRuwe2018}.  For example, \citet{kraus19} finds that few \Kepler\ target stars with RUWE $> 1.2$ are single stars.   As described in \S\ref{section:stellarPopulation}, we remove stars from our parent stellar population that have been identified as likely binaries in \citet{berger18} or have RUWE $> 1.2$, which is expected to remove many stars with undetected stellar multiplicity \citep[for details see][]{Bryson2020}.  We have not attempted to quantify the impact of undetected stellar multiplicity for our stellar population after this cut. 

\textbf{Planet radius and HZ limits}: 
There are several stellar, planetary and climate models dependent factors that could reduce the occurrence rates calculated in this work. It is quite possible that the uncertainties in stellar radii may alter the planet radii, moving some rocky planets into the mini-Neptune regime of $> 1.5$~$R_\oplus$. Or, it is possible that the upper limit of $1.5$~$R_\oplus$ is an overestimate of the rocky planet limit, and the rocky to gaseous transition may lie lower than 1.5~$R_\oplus$. Although, as pointed out in section \ref{section:habitability},   \cite{Otegi2020} indicate that the rocky regime can extend to as high as 2.5~$R_\oplus$, many of these large-radius regime planets are highly irradiated ones, so they may not be relevant to {\it HZ} rocky planets.

The HZ limits themselves may be uncertain, as they are model and atmospheric composition dependent. Several studies in recent years have calculated HZ limits with various assumptions (see \cite{Kopparapu2019} for review). In particular, the inner edge of the HZ could extend further in, closer to the star, due to slow rotation of the planet \citep{Yang2014b, Kopparapu2016, way2016}, and the outer edge of the HZ may shrink due to `limit cycling', a process where the planet near the outer edge of the HZ around FG stars may undergo cycles of globally glaciated and un-glaciated periods with no long-term stable climate state \citep{Kadoya2014, Kadoya2015, Menou2015, Haqq2016}. Consequently, the number of planets truly in the habitable zone remain uncertain. 

\subsection{Reducing Uncertainties} \label{section:reducingUncertainties}
Our computation of $\eta_\oplus$ has large uncertainties, with the 68\% credible interval spanning factors of 2 (see Tables~\ref{table:occStellarPop}, \ref{table:occStellarPopABC} and \ref{table:hab2AllOcc}).  The 99\% credible intervals in Table~\ref{table:occConfIntervals} span two orders of magnitude.  In \S\ref{section:computingEtaEarth} we discussed how comparing occurrence rates with and without input uncertainties in Tables~\ref{table:occStellarPop} and \ref{table:occStellarPopABC} indicates that these large uncertainties are present before considering the impact of uncertainties in the input data.  We also observed in Table~\ref{table:hab2AllOcc} that the uncertainties are considerably smaller for planets larger than those contributing to our $\eta_\oplus$.  We conclude that, while input uncertainties make a contribution, the dominant cause of our large uncertainties is Poisson uncertainty due to the very small number of habitable zone planets smaller than 1.5~$R_\oplus$ in very low completeness regions of the DR25 planet catalog (see Figure~\ref{figure:populations}).  Our uncertainties may be close to a noise floor induced by the small number of small habitable zone planets resulting from low completeness.

These large Poisson-driven uncertainties are unlikely to be reduced by resolving the issues discussed in \S\ref{section:unresolvedIssues}.  Only by increasing the small planet catalog completeness, resulting in a larger small-planet habitable zone population, can these uncertainties be reduced. 

There are two ways in which a well-characterized catalog with more small planets can be produced:
\begin{itemize}
    \item {\bf Develop improved planet vetting metrics} that produce a catalog that is both more complete and more reliable than the DR25 catalog.  There are several opportunities for such improved metrics, discussed in \citet{Bryson2020}, such as more fully exploiting pixel-level data and existing instrumental flags that can give more accurate reliability characterization than that given using DR25 products.  This approach requires new vetting metrics. \citet{Bryson2020b} has shown that varying the DR25 Robovetter thresholds does not significantly change occurrence rates or their uncertainties once completeness and reliability are taken into account.  In Appendix~\ref{app:robovetterVariations} we show that such changes in Robovetter metrics also do not significantly change the occurrence rates we find in this paper.  
    \item {\bf Obtain more data with a quality similar to \Kepler,} likely through more space-based observations.  In \S\ref{section:introduction} we described how the unrealized \Kepler\ extended mission, doubling the amount of data relative to DR25, was expected to significantly increase the yield of small planets in the habitable zone.  An additional 4 years of data observing the same stars as \Kepler\ with similar photometric precision would be sufficient.  8 years of observation on a different stellar population would also suffice.  As of this writing, plans for space-based missions such as TESS or PLATO do not include such long stares on a single field.  For example, PLATO currently plans no more than 3 years of continuous observation of a single field\footnote{\url{https://www.cosmos.esa.int/web/plato/observation-concept}}.
\end{itemize}

\section{Conclusions}
In this paper we compute the occurrence of rocky ($0.5 \ R_\oplus \leq r \leq 1.5 \ R_\oplus$) planets in the habitable zone for a range of main-sequence dwarf stars from the \Kepler\ DR25 planet candidate catalog and Gaia-based stellar properties.  We base our occurrence rates on differential population models dependent on radius, instellation flux and host star effective temperature (\S\ref{section:models}).  Our computations are corrected for completeness and reliability, making full use of the DR25 data products.  Using instellation flux instead of orbital period allows us to measure the occurrence in the habitable zone even though the habitable zone boundaries depend on stellar effective temperature (\S\ref{section:computeOccurrence}).  Instellation flux also allows us to transfer the unconstrained extrapolation required when extending analysis based on orbital period to a bounded extrapolation of detection completeness (\S\ref{section:completenessExtrap}), and we present our results in terms of these upper and lower bounds (\S\ref{section:results}).  The difference between the upper and lower bounds is smaller than the 68\% credible interval on these bounds.

We compute our occurrence rates using a range of models, stellar populations and computation methods.  We propagate uncertainties in the input data, account for detection completeness that depends on the stellar effective temperature, and check the dependence of our result on the population via a bootstrap study.  In all cases we find consistent results.  We take this as evidence that our occurrence rates are robust.

We find a likely, though not statistically compelling, dependence of our occurrence rates on stellar host effective temperature $T_\mathrm{eff}$ (\S\ref{section:hzOccurrence}, Figure~\ref{figure:occVsTeff}).  Much of this dependence can be understood as due to the habitable zone being larger for hotter stars (\S\ref{section:hzGeom}).  But we find that the $T_\mathrm{eff}$ dependence is weaker than would be expected on purely geometric grounds, implying a decreasing planet occurrence for longer-period orbits.   

Our occurrence rates for rocky planets have large uncertainties.  Comparing computations with and without input uncertainties, we find that these large uncertainties are not caused by the input uncertainties.  Comparing the uncertainties on our rocky planets with the uncertainties on the occurrence of larger planets (Table~\ref{table:hab2AllOcc}), we find that the larger planet occurrence has much lower uncertainty.  We conclude that the large uncertainties are due to the extremely low completeness of the DR25 catalog for small planets in the habitable zone, leading to few planet detections.  The only way we see to reduce these uncertainties is by generating more complete and reliable catalogs, either through improved analysis of existing data or through obtaining more data with quality comparable to \Kepler\ (\S\ref{section:reducingUncertainties}).

Conservative habitability considerations (\S\ref{section:habitability}) and the limited coverage of F stars in \Kepler\ data (\S\ref{figure:populations}) drive us to define $\eta_\oplus$ as the average number of habitable zone planets per star as planets with radii between 0.5~$R_\oplus$ and 1.5~$R_\oplus$ and host star effective temperatures between 4800~K and 6300~K.  Using this definition, we find that, for the conservative habitable zone, $\eta_\oplus$ is between $0.37^{+0.48}_{-0.21}$ and $0.60^{+0.90}_{-0.36}$ planets per star, while for the optimistic HZ $\eta_\oplus$ is between $0.58^{+0.73}_{-0.33}$ and $0.88^{+1.28}_{-0.51}$ planets per star. These occurrence rates imply that conservatively,  to $95\%$ confidence, the nearest rocky HZ planet around G and K-dwarfs is expected to be be within $\sim 6$~pc  (\S\ref{section:etaEarthImplications}). Furthermore, there could, on average, be 4 HZ rocky planets around G \& K dwarfs, respectively, within 10~pc from the Sun.

\acknowledgments
This research has made use of the NASA Exoplanet Archive, which is operated by the California Institute of Technology, under contract with the National Aeronautics and Space Administration under the Exoplanet Exploration Program. We thank our anonymous reviewer for many helpful comments that improved the manuscript.  We thank NASA, \Kepler\ management, and the Exoplanet Exploration Office for continued support of and encouragement for the analysis of \Kepler\ data. R.K.K  acknowledges support from the GSFC Sellers Exoplanet Environments Collaboration (SEEC), which is supported by NASA’s Planetary Science Division’s Research Program, and from NASA’s NExSS Virtual Planetary Laboratory funded by the NASA Astrobiology Program under grant 80NSSC18K0829.  Funding for the Stellar Astrophysics Centre is provided by The Danish National Research Foundation (Grant DNRF106).  V.S.A. acknowledges support from the Independent Research Fund Denmark (Research grant 7027-00096B) and the Carlsberg foundation (grant agreement CF19-0649).  E.B.F. was supported by a grant from the Simons Foundation/SFARI (675601).  E.B.F. acknowledges support from NASA Kepler Participating Scientist Program, grant \# NNX08AR04G, NNX12AF73G, \# and NNX14AN76G.  E.B.F. acknowledges support from the Penn State Eberly College of Science, Department of Astronomy \& Astrophysics, Institute for Computational \& Data Sciences, Center for Exoplanets and Habitable Worlds, and Center for Astrostatistics. E.S.D. acknowledges the National Council for Scientific and Technological Development (CNPQ) and the IFRJ for the financial support.  D.H. acknowledges support from the Alfred P. Sloan Foundation, the National Aeronautics and Space Administration (80NSSC19K0597), and the National Science Foundation (AST-1717000).  S. Mathur acknowledges support from the Spanish Ministry with the Ramon y Cajal fellowship number RYC-2015-17697.

\vspace{5mm}
\facilities{\Kepler}

\software{Python, Jupyter}


\appendixpage
\begin{appendices}
\section{Instellation Flux and Effective Temperature Population Rate Dependence from a Period Power Law} \label{app:InstellationDerivation}
We can qualitatively estimate the instellation flux portion of the differential rate function $\lambda$ by using $\mathrm{d} f / \mathrm{d} I = \left( \mathrm{d} f / \mathrm{d} p \right) \big/ \left( \mathrm{d} I / \mathrm{d} p \right)$.  From the formula for instellation flux and Kepler's third law, we have $I=R_*^2 T^4 \left( M_*^2 p^4 \right)^{-\frac{1}{3}}$, where $M_*$ is the stellar mass in Solar masses, $T = T_\mathrm{eff}/T_\odot$ is the effective temperature divided by the Solar effective temperature, and $p$ is the orbital period in years. Using the mass-radius relation for main-sequence dwarfs, this becomes $I \approx R_*^\mu T^4 p^{-\frac{4}{3}}$, where $\mu = 2 - \frac{2}{3 \xi}$.  When $M_* \leq M_\odot$, $\xi \approx 0.8$ and $\mu \approx 1.17$, while for $M_* > M_\odot$ and $\xi \approx 0.57$ and $\mu \approx 0.8$.  We make the crude ($\approx 20\%$ error) but convenient approximation that $\mu = 1$.  Then using the empirically linear relationship between radius and temperature for the main-sequence dwarfs in our stellar population, $I \approx \left(\tau T + R_0 \right) T^4 p^{-\frac{4}{3}}$ and, assuming $p$ and $T$ are independent, $\mathrm{d} I / \mathrm{d} p \approx -\frac{4}{3} \left(\tau T + R_0 \right) T^4 p^{-\frac{7}{3}}$.

Several studies, such as \citet{Burke2015} and \citet{Bryson2020},  studied planet occurrence in terms of the orbital period $p$ and have shown that $\mathrm{d} f / \mathrm{d} p$ is well-approximated by a power law $F p^\alpha$ (where $F$ is determined by the radius dependence and normalization).  Using this power law and $p \approx \left( \left(\tau T + R_0 \right) T^4 I^{-1} \right)^{\frac{3}{4}}$, we have
\begin{equation} \label{eqn:dfdIa}
\begin{split}
    \mathrm{d} f / \mathrm{d} I 
    & = \frac{\mathrm{d} f / \mathrm{d} p}{\mathrm{d} I / \mathrm{d} p} \\
    & \approx  \frac{3 F p^{\alpha+\frac{7}{3}}}{4\left(\tau T + R_0 \right) T^4} \\
    & \approx  \frac{3 F \left( \left(\tau T + R_0 \right) T^4 I^{-1} \right)^{\frac{3}{4} \left( \alpha+\frac{7}{3} \right)} }{4\left(\tau T + R_0 \right) T^4} \\
    & \approx  C I^\nu \left( \left(\tau T + R_0 \right) T^4 \right)^\delta
\end{split}
\end{equation}
where $\nu = -\frac{3}{4} \left( \alpha - \frac{7}{3} \right)$, $\delta = -\nu - 1$ and $C$ is independent of $I$.  Using the value $\alpha \approx -0.8$ from \citet{Bryson2020}, $\nu \approx -1.15$ and $\delta \approx 0.15$.

\section{Derivation of the Effective Temperature Dependent Likelihood} \label{app:likelihoodDerivation}

Our observed planet population is described by a point process with a instellation flux, radius and effective temperature dependent rate $\lambda(I,r,T)$ and completeness as a function of flux, radius and effective temperature for each star $s$ $\eta_s (I, r, T_s)$.  We assume that the probability that $n_i$ planets occur around an individual star in some region $B_i$ (say a grid cell) of flux-radius space is given by the Poisson probability
\begin{equation*}
    P \{ N \left( B_i \right) = n_i \} = \frac{\left( \Lambda(B_i) \right)^{n_i}}{n_i !} e^{-\Lambda(B_i)}
\end{equation*}
where
\begin{equation*}
    \Lambda(B_i) = \int_{B_i} \eta_s (I, r) \lambda(I, r, T_s) d I \, d r.
\end{equation*}
We do not integrate over $T_s$ because that is fixed to the effective temperature of the star.  We now cover our entire flux-radius range $D$ with a sufficiently fine regular grid with spacing $\Delta p$ and $\Delta r$ so that each grid cell $i$ centered at  flux and radius $(I_i, r_i)$ contains at most one planet.  Then in cell $i$
\begin{equation*}
\begin{split}
&P \{ N \left( B_i \right) = n_i \} \\ 
& \approx 
\begin{cases}
    \eta_s (I_i,r_i) \lambda(I_i,r_i,T_s) \Delta I \Delta r e^{-\Lambda(B_i)} & n_i = 1 \\
    e^{-\Lambda(B_i)} & n_i = 0.
\end{cases}
\end{split}
\end{equation*}
We now ask: what is the probability of a specific number $n_i$ of planets in each cell $i$?  We assume that the probability of a planet in different cells are independent, so 
\begin{equation}
\begin{split}
    &P \{ N \left( B_i \right) = n_i, i=1,\ldots,K\} \\
        &= \prod_{i=1}^K \frac{\left( \Lambda(B_i) \right)^{n_i}}{n_i !} e^{-\Lambda(B_i)} \\
        &\approx \left( \Delta I \Delta r \right)^{K_1} \, e^{-\sum_{i=1}^K \Lambda(B_i)} \\
        & \qquad \times \prod_{i=1}^{K_1} \eta_s (I_i,r_i) \lambda(I_i, r_i,T_s) \\
        &= \left( \Delta I \Delta r \right)^{K_1} \, e^{-\int_{D} \eta_s (I, r) \lambda(I, r,T_s) d I \, d r} \\
        & \qquad \times \prod_{i=1}^{K_1} \eta_s (I_i,r_i) \lambda(I_i, r_i,T_s) \label{equation:poisson1}
\end{split}
\end{equation}
because the $B_i$ cover $D$ and are disjoint.  Here $K$ is the number of grid cells and ${K_1}$ is the number of grid cells that contain a single planet.  So the grid has disappeared, and we only need to evaluate $\lambda(I, r, T_s)$ at the planet locations $(I_i, r_i, T_s)$ and integrate  $\eta_s \lambda$ over the entire domain.

We now consider the probability of detecting planets around a set of $N_*$ stars.  Assuming that the planet detections on different stars are independent of each other, then the joint probability of a specific set of detections specified by the set $\{n_i, i=1,\ldots,N_*\}$ in cell $i$ on on all stars indexed by $s$ is given by 

\begin{equation}
\begin{split}
    &P \{ N_s \left( B_i \right) = n_{s,i}, s=1,\ldots,N_*, i=1,\ldots,K \}  \\
        &= \prod_{s=1}^{N_*} \left( \Delta I \Delta r \right)^{K_1} \, e^{-\int_{D} \eta_s(I, r) \lambda(I, r, T_s) d I \, d r} \\
        & \qquad \times \prod_{i=1}^{K_1} \eta_s(I_i, r_i) \lambda(I_i, r_i, T_s).
        \label{equation:poisson2}
\end{split}
\end{equation}
When $\lambda$ does not depend on effective temperature, we are able to factor $\prod_{s=1}^{N_*} \exp \left[ -\int_{D} \eta_s(I, r) \lambda(I, r) d I \, d r \right]$ as $\exp \left[ -\int_{D} \eta(I, r) \lambda(I, r) d I \, d r \right]$,
where $\eta(I, r) = \sum_{s=1}^{N_*} \eta_s (I, r)$ is the sum of the completeness contours over all stars.  When $\lambda$ depends on effective temperature we partition the stars into effective temperature bins $S_{k}$, and approximate $T_s$ as the average temperature in each bin $\bar{T}_k$, so within each bin $\lambda$ does not depend on the star.  Then we can do the factoring within each bin:
\begin{equation}
\begin{split}
    & \prod_{s=1}^{N_*} e^{-\int_{D} \eta_s(I, r) \lambda(I, r, T_s) d I \, d r} \\
        &\approx \prod_{k} \prod_{s\in S_{k}} e^{-\int_{D} \eta_s(I, r) \lambda(I, r, \bar{T}_k) d I \, d r} \\
        &= \prod_{k}  e^{-\sum_{s\in S_{k}}\int_{D} \eta_s(I, r) \lambda(I, r, \bar{T}_k) d I \, d r} \\
        &= \prod_{k}  e^{-\int_{D} \eta_k(I, r) \lambda(I, r, \bar{T}_k) d I \, d r} \\
        &=   e^{-\sum_{k}\int_{D} \eta_k(I, r) \lambda(I, r, \bar{T}_k) d I \, d r} \\
\end{split}
\end{equation}
where $\eta_k(I, r) = \sum_{s\in S_{k}} \eta_s(I, r)$ is the sum of the completeness contours over the stars in bin $S_{k}$.  Note that we are not integrating over the effective temperature.
Therefore 
\begin{equation}
\begin{split}
    & P \{ N_s \left( B_i \right) = n_{s,i}, s=1,\ldots,N_*, i=1,\ldots,K \} \\
        &= V \, e^{-\sum_{k}\int_{D} \eta_k(I, r) \lambda(I, r, \bar{T}_k) d I \, d r} \\
        & \qquad \times \prod_{s=1}^{N_*} \prod_{i=1}^{K_1} \eta_s(I_i, r_i) \lambda(I_i, r_i, T_s)
        \label{equation:poisson3}
\end{split}
\end{equation}
where $V =  \left( \Delta I \Delta r \right)^{(K_1 N_*)}$.

We now let the rate function $\lambda(I, r, T, \boldsymbol{\theta})$ depend on a parameter vector $\boldsymbol{\theta}$, and consider the problem of finding the $\boldsymbol{\theta}$ that maximizes the likelihood
\begin{equation}
\begin{split}
    & P \{ N_s \left( B_i \right) = n_{s,i}, s=1,\ldots,N_*, i=1,\ldots,K | \boldsymbol{\theta} \} \\
        &= V \, e^{-\sum_{k}\int_{D} \eta_k(I, r) \lambda(I, r, \bar{T}_k, \boldsymbol{\theta}) d I \, d r} \\ 
        & \qquad \times \prod_{s=1}^{N_*}  \prod_{i=1}^{K_1} \eta_s(I_i, r_i) \lambda(I_i, r_i, T_s, \boldsymbol{\theta}) \\ 
        &= V \, \left( \prod_{s=1}^{N_*} \eta_s(I_i, r_i) \right) e^{-\sum_{k}\int_{D} \eta_k(I, r) \lambda(I, r, \bar{T}_k, \boldsymbol{\theta}) d I \, d r} \\
        & \qquad \times \prod_{i=1}^{K_1}  \lambda(I_i, r_i, T_s, \boldsymbol{\theta}). \\ 
        \label{equation:poisson4}
\end{split}
\end{equation}
Because we are maximizing with respect to $\boldsymbol{\theta}$, we can ignore all terms that do not depend on $\boldsymbol{\theta}$.  Therefore, maximizing equation~(\ref{equation:poisson4}) is equivalent to maximizing 
\begin{equation}
\begin{split}
    & P \{ N_s \left( B_i \right) = n_{s,i}, s=1,\ldots,N_*, i=1,\ldots,K | \boldsymbol{\theta} \} \\
    &= e^{-\sum_{k}\int_{D} \eta_k(I, r) \lambda(I, r, \bar{T}_k, \boldsymbol{\theta}) d I \, d r} \prod_{i=1}^{K_1}  \lambda(I_i, r_i, T_s, \boldsymbol{\theta}). \\ 
        \label{equation:poisson5}
\end{split}
\end{equation}

When we neglect the effective temperature dependence of $\lambda$ and have only one effective temperature partition containing all the stars, equation (\ref{equation:poisson5}) reduces to
\begin{equation}
\begin{split}
    & P \{ N_s \left( B_i \right) = n_{s,i}, s=1,\ldots,N_*, i=1,\ldots,K | \boldsymbol{\theta} \} \\
    & = e^{-\int_{D} \eta(I, r) \lambda(I, r, \boldsymbol{\theta}) d I \, d r} \prod_{i=1}^{K_1}  \lambda(I_i, r_i, \boldsymbol{\theta}). \nonumber \\ 
\end{split}
\end{equation}
used in \citet{Bryson2020}.

\section{Planet Candidate Properties} \label{app:pcProperties}
\begin{figure*}[ht]
  \centering
  \includegraphics[width=0.98\linewidth]{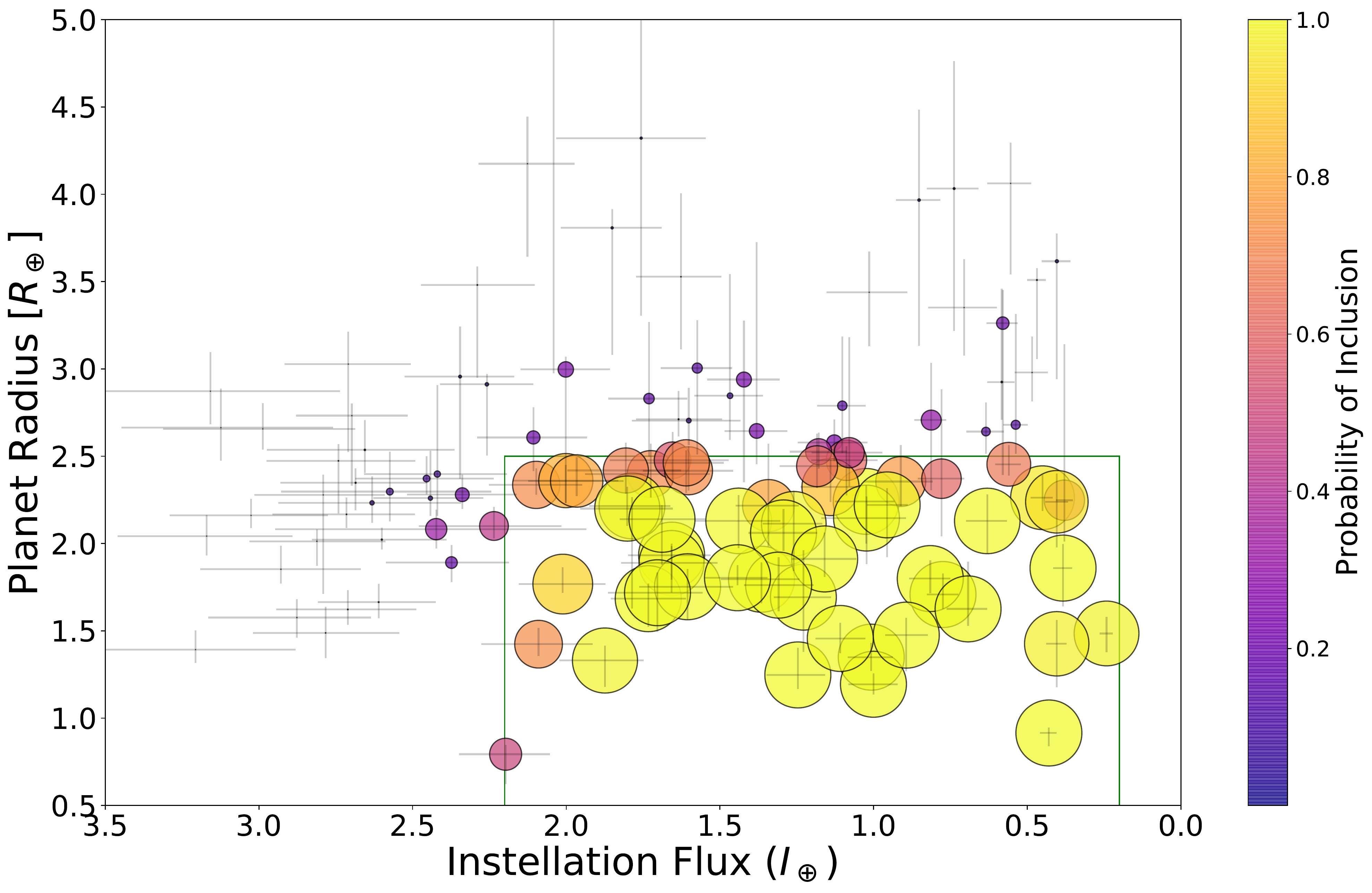}
   \caption{Planet candidates from Table~\ref{table:pcProperties}, sized and colored by their inclusion probability. The green box shows the computational domain $0.5  \ R_\oplus \leq r \leq 2.5 \ R_\oplus$ and $0.2 \ I_\oplus \leq I \leq 2.2 \ I_\oplus$.} \label{figure:probInclusion}
\end{figure*}

Figure~\ref{figure:probInclusion} and Table~\ref{table:pcProperties} give the properties of the DR25 candidates used in our study.  These planet candidates are detected on FGK host stars (of which hab and hab2 are subsets) after the cuts described in \S\ref{section:stellarPopulation}.  The basic PC population is that within our computation domain $0.5  \ R_\oplus \leq r \leq 2.5 \ R_\oplus$ and $0.2 \ I_\oplus \leq I \leq 2.2 \ I_\oplus$, defined by the planet radius and instellation central values.  When accounting for input uncertainties as described in \S\ref{section:inferenceMethods}, some of these planets exit our domain and other planets enter the domain.  In a particular realization, only those planets in the domain are involved in the computation of population models and occurrence rates.  The probability of a PC being in the domain in a particular realization is given by the ``Inclusion Probability'' column of Table~\ref{table:pcProperties}.  We list PCs with an inclusion probability $>1/4000$, which, if reliability = 1, have a 10\% chance of being included in one of the 400 realizations used in the computation with uncertainty.

\clearpage

\renewcommand{\arraystretch}{1}
\startlongtable
\begin{deluxetable*}{ r c c c c c c }
\centering
\tablecaption{Planet Candidate Properties.  Bold-faced KOIs have central values in the computational domain $0.5  \ R_\oplus \leq r \leq 2.5 \ R_\oplus$ and $0.2 \ I_\oplus \leq I \leq 2.2 \ I_\oplus$. \label{table:pcProperties}}
\startdata
\tablehead{KOI & Radius & Period & Instellation & Host Star $T_\mathrm{eff}$ & Reliability & Inclusion Probability \\ & ($R_{\oplus}$) & (Days) & ($I_{\oplus}$) & (K) & & }
{\bf4742.01} & $1.35^{+0.08}_{-0.08}$ & 112.30 & $1.01^{+0.08}_{-0.07}$ & $4602^{+84}_{-76}$ & 0.91 & 1.00000 \\
{\bf8107.01} & $1.19^{+0.06}_{-0.06}$ & 578.89 & $1.00^{+0.08}_{-0.08}$ & $5832^{+102}_{-103}$ & 0.62 & 1.00000 \\
{\bf7016.01} & $1.46^{+0.09}_{-0.08}$ & 384.85 & $1.11^{+0.08}_{-0.08}$ & $5900^{+102}_{-100}$ & 0.68 & 1.00000 \\
{\bf2719.02} & $1.25^{+0.15}_{-0.08}$ & 106.26 & $1.25^{+0.10}_{-0.09}$ & $4601^{+81}_{-76}$ & 0.96 & 1.00000 \\
{\bf701.03} & $1.80^{+0.07}_{-0.04}$ & 122.39 & $1.44^{+0.11}_{-0.10}$ & $4966^{+82}_{-82}$ & 1.00 & 1.00000 \\
{\bf4036.01} & $1.71^{+0.12}_{-0.08}$ & 168.81 & $0.77^{+0.05}_{-0.05}$ & $4697^{+76}_{-68}$ & 1.00 & 1.00000 \\
{\bf2194.03} & $1.80^{+0.10}_{-0.14}$ & 445.22 & $1.36^{+0.13}_{-0.12}$ & $5965^{+122}_{-116}$ & 0.68 & 1.00000 \\
{\bf4087.01} & $1.80^{+0.10}_{-0.08}$ & 101.11 & $0.82^{+0.07}_{-0.07}$ & $4171^{+56}_{-49}$ & 1.00 & 1.00000 \\
{\bf7923.01} & $0.91^{+0.03}_{-0.08}$ & 395.13 & $0.43^{+0.03}_{-0.03}$ & $5064^{+84}_{-73}$ & 0.40 & 1.00000 \\
{\bf8242.01} & $1.48^{+0.10}_{-0.21}$ & 331.55 & $0.89^{+0.07}_{-0.07}$ & $5736^{+105}_{-97}$ & 0.53 & 1.00000 \\
{\bf8047.01} & $1.86^{+0.14}_{-0.22}$ & 302.35 & $0.38^{+0.03}_{-0.03}$ & $4712^{+78}_{-74}$ & 0.72 & 1.00000 \\
{\bf8048.01} & $1.76^{+0.16}_{-0.15}$ & 379.67 & $1.31^{+0.11}_{-0.11}$ & $6058^{+108}_{-107}$ & 0.48 & 1.00000 \\
{\bf7894.01} & $1.91^{+0.15}_{-0.10}$ & 347.98 & $1.16^{+0.11}_{-0.10}$ & $5772^{+108}_{-106}$ & 0.86 & 0.99996 \\
{\bf2184.02} & $1.93^{+0.05}_{-0.21}$ & 95.91 & $1.66^{+0.14}_{-0.13}$ & $4820^{+88}_{-83}$ & 0.97 & 0.99993 \\
{\bf7749.01} & $1.68^{+0.09}_{-0.16}$ & 133.63 & $1.73^{+0.12}_{-0.12}$ & $5098^{+83}_{-78}$ & 0.01 & 0.99993 \\
{\bf7931.01} & $1.75^{+0.10}_{-0.19}$ & 242.04 & $1.61^{+0.16}_{-0.15}$ & $5843^{+106}_{-100}$ & 0.84 & 0.99992 \\
{\bf7915.01} & $2.14^{+0.08}_{-0.26}$ & 382.59 & $1.69^{+0.14}_{-0.14}$ & $6138^{+118}_{-117}$ & 0.47 & 0.99990 \\
{\bf7953.01} & $1.63^{+0.27}_{-0.10}$ & 432.97 & $0.69^{+0.07}_{-0.06}$ & $5421^{+107}_{-95}$ & 0.37 & 0.99940 \\
{\bf4450.01} & $2.06^{+0.14}_{-0.10}$ & 196.44 & $1.29^{+0.11}_{-0.11}$ & $5361^{+91}_{-89}$ & 0.99 & 0.99932 \\
{\bf8246.01} & $1.72^{+0.12}_{-0.22}$ & 425.65 & $1.70^{+0.16}_{-0.15}$ & $6091^{+125}_{-123}$ & 0.36 & 0.99898 \\
{\bf6971.01} & $1.69^{+0.27}_{-0.31}$ & 129.22 & $1.23^{+0.10}_{-0.09}$ & $4921^{+82}_{-83}$ & 0.98 & 0.99857 \\
{\bf87.01} & $2.22^{+0.10}_{-0.30}$ & 289.86 & $0.96^{+0.07}_{-0.07}$ & $5625^{+93}_{-93}$ & 0.96 & 0.99827 \\
{\bf7746.01} & $2.15^{+0.12}_{-0.26}$ & 393.96 & $1.02^{+0.15}_{-0.13}$ & $6135^{+118}_{-114}$ & 0.56 & 0.99799 \\
{\bf2992.01} & $2.24^{+0.09}_{-0.24}$ & 82.66 & $1.02^{+0.10}_{-0.09}$ & $4166^{+68}_{-57}$ & 0.66 & 0.99754 \\
{\bf2931.01} & $2.22^{+0.05}_{-0.59}$ & 99.25 & $1.79^{+0.14}_{-0.13}$ & $4806^{+84}_{-76}$ & 0.99 & 0.99682 \\
{\bf3344.03} & $2.13^{+0.15}_{-0.16}$ & 208.54 & $1.44^{+0.14}_{-0.14}$ & $5495^{+96}_{-95}$ & 0.97 & 0.99409 \\
{\bf8063.01} & $2.13^{+0.15}_{-0.19}$ & 405.35 & $0.63^{+0.07}_{-0.06}$ & $5455^{+103}_{-102}$ & 0.78 & 0.99238 \\
{\bf8159.02} & $2.20^{+0.13}_{-0.10}$ & 353.02 & $1.80^{+0.15}_{-0.15}$ & $6290^{+121}_{-118}$ & 0.84 & 0.98719 \\
{\bf7882.01} & $1.33^{+0.08}_{-0.15}$ & 65.42 & $1.87^{+0.15}_{-0.13}$ & $4390^{+81}_{-74}$ & 0.90 & 0.98612 \\
{\bf3282.01} & $1.89^{+0.11}_{-0.10}$ & 49.28 & $1.66^{+0.16}_{-0.15}$ & $4050^{+64}_{-69}$ & 1.00 & 0.98445 \\
{\bf4622.01} & $1.48^{+0.09}_{-0.11}$ & 207.25 & $0.24^{+0.02}_{-0.02}$ & $4147^{+67}_{-45}$ & 0.98 & 0.98121 \\
{\bf5067.01} & $2.11^{+0.19}_{-0.27}$ & 219.93 & $1.26^{+0.10}_{-0.09}$ & $5526^{+93}_{-87}$ & 0.21 & 0.97935 \\
{\bf571.05} & $1.43^{+0.14}_{-0.25}$ & 129.95 & $0.40^{+0.03}_{-0.03}$ & $4023^{+58}_{-62}$ & 0.92 & 0.97528 \\
{\bf2770.01} & $2.26^{+0.13}_{-0.08}$ & 205.39 & $0.45^{+0.04}_{-0.03}$ & $4475^{+80}_{-75}$ & 0.99 & 0.96615 \\
{\bf8033.01} & $2.24^{+0.16}_{-0.26}$ & 362.13 & $0.40^{+0.04}_{-0.04}$ & $5035^{+90}_{-83}$ & 0.55 & 0.94736 \\
{\bf2290.01} & $1.77^{+0.10}_{-0.05}$ & 91.50 & $2.01^{+0.14}_{-0.14}$ & $4944^{+75}_{-74}$ & 1.00 & 0.90532 \\
{\bf4084.01} & $2.32^{+0.16}_{-0.09}$ & 214.88 & $1.14^{+0.10}_{-0.09}$ & $5288^{+94}_{-89}$ & 0.99 & 0.86571 \\
{\bf250.04} & $2.36^{+0.09}_{-0.13}$ & 46.83 & $2.00^{+0.17}_{-0.17}$ & $4124^{+43}_{-68}$ & 1.00 & 0.82052 \\
{\bf4005.01} & $2.36^{+0.14}_{-0.09}$ & 178.14 & $1.97^{+0.16}_{-0.15}$ & $5545^{+94}_{-94}$ & 0.99 & 0.79403 \\
{\bf4054.01} & $2.22^{+0.35}_{-0.23}$ & 169.14 & $1.34^{+0.11}_{-0.10}$ & $5216^{+91}_{-86}$ & 1.00 & 0.78695 \\
{\bf5276.01} & $2.36^{+0.21}_{-0.14}$ & 220.72 & $0.91^{+0.12}_{-0.10}$ & $5086^{+95}_{-88}$ & 0.96 & 0.75601 \\
{\bf4015.01} & $2.42^{+0.14}_{-0.11}$ & 133.30 & $1.60^{+0.15}_{-0.15}$ & $5051^{+90}_{-85}$ & 1.00 & 0.72676 \\
{\bf2162.02} & $1.42^{+0.09}_{-0.07}$ & 199.67 & $2.09^{+0.19}_{-0.18}$ & $5814^{+116}_{-112}$ & 0.99 & 0.72275 \\
{\bf1989.01} & $2.34^{+0.09}_{-0.06}$ & 201.12 & $2.10^{+0.15}_{-0.15}$ & $5756^{+97}_{-96}$ & 1.00 & 0.71591 \\
{\bf2028.03} & $2.40^{+0.17}_{-0.13}$ & 142.54 & $1.72^{+0.21}_{-0.19}$ & $5213^{+97}_{-91}$ & 1.00 & 0.71120 \\
{\bf5874.01} & $2.46^{+0.07}_{-0.17}$ & 287.33 & $1.61^{+0.13}_{-0.12}$ & $5432^{+109}_{-102}$ & 0.03 & 0.70033 \\
{\bf5433.01} & $2.42^{+0.16}_{-0.13}$ & 237.82 & $1.81^{+0.19}_{-0.18}$ & $5798^{+112}_{-110}$ & 0.97 & 0.68318 \\
{\bf518.03} & $2.45^{+0.11}_{-0.07}$ & 247.35 & $0.56^{+0.04}_{-0.04}$ & $4918^{+90}_{-88}$ & 1.00 & 0.66314 \\
{\bf7345.01} & $2.44^{+0.19}_{-0.13}$ & 377.50 & $1.18^{+0.12}_{-0.11}$ & $5883^{+113}_{-111}$ & 0.88 & 0.62068 \\
{\bf2834.01} & $2.48^{+0.08}_{-0.20}$ & 136.21 & $1.09^{+0.11}_{-0.10}$ & $4775^{+91}_{-83}$ & 1.00 & 0.61320 \\
{\bf8201.01} & $2.25^{+0.89}_{-0.24}$ & 392.60 & $0.38^{+0.03}_{-0.03}$ & $5141^{+91}_{-88}$ & 0.04 & 0.61117 \\
{\bf4745.01} & $2.37^{+0.51}_{-0.33}$ & 177.67 & $0.78^{+0.08}_{-0.07}$ & $4790^{+84}_{-78}$ & 0.99 & 0.59915 \\
{\bf2841.01} & $2.48^{+0.16}_{-0.13}$ & 159.39 & $1.65^{+0.20}_{-0.18}$ & $5397^{+103}_{-100}$ & 0.99 & 0.55438 \\
{\bf7673.01} & $0.79^{+0.05}_{-0.17}$ & 80.77 & $2.20^{+0.15}_{-0.14}$ & $4747^{+78}_{-70}$ & 0.74 & 0.48552 \\
4121.01 & $2.52^{+0.66}_{-0.19}$ & 198.09 & $1.08^{+0.12}_{-0.11}$ & $5237^{+90}_{-86}$ & 0.99 & 0.45819 \\
812.03 & $2.10^{+0.11}_{-0.07}$ & 46.18 & $2.24^{+0.24}_{-0.22}$ & $4293^{+82}_{-90}$ & 1.00 & 0.43624 \\
2757.01 & $2.53^{+0.11}_{-0.10}$ & 234.64 & $1.18^{+0.09}_{-0.09}$ & $5437^{+96}_{-97}$ & 0.96 & 0.39568 \\
238.03 & $2.08^{+0.11}_{-0.11}$ & 362.98 & $2.42^{+0.52}_{-0.49}$ & $6572^{+272}_{-320}$ & 0.90 & 0.32274 \\
4016.01 & $2.71^{+0.33}_{-0.40}$ & 125.41 & $0.81^{+0.06}_{-0.05}$ & $4444^{+78}_{-76}$ & 0.99 & 0.30347 \\
612.03 & $3.00^{+0.07}_{-0.77}$ & 122.08 & $2.00^{+0.15}_{-0.14}$ & $5192^{+94}_{-89}$ & 0.73 & 0.23460 \\
1876.01 & $2.58^{+0.13}_{-0.11}$ & 82.53 & $1.13^{+0.12}_{-0.11}$ & $4269^{+81}_{-76}$ & 1.00 & 0.23173 \\
8156.01 & $2.94^{+0.34}_{-0.59}$ & 364.98 & $1.42^{+0.12}_{-0.12}$ & $6214^{+114}_{-108}$ & 0.41 & 0.22789 \\
1353.03 & $2.64^{+1.08}_{-0.19}$ & 330.07 & $1.38^{+0.10}_{-0.10}$ & $6081^{+102}_{-101}$ & 0.30 & 0.22344 \\
427.03 & $2.28^{+0.11}_{-0.08}$ & 117.03 & $2.34^{+0.18}_{-0.18}$ & $5208^{+90}_{-84}$ & 1.00 & 0.21061 \\
7880.01 & $2.61^{+0.17}_{-0.19}$ & 623.71 & $2.11^{+0.18}_{-0.18}$ & $6753^{+156}_{-139}$ & 0.47 & 0.20068 \\
8238.01 & $3.26^{+0.19}_{-0.87}$ & 495.66 & $0.58^{+0.05}_{-0.05}$ & $5540^{+108}_{-106}$ & 0.74 & 0.18981 \\
4076.01 & $1.89^{+0.10}_{-0.11}$ & 124.83 & $2.37^{+0.21}_{-0.19}$ & $5552^{+94}_{-89}$ & 0.97 & 0.17777 \\
1430.03 & $2.83^{+0.44}_{-0.33}$ & 77.47 & $1.73^{+0.13}_{-0.12}$ & $4543^{+79}_{-75}$ & 1.00 & 0.15944 \\
1871.01 & $3.00^{+0.27}_{-0.49}$ & 92.73 & $1.57^{+0.12}_{-0.11}$ & $4589^{+75}_{-72}$ & 1.00 & 0.15384 \\
2762.01 & $2.79^{+0.40}_{-0.27}$ & 133.00 & $1.10^{+0.08}_{-0.08}$ & $4694^{+80}_{-75}$ & 1.00 & 0.14364 \\
5581.01 & $2.68^{+0.64}_{-0.17}$ & 374.88 & $0.54^{+0.04}_{-0.04}$ & $5311^{+89}_{-89}$ & 0.92 & 0.13855 \\
4356.01 & $2.64^{+0.17}_{-0.13}$ & 174.51 & $0.63^{+0.06}_{-0.06}$ & $4577^{+85}_{-80}$ & 0.99 & 0.13305 \\
7889.01 & $2.30^{+0.23}_{-0.17}$ & 130.24 & $2.57^{+0.35}_{-0.33}$ & $5494^{+105}_{-102}$ & 0.94 & 0.10455 \\
581.02 & $2.37^{+0.13}_{-0.09}$ & 151.86 & $2.45^{+0.23}_{-0.22}$ & $5669^{+100}_{-94}$ & 0.99 & 0.10319 \\
2529.02 & $2.40^{+0.51}_{-0.24}$ & 64.00 & $2.42^{+0.24}_{-0.23}$ & $4607^{+89}_{-84}$ & 0.96 & 0.09608 \\
1596.02 & $2.85^{+0.70}_{-0.25}$ & 105.36 & $1.47^{+0.12}_{-0.11}$ & $4626^{+74}_{-69}$ & 0.79 & 0.08588 \\
3086.01 & $2.70^{+0.19}_{-0.14}$ & 174.73 & $1.60^{+0.19}_{-0.17}$ & $5480^{+104}_{-104}$ & 0.98 & 0.07218 \\
4636.01 & $5.02^{+5964.94}_{-2.05}$ & 122.75 & $2.04^{+0.24}_{-0.22}$ & $5158^{+92}_{-86}$ & 0.01 & 0.07088 \\
4009.01 & $2.23^{+0.15}_{-0.11}$ & 175.14 & $2.63^{+0.31}_{-0.29}$ & $5870^{+120}_{-117}$ & 0.99 & 0.06576 \\
1938.01 & $2.26^{+0.27}_{-0.10}$ & 96.92 & $2.44^{+0.19}_{-0.17}$ & $5086^{+86}_{-82}$ & 1.00 & 0.06496 \\
505.05 & $2.91^{+0.06}_{-0.41}$ & 87.09 & $2.26^{+0.15}_{-0.15}$ & $4868^{+75}_{-70}$ & 1.00 & 0.05449 \\
5622.01 & $3.62^{+0.16}_{-0.68}$ & 469.61 & $0.40^{+0.05}_{-0.04}$ & $5260^{+98}_{-90}$ & 0.83 & 0.04917 \\
4014.01 & $2.96^{+0.29}_{-0.59}$ & 234.24 & $2.35^{+0.18}_{-0.18}$ & $5993^{+103}_{-101}$ & 0.93 & 0.04457 \\
5790.01 & $3.97^{+0.52}_{-0.84}$ & 178.27 & $0.85^{+0.08}_{-0.07}$ & $4797^{+80}_{-75}$ & 0.98 & 0.03965 \\
1527.01 & $3.81^{+0.11}_{-0.73}$ & 192.67 & $1.85^{+0.17}_{-0.16}$ & $5603^{+105}_{-105}$ & 0.82 & 0.03565 \\
1707.02 & $4.32^{+11.24}_{-1.02}$ & 265.48 & $1.76^{+0.28}_{-0.21}$ & $5766^{+142}_{-130}$ & 0.50 & 0.03458 \\
8193.01 & $4.03^{+0.73}_{-0.82}$ & 367.95 & $0.74^{+0.09}_{-0.08}$ & $5546^{+95}_{-92}$ & 0.36 & 0.03023 \\
2210.02 & $2.92^{+0.54}_{-0.22}$ & 210.63 & $0.58^{+0.05}_{-0.04}$ & $4779^{+80}_{-78}$ & 1.00 & 0.02503 \\
4202.01 & $2.54^{+0.17}_{-0.13}$ & 153.98 & $2.66^{+0.32}_{-0.29}$ & $5741^{+110}_{-104}$ & 1.00 & 0.02312 \\
947.01 & $2.02^{+0.07}_{-0.06}$ & 28.60 & $2.60^{+0.23}_{-0.21}$ & $3926^{+60}_{-61}$ & 1.00 & 0.01918 \\
2828.01 & $2.35^{+0.08}_{-0.16}$ & 59.50 & $2.69^{+0.24}_{-0.23}$ & $4629^{+88}_{-80}$ & 1.00 & 0.01785 \\
4051.01 & $2.71^{+0.16}_{-0.10}$ & 163.69 & $1.63^{+0.14}_{-0.14}$ & $5351^{+100}_{-98}$ & 0.97 & 0.01523 \\
4242.01 & $1.66^{+0.10}_{-0.09}$ & 145.79 & $2.61^{+0.20}_{-0.19}$ & $5725^{+89}_{-92}$ & 0.90 & 0.01379 \\
2686.01 & $3.51^{+0.07}_{-0.45}$ & 211.03 & $0.47^{+0.03}_{-0.03}$ & $4475^{+73}_{-69}$ & 1.00 & 0.01289 \\
3508.01 & $1.62^{+0.11}_{-0.09}$ & 190.80 & $2.71^{+0.23}_{-0.22}$ & $6067^{+106}_{-106}$ & 0.98 & 0.01120 \\
172.02 & $2.17^{+0.10}_{-0.08}$ & 242.47 & $2.71^{+0.24}_{-0.22}$ & $5890^{+118}_{-114}$ & 0.98 & 0.01037 \\
8276.01 & $3.48^{+0.11}_{-0.53}$ & 385.86 & $2.29^{+0.18}_{-0.19}$ & $6618^{+128}_{-123}$ & 0.63 & 0.01028 \\
2172.02 & $2.47^{+0.10}_{-0.30}$ & 116.58 & $2.74^{+0.23}_{-0.25}$ & $5420^{+97}_{-93}$ & 0.98 & 0.00927 \\
4926.01 & $1.49^{+0.15}_{-0.14}$ & 69.09 & $2.78^{+0.24}_{-0.24}$ & $4831^{+92}_{-86}$ & 0.39 & 0.00750 \\
1986.01 & $3.53^{+0.48}_{-0.42}$ & 148.46 & $1.63^{+0.15}_{-0.13}$ & $5228^{+107}_{-100}$ & 0.99 & 0.00673 \\
1608.03 & $2.01^{+0.07}_{-0.14}$ & 232.04 & $2.81^{+0.22}_{-0.22}$ & $6128^{+111}_{-111}$ & 0.89 & 0.00310 \\
2525.01 & $1.85^{+0.13}_{-0.08}$ & 57.29 & $2.93^{+0.26}_{-0.26}$ & $4617^{+87}_{-79}$ & 1.00 & 0.00253 \\
8249.01 & $1.58^{+0.11}_{-0.12}$ & 309.19 & $2.88^{+0.29}_{-0.24}$ & $6153^{+129}_{-113}$ & 0.64 & 0.00249 \\
4385.02 & $2.98^{+0.21}_{-0.16}$ & 386.37 & $0.48^{+0.06}_{-0.05}$ & $5215^{+99}_{-93}$ & 0.90 & 0.00180 \\
3266.01 & $2.28^{+0.12}_{-0.57}$ & 54.51 & $2.79^{+0.22}_{-0.20}$ & $4459^{+75}_{-70}$ & 1.00 & 0.00158 \\
8275.01 & $4.06^{+0.23}_{-0.52}$ & 389.88 & $0.55^{+0.08}_{-0.07}$ & $5370^{+113}_{-109}$ & 0.15 & 0.00138 \\
7982.01 & $3.44^{+0.23}_{-0.31}$ & 376.38 & $1.01^{+0.14}_{-0.12}$ & $5814^{+107}_{-103}$ & 0.56 & 0.00119 \\
7798.01 & $2.66^{+0.22}_{-0.19}$ & 309.89 & $3.12^{+0.32}_{-0.37}$ & $6258^{+133}_{-127}$ & 0.54 & 0.00110 \\
6786.01 & $3.35^{+0.28}_{-0.28}$ & 455.62 & $0.71^{+0.12}_{-0.11}$ & $5622^{+118}_{-115}$ & 0.80 & 0.00099 \\
2650.01 & $1.39^{+0.11}_{-0.08}$ & 34.99 & $3.21^{+0.34}_{-0.33}$ & $4096^{+69}_{-80}$ & 1.00 & 0.00099 \\
416.02 & $3.03^{+0.19}_{-0.50}$ & 88.26 & $2.71^{+0.21}_{-0.20}$ & $5083^{+91}_{-85}$ & 1.00 & 0.00091 \\
1980.01 & $2.73^{+0.07}_{-0.40}$ & 122.88 & $2.70^{+0.18}_{-0.18}$ & $5441^{+85}_{-85}$ & 1.00 & 0.00087 \\
401.02 & $4.17^{+0.27}_{-0.53}$ & 160.02 & $2.13^{+0.16}_{-0.15}$ & $5516^{+90}_{-87}$ & 0.95 & 0.00057 \\
1078.03 & $2.16^{+0.09}_{-0.07}$ & 28.46 & $3.03^{+0.27}_{-0.25}$ & $4015^{+58}_{-62}$ & 1.00 & 0.00047 \\
1970.02 & $2.66^{+0.15}_{-0.12}$ & 125.60 & $2.99^{+0.32}_{-0.30}$ & $5585^{+98}_{-96}$ & 1.00 & 0.00040 \\
4856.01 & $2.87^{+0.22}_{-0.19}$ & 147.39 & $3.16^{+0.47}_{-0.42}$ & $5773^{+108}_{-104}$ & 1.00 & 0.00029 \\
775.03 & $2.04^{+0.12}_{-0.11}$ & 36.45 & $3.17^{+0.29}_{-0.28}$ & $4164^{+49}_{-62}$ & 1.00 & 0.00026 \\
\enddata
\end{deluxetable*}

\clearpage

\section{Robovetter Variations} \label{app:robovetterVariations}
\begin{figure*}[ht]
  \centering
  \Large No Reliability Correction \hspace{1 in} Corrected for Reliability\\
  \includegraphics[width=0.48\linewidth]{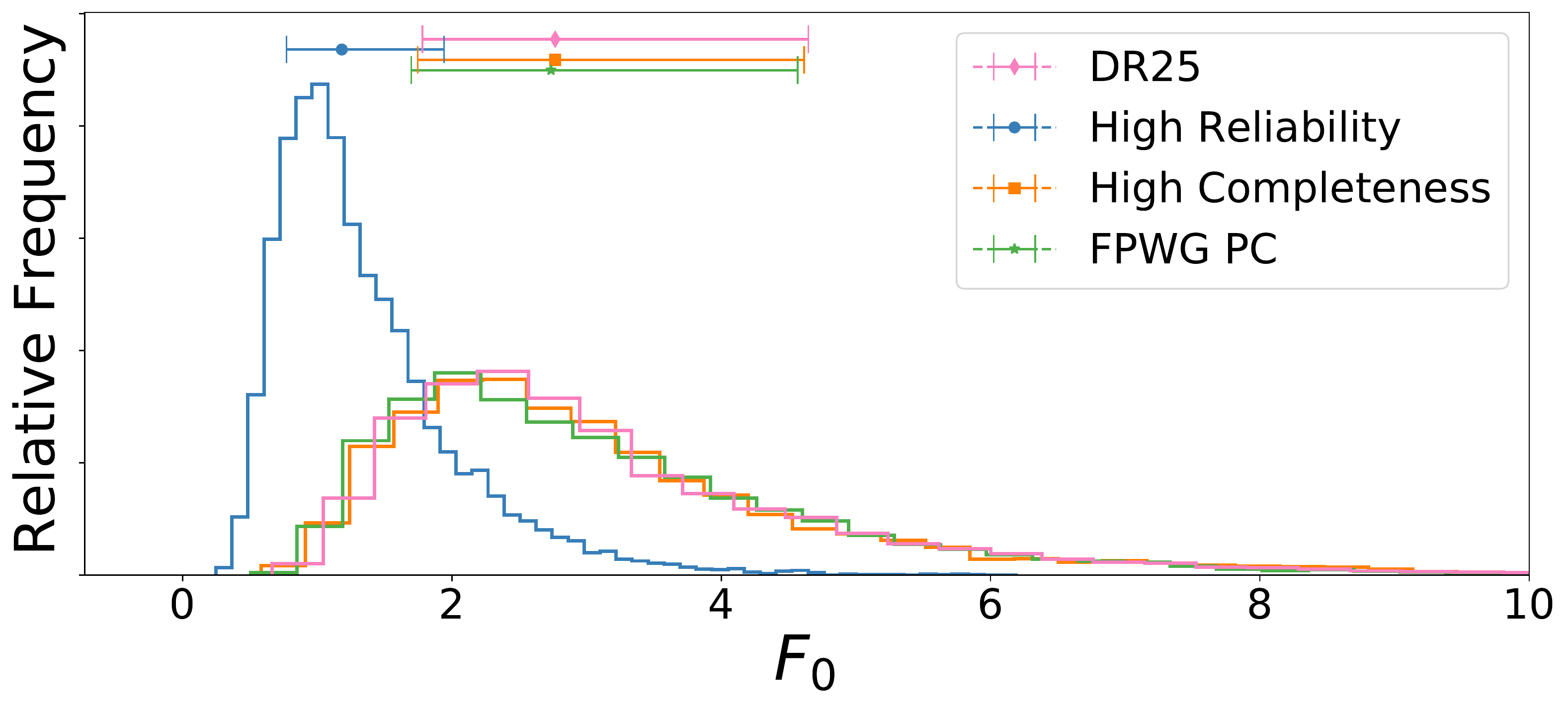}
  \includegraphics[width=0.48\linewidth]{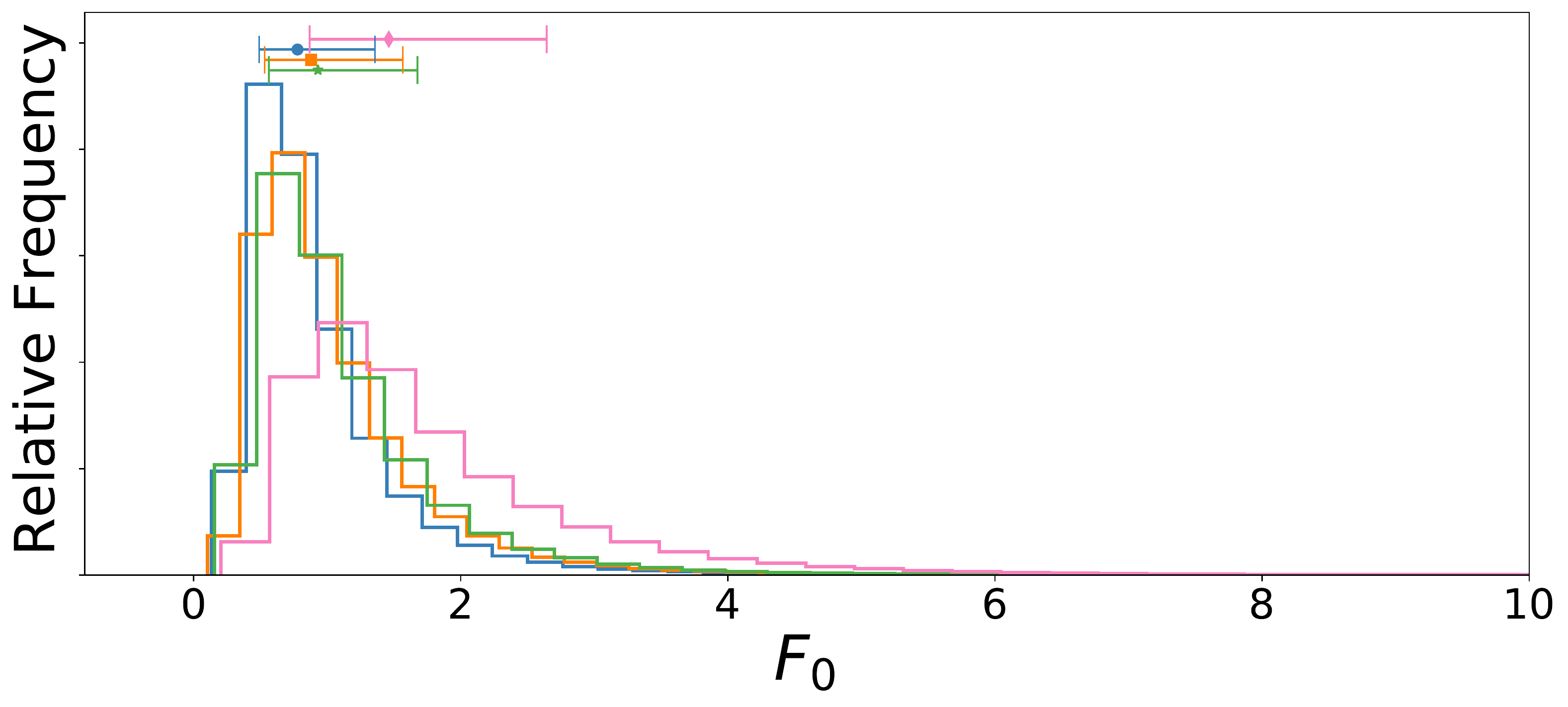} \\
   \caption{Distributions of the parameter $F_0$ in model 1 (see Equation (\ref{eqn:models})), the occurrence for the hab2 stellar population for $0.5  \ R_\oplus \leq r \leq 2.5 \ R_\oplus$ and instellation flux range $0.2 \leq I \ I_\oplus \leq 2.2 \ I_\oplus$ for the  high reliability (blue), DR25 (pink), FPWG PC (green) and high completeness (orange) vetting, computed with the Poisson method.  Left: without correcting for reliability.  Right: corrected for reliability.} \label{figure:rvVariations}
\end{figure*}

\citet{Bryson2020b} provides alternative planet candidate catalogs based on the \Kepler\ data, created by changing automated vetting thresholds.  They argue that occurrence rate estimates should roughly agree between these alternative catalogs.  
Figure~\ref{figure:rvVariations} shows the distribution of the model parameter $F_0$ for model 1 for these catalogs, computed without input uncertainties and zero completeness extrapolation with the hab2 stellar population.  We see reasonable agreement between the Robovetter variations when reliability corrections are applied.

\renewcommand{\arraystretch}{1.25}
\begin{table*}[ht]
\centering
\caption{Fit coefficients for the alternative planet candidate catalogs using the hab2 population with zero completeness extrapolation}\label{table:fitResults}
\begin{tabular}{ r r r r r c}
\hline
\hline
 & DR25 & High Reliability & High Completeness & FPWG PC & Max Separation $(\sigma)$ \\
\hline
& \multicolumn{5}{c}{With Reliability Correction} \\
$F_0$
& $1.46^{+1.18}_{-0.59}$
& $0.78^{+0.58}_{-0.29}$
& $0.88^{+0.69}_{-0.35}$
& $0.93^{+0.74}_{-0.37}$
& 0.82
\\
$\alpha$
& $-1.03^{+0.83}_{-0.77}$
& $0.10^{+1.08}_{-0.98}$
& $-0.65^{+0.97}_{-0.87}$
& $-0.66^{+0.94}_{-0.86}$
& 0.88
\\
$\beta$
& $-1.15^{+0.34}_{-0.33}$
& $-1.17^{+0.36}_{-0.34}$
& $-0.95^{+0.38}_{-0.35}$
& $-1.00^{+0.37}_{-0.35}$
& 0.44
\\
$\gamma$
& $-1.03^{+1.66}_{-1.64}$
& $-2.12^{+1.77}_{-1.77}$
& $-2.83^{+1.73}_{-1.68}$
& $-2.42^{+1.72}_{-1.73}$
& 0.76
\\
& \multicolumn{5}{c}{No Reliability Correction} \\
$F_0$
& $2.77^{+1.88}_{-0.99}$
& $1.18^{+0.76}_{-0.41}$
& $2.77^{+1.85}_{-1.02}$
& $2.74^{+1.83}_{-1.04}$
& 1.27
\\
$\alpha$
& $-1.35^{+0.59}_{-0.63}$
& $-0.36^{+0.79}_{-0.80}$
& $-1.39^{+0.62}_{-0.62}$
& $-1.36^{+0.64}_{-0.62}$
& 1.01
\\
$\beta$
& $-1.39^{+0.28}_{-0.28}$
& $-1.26^{+0.32}_{-0.30}$
& $-1.38^{+0.28}_{-0.27}$
& $-1.39^{+0.28}_{-0.28}$
& 0.32
\\
$\gamma$
& $0.54^{+1.43}_{-1.34}$
& $-1.18^{+1.56}_{-1.55}$
& $0.44^{+1.38}_{-1.34}$
& $0.46^{+1.45}_{-1.46}$
& 0.84
\\
\end{tabular}
\tablecomments{The posteriors of model 1 for the DR25, high reliability, high completeness, and FPWG PC catalogs from \citet{Bryson2020b} using the hab2 stellar population and the Poisson likelihood method with zero completeness extrapolation. The maximum separation in each row is the maximum over each row of the difference in medians divided by the propagated uncertainty of that distance.
}
\end{table*}

\end{appendices}

\clearpage

\bibliography{refs}


\end{document}